\documentclass[
 superscriptaddress,
 onecolumn
]{revtex4-2}

\usepackage{amsmath, amssymb, amsthm, wasysym}
\usepackage{graphicx}
\usepackage{dcolumn}
\usepackage{bm}
\usepackage{xcolor}
\usepackage{relsize}
\usepackage{accents}

\usepackage[english]{babel}
\usepackage[utf8]{inputenc}
\usepackage[colorinlistoftodos, color=green!40, prependcaption]{todonotes}
\usepackage{amsthm}
\usepackage{mathtools}
\usepackage{physics}
\usepackage{xcolor}
\usepackage{graphicx}
\usepackage[left=23mm,right=13mm,top=35mm,columnsep=15pt]{geometry} 
\usepackage{adjustbox}
\usepackage{placeins}
\usepackage[T1]{fontenc}
\usepackage{lipsum}
\usepackage{csquotes}
\usepackage{amsmath}
\usepackage{amsfonts}
\usepackage{cases}
\usepackage{xspace}
\usepackage{psfrag}
\usepackage{color}
\usepackage{xcolor}

\renewcommand{\bra}[1]{\left\langle#1\right|}

\renewcommand{\ket}[1]{\left|#1\right\rangle}

\newcommand{\ave}[1]{\mathchoice{\left\langle #1 \right\rangle}{\langle #1 \rangle}{\langle #1 \rangle}{\langle #1 \rangle}}

\newcommand{\imag}{\mathring{\imath}}

\newcommand{\union}{\cup}
\newcommand{\intersect}{\cap}

\newcommand{\plaind}{\mathrm{d}}
\newcommand{\dint}[1]{\mathchoice{\!\plaind#1\,}{\!\plaind#1\,}{\!\plaind#1\,}{\!\plaind#1\,}}
\newcommand{\ddint}[1]{\ddintx{#1}{d}}
\newcommand{\ddintx}[2]{\mathchoice{\!\plaind^{#2}#1\,}{\!\plaind^{#2}#1\,}{\!\plaind^{#2}#1\,}{\!\plaind^{#2}#1\,}}

\newcommand{\dbar}{\plaind\mkern-7mu\mathchar'26}
\newcommand{\deltabar}{\delta\mkern-8mu\mathchar'26}
\newcommand{\dintbar}[1]{\mathchoice{\!\dbar#1\,}{\!\dbar#1\,}{\!\dbar#1\,}{\!\dbar#1\,}}
\newcommand{\ddintbar}[1]{\mathchoice{\!\dbar^d#1\,}{\!\dbar^d#1\,}{\!\dbar^d#1\,}{\!\dbar^d#1\,}}

\newcommand{\Dint}[1]{\mathcal{D}\!#1\,}

\DeclareMathOperator*{\SumInt}{%
\mathchoice%
  {\ooalign{$\displaystyle\sum$\cr\hidewidth$\displaystyle\int$\hidewidth\cr}}
  {\ooalign{\raisebox{.14\height}{\scalebox{.7}{$\textstyle\sum$}}\cr\hidewidth$\textstyle\int$\hidewidth\cr}}
  {\ooalign{\raisebox{.2\height}{\scalebox{.6}{$\scriptstyle\sum$}}\cr$\scriptstyle\int$\cr}}
  {\ooalign{\raisebox{.2\height}{\scalebox{.6}{$\scriptstyle\sum$}}\cr$\scriptstyle\int$\cr}}
}

\usepackage{dsfont}

\newcommand{\gpset}[1]{\mathds{#1}}

\newcommand{\canetset}[1]{{\mathchoice {\hbox{$\sf\textstyle #1\kern-0.4em #1$}}
{\hbox{$\sf\textstyle #1\kern-0.4em #1$}}
{\hbox{$\sf\scriptstyle #1\kern-0.3em #1$}}
{\hbox{$\sf\scriptscriptstyle #1\kern-0.2em #1$}}}}

\newcommand{\Zset}{\gpset{Z}}

\newcommand{\Rset}{\gpset{R}}
\newcommand{\Cset}{\gpset{C}}

\def\nbZ{{\mathchoice {\hbox{$\sf\textstyle Z\kern-0.4em Z$}}
{\hbox{$\sf\textstyle Z\kern-0.4em Z$}}
{\hbox{$\sf\scriptstyle Z\kern-0.3em Z$}}
{\hbox{$\sf\scriptscriptstyle Z\kern-0.2em Z$}}}}

\newcommand{\gpvec}[1]{\mathbf{#1}}

\newcommand{\kvec}{\gpvec{k}}

\newcommand{\xvec}{\gpvec{x}}
\newcommand{\yvec}{\gpvec{y}}

\newcommand{\OC}{\mathcal{O}}

\newcommand{\psitilde}{\tilde{\psi}}

\newcommand{\psidagger}{\psi^{\dagger}}

\newcommand{\Exp}[1]{\operatorname{exp}\left(#1\right)}
\newcommand{\ExpNB}[1]{\operatorname{exp}}
\renewcommand{\exp}[1]{\mathchoice{\mathrm{e}^{#1}}{\operatorname{exp}\left(#1\right)}{\operatorname{exp}\left(#1\right)}{\operatorname{exp}\left(#1\right)}}

\newcommand{\elabel}[1]{\label{eq:#1}}
\newcommand{\eref}[1]{(\ref{eq:#1})}
\newcommand{\Eref}[1]{Eq.~(\ref{eq:#1})}
\newcommand{\Erefs}[1]{Eqs.~(\ref{eq:#1})}

\newcommand{\seclabel}[1]{\label{sec:#1}}

\newcommand{\Sref}[1]{Section~\ref{sec:#1}}
\newcommand{\Srefs}[1]{Sections~\ref{sec:#1}}

\newcommand{\latin}[1]{{\it #1}}
\newcommand{\ie}{\latin{i.e.}\@\xspace}

\newcommand{\cf}{\latin{cf.}\@\xspace}

\newlength \standardfigwidth
\DeclareMathAlphabet{\matheub}{U}{eur}{m}{n}

\newcounter{exercise}
{\addtocounter{exercise}{1}\begin{center}\begin{minipage}{0.8\linewidth}\textbf{Exercise
\arabic{exercise}:}\begin{itshape}}
{\end{itshape}\end{minipage}\end{center}}

\makeatletter
\newcommand{\creat}[3][]{\@ifempty{#1}{#2^{\dagger}}{\left(#2^{\dagger}\right)^{#1}}\@ifempty{#3}{}{\!(#3)}}

\newcommand{\creatDoi}[3][]{\@ifempty{#1}{\tilde{#2}}{\left(\tilde{#2}\right)^{#1}}\@ifempty{#3}{}{(#3)}}

\newcommand{\annih}[3][]{#2\@ifempty{#1}{}{^{#1}}\@ifempty{#3}{}{(#3)}}

\makeatother

% ...

\newlength{\bibmarkkeyAleft}

\newlength{\bibmarkkeyBleft}

\newlength{\bibmarkkeyCleft}

\newlength{\bibmarkkeyDleft}

\usepackage{wasysym}
\newcommand{\abyss}{\sun}
\newcommand{\stirling}[2]{\genfrac\{\}{0pt}{}{#1}{#2}}
\usepackage[colorlinks=true,linkcolor=black, urlcolor = black, citecolor = blue, anchorcolor = blue]{hyperref}

\usepackage{graphicx}
\usepackage[tight]{subfigure}
\usepackage{multirow}
\usepackage{subeqnarray}
\usepackage{wasysym}
\usepackage{eufrak}
\usepackage{amsopn}

\usepackage{amssymb}

\usepackage{framed}

\usepackage{vmargin}
\usepackage{epsfig}
\usepackage{color}
\usepackage{alltt}
\usepackage{pstricks,pst-node,pst-tree,pst-grad,pst-text,graphics}
\usepackage{dcolumn}
\usepackage{url}

\usepackage{cases}

\usepackage{ifthen}

\usepackage{afterpage}

\usepackage{comment}

\usepackage{supertabular}

\usepackage{xspace}

\usepackage{setspace}

\usepackage[bf,footnotesize]{caption}

\usepackage{textcomp}

\usepackage{tikz}
\usetikzlibrary{decorations.pathmorphing}
\usetikzlibrary{decorations.markings}
\usetikzlibrary{arrows,shapes,snakes,automata,backgrounds,petri}
\usetikzlibrary{calc}
\usetikzlibrary{patterns}
\usetikzlibrary{decorations.text}
\tikzset{
xxtsubstrate/.style={decorate, 
line width=1pt,
draw=olive, 
decoration=snake, 
segment amplitude=0.75mm, 
line after snake=0.25mm,
line before snake=0.25mm
},
tsubstrate/.style={decorate, 
line width=1pt,
draw=olive, 
decoration=snake, 
segment amplitude=0.5mm, 
segment length=5pt,
segment amplitude=0.2mm, 
line after snake=1mm,
line before snake=1mm
},
Bsubstrate/.style={decorate, 
line width=1pt,
draw=orange, 
decoration=snake,
segment length=5pt,
segment aspect=0,
segment amplitude=0.5mm, 
line after snake=0mm,
line before snake=0mm
},
substrate/.style={decorate, 
line width=1pt,
draw=orange,
decoration=snake, 
segment length=5pt,
segment amplitude=0.5mm, 
line after snake=0.5mm,
line before snake=0.5mm
},
activity/.style={very thick,draw=red,postaction={decorate},
decoration={markings,mark=at position .5 with
{\arrow[draw=red]{>}}}},
tactivity/.style={thick,draw=red,postaction={decorate},
decoration={markings,mark=at position .5 with
{\arrow[draw=red]{>}}}},
tEPSactivity/.style={thick,draw=red,postaction={decorate},
decoration={markings,mark=at position .55 with
{\arrow[draw=red]{>}}}},
tAactivity/.style={thick,draw=red},
Aactivity/.style={thick,draw=black},
tSactivity/.style={thick,draw=red,postaction={decorate},
decoration={markings,mark=at position .7 with
{\arrow[draw=red]{>}}}},
Sactivity/.style={very thick,draw=red,postaction={decorate},
decoration={markings,mark=at position .7 with
{\arrow[draw=red]{>}}}},
polarity/.style={decorate, 
line width=1pt,
draw=black,
decoration={markings,mark=between positions 0 and 1 step 1.6mm with {\draw[black,thick,fill=black]  (0,0) circle (0.02)} },
segment length=5pt,
segment amplitude=0.5mm, 
},
Bpolarity/.style={decorate, 
line width=1.5pt,
draw=red,
segment length=5pt,
segment amplitude=0.5mm, 
},
density/.style={ 
line width=1.5pt,
draw=magenta,
densely dashed,
segment length=5pt,
segment amplitude=0.5mm, 
},
}

\pgfdeclarepatternformonly{densedots}{\pgfqpoint{-1pt}{-1pt}}{\pgfqpoint{1pt}{1pt}}{\pgfqpoint{2pt}{2pt}}
{%
  \pgfpathcircle{\pgfqpoint{0pt}{0pt}}{.5pt}%
  \pgfusepath{fill}%
}%

\newcommand{\fourPointConnectedA}[4]{
\tikz[baseline=-2.5pt]{
      \draw[Aactivity] (0,0.0) -- (0.25,0.0) +(2pt,0) circle (2pt)[fill];
      \begin{scope}[rotate=0]
      \draw[Aactivity] (220:0.7) -- (0,0.0);
      \draw[dotted,thin] (140:0.25) arc (140:220:0.25); 
      \draw[Aactivity] ($(-0.07,0.07)+(220:0.2)$) -- ($(0.07,-0.07)+(220:0.2)$);
      \end{scope}
      \draw[Aactivity] (140:0.7) -- (0,0.0);
      \draw[Aactivity] ($(0.07,0.07)+(140:0.2)$) -- ($(-0.07,-0.07)+(140:0.2)$);
      \begin{scope}[xshift=-0.54cm,yshift=0.45cm]
      \draw[Aactivity] (220:0.6) -- (0,0.0);
      \draw[Aactivity] (140:0.6) -- (0,0.0);
      \draw[dotted,thin] (140:0.25) arc (140:220:0.25);
      \draw[Aactivity] ($(0.07,0.07)+(140:0.2)$) -- ($(-0.07,-0.07)+(140:0.2)$);
      \draw[Aactivity] ($(-0.07,0.07)+(220:0.2)$) -- ($(0.07,-0.07)+(220:0.2)$);
     \end{scope}
     \begin{scope}[xshift=-0.54cm,yshift=-0.45cm]
      \draw[Aactivity] (220:0.6) -- (0,0.0);
      \draw[Aactivity] (140:0.6) -- (0,0.0);
      \draw[dotted,thin] (140:0.25) arc (140:220:0.25);
      \draw[Aactivity] ($(0.07,0.07)+(140:0.2)$) -- ($(-0.07,-0.07)+(140:0.2)$);
      \draw[Aactivity] ($(-0.07,0.07)+(220:0.2)$) -- ($(0.07,-0.07)+(220:0.2)$);
     \end{scope}
    \node [yshift=0.8cm,xshift=-1.2cm] {$#1$};
    \node [yshift=0.25cm,xshift=-1.2cm] {$#2$};
    \node [yshift=-0.15cm,xshift=-1.2cm] {$#3$};
    \node [yshift=-0.8cm,xshift=-1.2cm] {$#4$};
    }
}

\newcommand{\fourPointConnectedB}[4]{
\tikz[baseline=-2.5pt]{
      \begin{scope}[rotate=0]
      \draw[Aactivity] (220:1) -- (0,0.0);
      \draw[Aactivity] ($(-0.07,0.07)+(220:0.2)$) -- ($(0.07,-0.07)+(220:0.2)$);
      \end{scope}
      \draw[Aactivity] (140:0.5) -- (0,0.0);
      \draw[Aactivity] ($(0.07,0.07)+(140:0.2)$) -- ($(-0.07,-0.07)+(140:0.2)$);
      \draw[dotted,thin] (140:0.25) arc (140:220:0.25);
      \begin{scope}[xshift=-0.35cm,yshift=0.3cm]
      \draw[Aactivity] (220:0.6) -- (0,0.0);
      \draw[Aactivity] (140:0.6) -- (0,0.0);
      \draw[Aactivity] ($(0.07,0.07)+(140:0.2)$) -- ($(-0.07,-0.07)+(140:0.2)$);
      \draw[Aactivity] ($(-0.07,0.07)+(220:0.2)$) -- ($(0.07,-0.07)+(220:0.2)$);
      \draw[dotted,thin] (140:0.25) arc (140:220:0.25);
     \end{scope}
     \begin{scope}[xshift=0.45cm,yshift=-0.38cm]
      \draw[Aactivity] (220:0.8) -- (0,0.0);
      \draw[Aactivity] (140:0.6) -- (0,0.0);
      \draw[Aactivity] (0,0.0) -- (0.4,0.0) +(2pt,0) circle (2pt)[fill];
      \draw[Aactivity] ($(0.07,0.07)+(140:0.2)$) -- ($(-0.07,-0.07)+(140:0.2)$);
      \draw[Aactivity] ($(-0.07,0.07)+(220:0.2)$) -- ($(0.07,-0.07)+(220:0.2)$);
      \draw[dotted,thin] (140:0.25) arc (140:220:0.25);
     \end{scope}
    \node [yshift=0.8cm,xshift=-1.cm] {$#1$};
    \node [yshift=0.1cm,xshift=-1.cm] {$#2$};
    \node [yshift=-0.5cm,xshift=-1.cm] {$#3$};
    \node [yshift=-0.9cm,xshift=-0.4cm] {$#4$};
    }
}
\newcommand{\partitionX}[2]{\mathcal{P}\left(#1,#2\right)}
\newcommand{\partition}[2]{\partitionX{\Pset{#1}}{#2}}
\newcommand{\Pset}[1]{\mathbb{#1}} 
\newcommand{\cardinalityX}[1]{|#1|} 
\newcommand{\setvecX}[1]{\mathbf{K}\!\left(#1\right)}
\newcommand{\cardinality}[1]{\cardinalityX{\Pset{#1}}}
\newcommand{\setvec}[1]{\setvecX{\Pset{#1}}}
\newcommand{\fguts}{\mathcal{F}} 
\newcommand{\KBbar}{\bar{\mathbf{K}}}
\newcommand{\knullfix}[1]{\exp{-\imag(#1)\cdot\xvec_0}}

\begin{document}

\title{Particle Entity in the Doi-Peliti and Response Field Formalisms}

\author{Marius Bothe}
    \affiliation{Department of Mathematics, Imperial College London, London SW7 2AZ}
\author{Luca Cocconi}
    \affiliation{Department of Mathematics, Imperial College London, London SW7 2AZ}
    \affiliation{Department of Genetics and Evolution, University of Geneva, 1205 Geneva}
    \affiliation{The Francis Crick Institute, NW1 1AT London}
\author{Zigan Zhen}
    \affiliation{Department of Mathematics, Imperial College London, London SW7 2AZ}
\author{Gunnar Pruessner}
    \affiliation{Department of Mathematics, Imperial College London, London SW7 2AZ}
    \email[Correspondence email address: ]{g.pruessner@imperial.ac.uk}
\date{\today}

\begin{abstract}
We introduce a procedure to test a theory for point particle entity, that is, whether said theory takes into account the discrete nature of the constituents of the system. We then identify the mechanism whereby particle entity is enforced in the context of two field-theoretic frameworks designed to incorporate the particle nature of the degrees of freedom, namely the Doi-Peliti field theory and the response field theory that derives from Dean’s equation. While the Doi-Peliti field theory encodes the particle nature at a very fundamental level that is easily revealed, demonstrating the same for Dean’s equation is more involved and results in a number of surprising diagrammatic identities. We derive those and discuss their implications. These results are particularly pertinent in the context of active matter, whose surprising and often counterintuitive phenomenology rests wholly on the particle nature of the agents and their degrees of freedom as particles. 
\end{abstract}


\maketitle

\section{Introduction}

The mathematical description of non-equilibrium many-particle systems typically requires a choice of scale at which their behaviour is resolved. When the focus is on the collective dynamics of a large ensemble of particles, it can be convenient to disregard some of the microscopic information and to rely on a coarse-grained description in terms of densities $\rho(x,t)$, which are continuous in space. What is generally lost upon such coarse-graining is ``particle entity'', namely the familiar attribute of classical point particles whose initial property of being localised at one point only is preserved under the dynamics, in other words that individual particles can only exist at one position in space at any given time.
The distinction between effective and microscopically resolved theories has recently been debated in the context of active matter and, more specifically, entropy production \cite{nardini_entropy_2017,cocconi_en,garcia2020rnt,Busiello_2019}, where different levels of description grant access to different types of information about the degree of irreversibility of a stochastic process \cite{fodor2021}.
More generally, the study of sparse collections of interacting particles \cite{gompper20202020,soto2015self,slowman2016jamming} can make it necessary to equip theories with a notion of ``granularity'' of their constituents. 
Field theories have traditionally been the most successful approach to capture the physics and mathematics of phenomena emerging from the
interaction of many degrees of freedom \cite{lebellac1991,Hohenberg_1977,tauber_2014}. The Doi-Peliti formalism, which has a discrete number-state master equation as its starting point, is perhaps the best known example of a path-integral approach that preserves particle entity \cite{cardynotes,gunnarnotes}. Another, less familiar example is the response field or Martin-Siggia-Rose-Janssen-De Dominicis \cite{MartinSiggiaRose:1973, Janssen:1976, DeDominicis:1976} field theory \cite{hertz_path_2016} that derives from Dean's equation \cite{Dean_1996,gemlinson_2015,Velenich_2008}. While it is generally accepted that these theories correctly describe the behaviour of physical point particles by construction and that they are, in fact, equivalent \cite{lefevre_2007}, the precise mechanism whereby this property is enforced, as well as a general procedure to determine whether a given field theory possesses particle entity, have not been identified. We fill this gap in the following by introducing a signature of particle entity, \Eref{eq:PI_signature}, that draws solely on the moments of the integrated number density in a patch $\Omega$ of space. These moments can be computed by standard Feynman diagrammatic techniques. 

This work is organised as follows. In \Sref{setup_dp_dean}, we set the scene by introducing the Doi-Peliti field theory and the response field formalism. As an illustrative example, we compute the two-point correlation function of the number density of $n_0$ non-interacting diffusive particles, thus highlighting some of the key similarities and differences between the two approaches. In \Sref{prob_PI} we formalise the concept of single-particle entity and derive different observables to probe it. This signature of particle entity is then applied to the Doi-Peliti field theory (\Sref{dp_PI}) and the response field formalism of Dean's equation without interaction (\Sref{dean_PI}), confirming that both are indeed valid descriptions of physical point particles. 
In this last section we also discuss the role of integer particle numbers and relate some of the results to a more intuitive probabilistic picture. Finally, in \Sref{conclusion}, we summarise our findings and highlight some open questions. Some of the technical details are relegated to the appendices.

\section{Setting up the formalisms \seclabel{setup_dp_dean}}
\subsection{Doi-Peliti field theory}
A Doi-Peliti field theory, sometimes referred to as a coherent-state path integral, is a standard procedure to cast the discrete-state, continuous-time master equation of reaction-diffusion processes in a second quantised form that is amenable to a perturbative treatment \cite{cardynotes,gunnarnotes,tauber2005applications}. Its derivation starts from the master equation for the probability $P(\{n_i\},t)$ to find the system in state $\{n_i\} = \{n_0,n_1,...\}$, that is to find precisely $n_i$ particles at each site $i$, which is then written in a second quantised form by introducing a Fock space vector $|\{n_i\}\rangle$, together with the ladder operators 
$a_i^\dagger$ and $a_i$ for creation and annihilation on each lattice site $i$.
The operators satisfy the commutation relations
\begin{equation}\elabel{commutation_rel}
    [a_i,a_j^\dagger] = \delta_{ij}, \quad [a_i,a_j] = [a_i^\dagger,a_j^\dagger] = 0
\end{equation}
and act on $|\{n_i\}\rangle$ according to
\begin{equation}
    a_j |\{n_i\}\rangle = n_j |\{n_j - 1\} \rangle, \quad a_j^\dagger |\{n_i\}\rangle = |\{n_j + 1\}\rangle ~,
\end{equation}
so that $a^\dagger_i a_i$ is the number operator counting the number of particles at site $i$.
The notation $\{n_j + 1\}$ and similar is a suggestive shorthand to indicate that this is the same particle number state as $\{n_i\}$ except that the count at site $j$ is increased by one.
The state of the system is thus described by the mixed state
\begin{equation}
    | \Psi(t) \rangle = \sum_{\{n_i\}} P(\{n_i\},t) |\{n_j\}\rangle~,
\elabel{eq:def_DP_state}
\end{equation}
which evolves in time according to an imaginary-time Schr\"{o}dinger equation of the form \cite{cardynotes,gunnarnotes}
\begin{equation}
\elabel{eq:shroed_im_t}
    \partial_t | \Psi(t) \rangle = \hat{A}(a,a^\dagger)| \Psi(t) \rangle~.
\end{equation}
For a simple diffusive process on a one-dimensional lattice with homogeneous hopping rate $h$ and extinction rate $r$, the operator $\hat{A}$ reads
\begin{equation}
\elabel{eq:ham_diff}
    \hat{A}(a,a^\dagger) = \sum_i h (a^\dagger_{i+1} + a^\dagger_{i-1} - 2 a^\dagger_{i} )a_i - r (a^\dagger_{i}  - 1)a_i  ~.
\end{equation}
The formal solution of \Eref{eq:shroed_im_t}, $| \Psi(t) \rangle = e^{\hat{A}t} | \Psi(0) \rangle$, can then be cast into path-integral form, whereby the creation and annihilation operators are converted to time-dependent fields, denoted $\psi_i^\dagger(t)$ and $\psi_i(t)$, respectively. For technical reasons discussed extensively elsewhere \cite{cardynotes,gunnarnotes}, it is convenient at this stage to introduce the so-called Doi-shifted creation field, $\tilde{\psi}_i(t)$, according to the convention $\psi_i^\dagger(t) = 1 + \tilde{\psi}_i(t)$. 
For the case of simple diffusion, \Eref{eq:ham_diff}, generalised to $d$ dimensions, the action functional of the resulting field theory reads, upon taking the continuum limit,
\begin{equation}
\elabel{eq:dp_action}
    A[\tilde{\psi}(\xvec,t),\psi(\xvec,t)] = \int \ddint{x} \dint{t}  \tilde{\psi}(\xvec,t) ( \partial_t - D \Delta + r) \psi(\xvec,t)
\end{equation}
and is fully bilinear. 
In momentum and frequency space it reads
\begin{equation}
    A[\tilde{\psi}(\kvec,\omega),\psi(\kvec,\omega)] = \int \ddintbar{k} \dintbar{\omega} \tilde{\psi}(\kvec,\omega) (-\imag \omega + D \kvec^2 + r) \psi(-\kvec,-\omega)
\end{equation}
where we have used the convention 
\begin{equation}
    \psi (\xvec,t) = \int \ddintbar{k} \dintbar{\omega} \ e^{\imag \kvec \cdot \xvec} e^{-\imag \omega t} \psi(\kvec,\omega) 
\quad \text{and} \quad
    \psi (\kvec,\omega) = \int \ddint{x} \dint{t} \ e^{-\imag \kvec \cdot \xvec} e^{\imag \omega t} psi(\xvec,t) \, ,
\end{equation}

with $\ddintbar{k} = \ddint{k}/(2\pi)^d$ and $\dintbar{\omega} = d\omega/(2\pi)$ (similarly for $\tilde{\psi})$. We will change freely between different representations. 

The diffusive propagator can be obtained by Gaussian integration and reads in $k,\omega$
\begin{equation} \elabel{DP_propagator} 
    \langle \psi(\kvec,\omega) \tilde{\psi}(\kvec',\omega') \rangle = \frac{\deltabar(\omega+\omega') \deltabar(\kvec + \kvec')}{-\imag \omega + D \kvec^2 + r} \corresponds
    \ 
    \tikz[baseline=-2.5pt]{
      \draw[Aactivity] (0,0.0) node[above] {$\kvec,\omega$} -- (1.2,0.0) node[above] {$\kvec',\omega'$};
    }~,
\end{equation}
with $\deltabar(\kvec) = (2\pi) \delta(\kvec)$ and $\deltabar(\omega) = (2\pi) \delta(\omega)$. All diagrams are to be read from right to left. Expressing fields in $x,t$, the propagator reads
\begin{equation}
\elabel{dp_prop}
    \langle \psi(\xvec,t) \tilde{\psi}(\xvec',t')\rangle = \theta(t-t') \left(\frac{1}{4\pi D (t-t')}\right)^{d/2}\Exp{-\frac{(\xvec-\xvec')^2}{4D(t-t')}} ~,
\end{equation}
for $r \to 0^+$, with the Heaviside theta function $\theta(t)$ enforcing causality. 
The mass $r$ has solely the role to regularise the large $t$ behaviour and establish causality. In the following, we may take the limit $r \to 0^+$ whenever convenient. For completeness, the propagator in mixed momentum-time representation reads
\begin{equation} \elabel{DP_mixed_propagator}
    \langle \psi(\kvec,t) \tilde{\psi}(\kvec',t') \rangle = \theta(t-t') \deltabar(\kvec+\kvec') e^{-D k^2 (t-t')} 
\end{equation}

A general observable $\mathcal{O}(\{n_i\})$ in the Doi-Peliti formalism corresponds to a composite operator $\hat{\mathcal{O}}(a_i,a_i^\dagger)$, which we assume to be normal ordered, and which is defined by acting on the pure state $ | \{ n_i\} \rangle$ according to $\hat{\mathcal{O}}(a_i,a_i^\dagger) | \{ n_i\} \rangle = \mathcal{O}(\{n_i\}) | \{ n_i\} \rangle$. Its expectation translates into a path integral according to the following procedure \cite{pausch2019topics}
\begin{align}
    \langle \mathcal{O} \rangle 
    &= \sum_{\{n_i\}} \mathcal{O}(\{n_i\}) P(\{n_i\},t)  |\{n_j\}\rangle \\
    &= \langle \abyss | \hat{\mathcal{O}}(a_i,a_i^\dagger) e^{\tilde{A}t}| \Psi(0) \rangle \\
    &= \int \mathcal{D} \psi \mathcal{D} \tilde{\psi} \ \hat{\mathcal{O}}(\psi(t),\tilde{\psi}(t) + 1) \
    e^{A[\tilde{\psi},\psi] } \mathbb{I}(\tilde{\psi}(0) + 1) \elabel{eq:dp_path_in}
\end{align}
where we have introduced the coherent state, 
\begin{equation}
    \langle \abyss | = \sum_{\{n_i\}} \langle \{n_i\} | 
\end{equation}
with $\sum_{\{n_i\}} \langle \{n_i\}|$ summing over all $n$-particle occupation number states, as well as the initialisation operator $\mathbb{I}(a_i^\dagger)$, which satisfies $\mathbb{I}(a_i^\dagger)|0\rangle = | \Psi(0) \rangle$, with $| 0 \rangle$ the vacuum state. For an initial condition where $m_i$ particles are placed at each site $i$ at time $t=0$, the initialisation appears within the path integral \Eref{eq:dp_path_in} as 
\begin{equation}
\mathbb{I}(\tilde{\psi}(0) + 1) = \prod_i (\tilde{\psi}_i(0) + 1)^{m_i} = \prod_i \sum_{k=0}^{m_i} {m_i \choose k} \tilde{\psi}_i^k(0) ~.
\end{equation}

\subsection{Dean's equation in the response field formalism}
Dean's equation \cite{Dean_1996} is a stochastic differential equation of the It{\^o} type obeyed by the number density function $\rho(\xvec,t)$ for a system of Langevin processes interacting via a pairwise potential. It is an exact mapping of, and thus contains the same information as, the full set of Langevin equations for the individual ``single particle'' processes. It reads
\begin{equation}
\elabel{eq:deans}
    \partial_t \rho (\xvec,t) = \nabla \cdot \left( \rho \nabla \left. \frac{\delta F[\rho]}{\delta \rho}\right|_{\rho(\xvec,t)} \right) + \nabla \cdot(\rho^{1/2}\bm{\eta}(\xvec,t)) + \sum_i n_i \delta(t-t_i) \delta(\xvec-\xvec_i)
\end{equation}
where $F[\rho]$ denotes the free energy functional, defined as 
\begin{equation}
    F[\rho(\xvec)] = \int \ddint{x}  \rho(\xvec) \left( V(\xvec) + 
    D \log(\rho(\xvec)) + \int \ddint{y}  U(\xvec-\mathbf{y}) \rho(y) \right) ~,
\end{equation}
with $V(\xvec)$ a general single-particle potential and $U(\xvec-\mathbf y)$ a translationally invariant pairwise interaction potential. The last term on the right-hand side of \Eref{eq:deans} describes the initialisation of $n_i \in \mathbb{Z}$ particles in state $\xvec_i$ at time $t_i$ so that $\lim_{t\to-\infty} \rho(\xvec,t) = 0$. We will make the simplifying assumption of having only a single non-zero $n_i$, namely $n_0$, and generalise our result in Appendix~\ref{a:mult_strt_pt}. The vector-valued noise $\bm{\eta}(\xvec,t) \in \mathbb{R}^d$ is an uncorrelated white noise with covariance
\begin{equation}
    \langle \eta_\mu(\xvec,t) \eta_\nu(\xvec',t')\rangle = 2 D \delta_{\mu \nu}\delta(t-t')\delta(\xvec-\xvec') ~,
\end{equation}
for $\mu,\nu = 1,2,...,d$.
The unique feature of Dean's formalism is the nature of the noise term in \Eref{eq:deans}, $\nabla \cdot(\rho^{1/2} \bm{\eta})$, which is both conservative and It{\^o}-multiplicative, thus conserving the total particle number while preventing fluctuations from producing regions of negative density. Following the standard procedure \cite{hertz_path_2016,tauber_2014}, which requires special attention due to the multiplicative nature of the noise \cite{honkonen,Velenich_2008}, Dean's equation \eref{eq:deans} for the time and space dependent field $\rho(\xvec,t)$ can be cast as a response field, or Martin-Siggia-Rose-Janssen-De Dominicis, field theory with action
\begin{equation}
A[\rho,\tilde{\rho}]=\int \ddint{x} \dint{t} \tilde\rho\left( \partial_t \rho-\nabla\cdot\rho\nabla \left. \frac{\delta F[\rho]}{\delta \rho}\right|_{\rho(\xvec,t)} \right)-\rho D(\nabla \tilde \rho)^2 - \tilde{\rho} \sum_i n_i \delta(t-t_i) \delta(\xvec-\xvec_i)~,
\elabel{dean_action_raw}
\end{equation}
which simplifies to
\begin{align}
    A[\rho,\tilde{\rho}] 
    &=\int \ddint{x} \dint{t} \tilde\rho(\xvec,t)\left( \partial_t \rho(\xvec,t) -D \Delta \rho(\xvec,t) \right) - \tilde{\rho} \sum_i n_i \delta(t-t_i) \delta(\xvec-\xvec_i) -\rho D(\nabla \tilde \rho)^2 \elabel{dean_action}  \\
    &= \int \ddintbar{k} \dintbar{\omega} \tilde{\rho}(-\kvec,-\omega) (-\imag \omega + D \kvec^2) \rho(\kvec,\omega) 
    - \tilde{\rho}(\kvec,\omega) \sum_i n_i e^{\imag \kvec \cdot \xvec_i} e^{-\imag \omega t_i} \nonumber \\
    &\quad + \int \ddintbar{k} \ddintbar{k'} \dintbar{\omega} \dintbar{\omega'} D (\kvec \cdot \kvec') \tilde{\rho}(\kvec,\omega) \tilde{\rho}(\kvec',\omega') \rho(-(\kvec+\kvec'),-(\omega+\omega'))
    \elabel{Dean_action_simple_komega}
\end{align}
in the case of non-interacting particles undergoing simple diffusion without external potential. 
Unlike the Doi-Peliti path integral, \Eref{eq:dp_path_in},
the initialisation here shows up as a term in the action.  In a diagrammatic perturbation theory, these $n_i$ particles starting from positions $\xvec_i$, or, as a matter of fact, only one such position, $\xvec_0$ with $n_0$ particles starting from there, will be shown as a small, filled circle acting as a source,
\begin{equation}
\tikz[baseline=-2.5pt]{
    \draw[Aactivity] (0:0.5) -- (0:0.8) +(2pt,0) circle (2pt)[fill]; 
    }
    \ .
\end{equation}
The presence of the source spoils translational invariance and as a result, the hallmark $\delta$-function as it normally multiplies any correlation function, say $\deltabar(\kvec_0+\kvec_1+\ldots+\kvec_n)$ will be replaced by 
\begin{equation}\elabel{def_knullfix}
\int\ddintbar{k_0} \exp{\imag\kvec_0\cdot\xvec_0}
\delta(\kvec_0+\kvec_1+\ldots+\kvec_n)
=
\knullfix{\kvec_1+\ldots+\kvec_n} \ .
\end{equation}
Where readability is improved by it, we will retain the integral.

\begin{figure}
    \centering
    \includegraphics[scale=0.65]{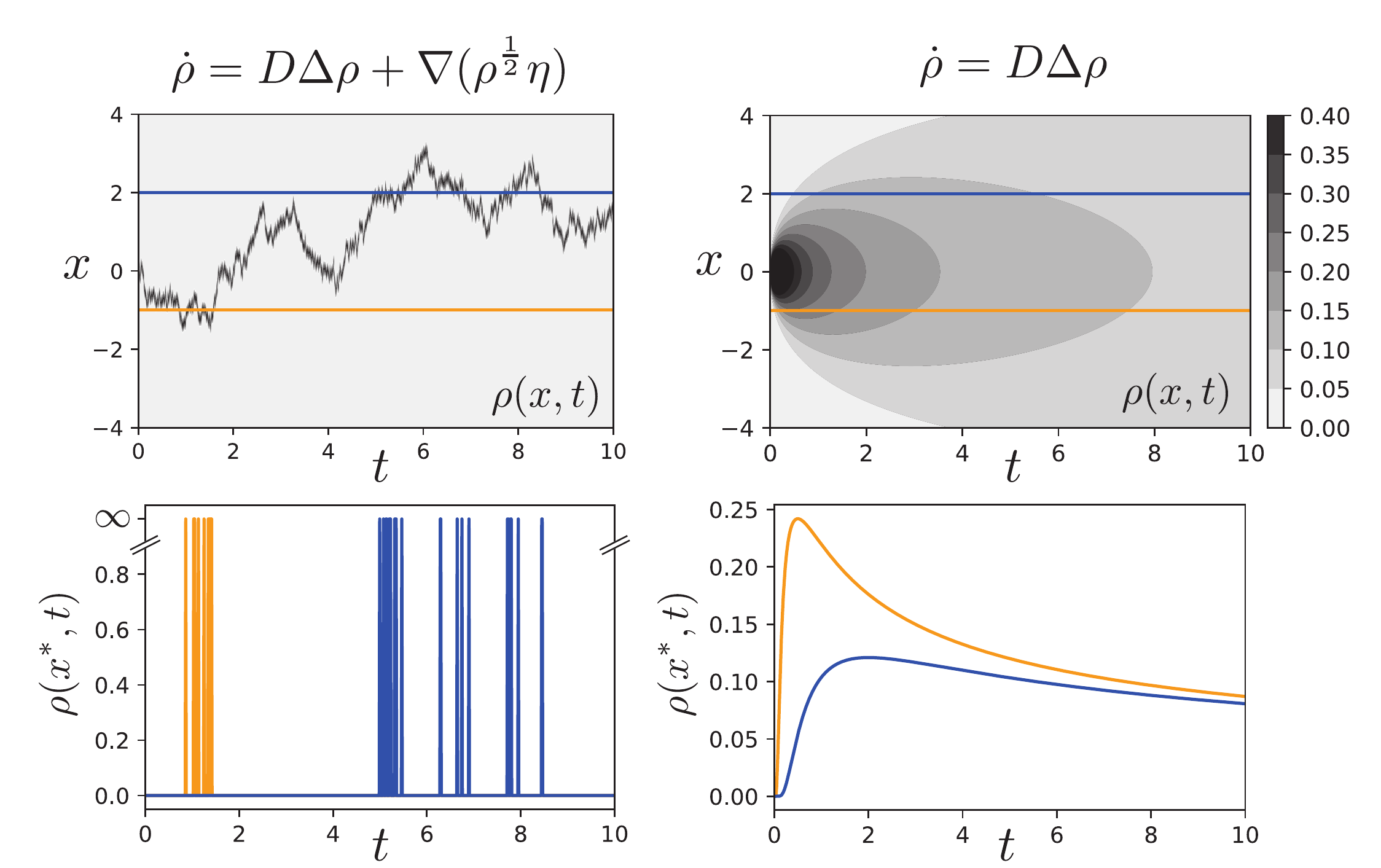}
    \caption{The time-dependent number density $\rho(x,t)$ for a physical point particle undergoing diffusion is expected to remain localised under the dynamics, indicating that the particle can only occupy one position in space at any given time. 
    While this property is preserved under Dean's dynamics (left column), it is generally lost when resorting to effective descriptions, such as the classical diffusion equation (right column). This difference is most obvious when measuring the instantaneous particle number density at two points a finite distance away from each other (bottom row).}
    \label{fig:schematic}
\end{figure}

The expectation value of a field-dependent observable $\mathcal{O}[\rho]$ can then be computed via the path integral
\begin{equation}
    \langle \mathcal{O}[\rho] \rangle = \int \mathcal{D}\rho \mathcal{D} \tilde{\rho} \  \mathcal{O}[\rho] \exp{- A[\rho,\tilde{\rho}]} 
\end{equation}
where $\tilde{\rho}$ is the purely imaginary response field. The normalisation is chosen such that $\langle 1 \rangle = 1$. The action $A$ is then split into a bilinear and an interacting part, denoted $A_0$ and $A_{\rm int}$ respectively, according to
\begin{equation}\elabel{bilin_Dean}
    A_0[\rho,\tilde{\rho}] = \int \ddint{x} \dint{t} \tilde\rho\left( \partial_t \rho-D \Delta \rho \right) 
\end{equation}
and 
\begin{align}
    A_{\rm int}[\rho,\tilde{\rho}] 
    = - \int \ddint{x} \dint{t} \Bigg\{&\rho(\xvec,t) D(\nabla \tilde \rho(\xvec,t))^2 
    \\ 
\nonumber    
   &
    + \tilde{\rho}(\xvec,t)  \nabla_{\xvec}  \cdot \left(  \rho(\xvec,t) \nabla_{\xvec} \left[  V(\xvec) +  \int \ddint{y} \ U(\xvec-\yvec) \rho(\yvec,t)  \right] \right)
     \\
     \nonumber
    &
     + \tilde{\rho}(\xvec,t) \sum_i n_i  \delta(\xvec-\xvec_i) \delta(t-t_i) \Bigg\}
\end{align}
\begin{align}
     &=
     \int \ddintbar{k_{1,2,3}}\,\, \dintbar{\omega_{1,2,3}}\deltabar(\kvec_1+\kvec_2+\kvec_3)\deltabar(\omega_1+\omega_2+\omega_3) 
     \ \bigg\{
     \rho(\kvec_1,\omega_1) D (\kvec_2\cdot\kvec_3)\tilde{\rho}(\kvec_2,\omega_2)\tilde{\rho}(\kvec_3,\omega_3)
     \\ \nonumber
     &\quad\qquad
     + \tilde{\rho}(\kvec_1,\omega_1)  ((\kvec_2+\kvec_3)\cdot\kvec_3)\rho(\kvec_2,\omega_2)\bigg[V(\kvec_3)\deltabar(\omega_3) + U(\kvec_3)\rho(\kvec_3,\omega_3) \bigg] \bigg\}
          \\ \nonumber
     &
     \quad-
\int \ddintbar{k}\,\dintbar{\omega} \tilde{\rho}(\kvec,\omega) \sum_i n_i \exp{\imag\kvec\cdot\xvec_i} \exp{-\imag\omega t_i}
\end{align}
Finally, expectations are computed in a perturbation theory about the bilinear theory using
\begin{equation}
\elabel{eq:exp_gauss_pert}
    \langle \mathcal{O}[\rho] \rangle = \sum_{n=0}^\infty \left\langle \frac{\left(- A_{\rm int}[\rho,\tilde{\rho}] \right)^n}{n!} \mathcal{O}[\rho] \right\rangle_0 ~,
\end{equation}
where 
\begin{equation}
    \langle\bullet\rangle_0=\int\Dint{\rho}\Dint{\tilde{\rho}} \bullet \exp{-A_0[\rho,\tilde{\rho}]}
\end{equation}
denotes expectation with respect to the bilinear action, \Eref{bilin_Dean}. The right hand side of \Eref{eq:exp_gauss_pert} involves products of fields and the Wick-Isserlis theorem \cite{lebellac1991} can be invoked to express these in terms of the bare propagator,
\begin{equation}
    G(\xvec-\xvec',t-t') = \langle \rho(\xvec,t) \tilde{\rho}(\xvec',t')\rangle_0 \ 
    \corresponds
    \ \
    \tikz[baseline=-2.5pt]{
      \draw[Aactivity] (0,0.0) node[above] {$\xvec,t$} -- (1.2,0.0) node[above] {$\xvec',t'$};
    }~,
\end{equation}
obtained from the bilinear action. 
Henceforth we will use the symbol $\corresponds$ to indicate equivalence between diagrams and other mathematical expressions.
In $d$ dimensions, the bare propagator reads
\begin{equation}
    G(\xvec-\xvec',t-t')
    =
    \theta(t-t') \left( \frac{1}{4\pi D (t-t')}\right)^{d/2} \Exp{-\frac{(\xvec-\xvec')^2}{4D(t-t')}}~, \elabel{dean_prop}
\end{equation}
with the Heaviside theta function $\theta(t)$ enforcing causality. This propagator is identical to that of the corresponding Doi-Peliti field theory, \Eref{dp_prop}. For later use, we recall the form of the propagator in momentum-frequency representation,
\begin{equation}
\elabel{kw_prop}
    \langle \rho(\kvec,\omega) \tilde{\rho}(\kvec',\omega') \rangle_0 = \frac{\delta(\kvec+\kvec') \delta(\omega+\omega')}{-\imag \omega + D \kvec^2+r} ~,
\end{equation}
which we have amended by a mass $r \to 0^+$ to enforce causality, as \Eref{DP_propagator}. Further, we introduce
the mixed momentum-time representation,
\begin{equation}\elabel{Dean_mixed_propagator}
    \langle \rho(\kvec,t) \tilde{\rho}(\kvec',t')\rangle_0 = \theta(t-t') \delta(\kvec+\kvec') \exp{-D 
    (t-t')\kvec^2}  ~,
\end{equation}
see \Eref{DP_mixed_propagator}.
For non-interacting particles in a flat potential, $\nabla V(\xvec) = 0$,  $\nabla U(\xvec) = 0$, 
the bare propagator equals the full propagator,
\begin{equation}
    \langle \rho(\kvec,t) \tilde{\rho}(\kvec',t')\rangle = \langle \rho(\kvec,t) \tilde{\rho}(\kvec',t')\rangle_0
\end{equation}
as
the only non-linear term
in the action is the amputated three-point vertex
\begin{equation}
\elabel{dean_vertex}
-\rho D(\nabla \tilde \rho)^2 
\ \corresponds\ 
    \tikz[baseline=-2.5pt]{
      
      \draw[Aactivity] (0,0.0) -- (0.45,0.0);
      \draw[Aactivity] (220:0.45) -- (0,0.0);
      \draw[Aactivity] (140:0.45) -- (0,0.0);
      \draw[thin] ($(0.07,0.07)+(140:0.3)$) -- ($(-0.07,-0.07)+(140:0.3)$);
      \draw[thin] ($(-0.07,0.07)+(220:0.3)$) -- ($(0.07,-0.07)+(220:0.3)$);
      \draw[dotted,thin] (140:0.3) arc (140:220:0.3);
    }
\end{equation}
with the dashes on the propagators denoting spatial derivatives acting on the response fields and the dotted line the scalar product of these derivatives. 
The presence of such a vertex in the free particle case is a non-trivial feature of Dean's equation and clashes somewhat with the notion of `interaction' associated with terms of order higher than bilinear \cite{tauber_2014}. As we will demonstrate below, \Eref{dean_vertex}, which we will refer to interchangeably as \emph{Dean's vertex} or a \emph{virtual branching} vertex, is the term that implements the particle nature of the degrees of freedom within the Dean framework. In contrast to Doi-Peliti, particle entity in the response field formalism of Dean's equation is a perturbative feature. The effect of Dean's vertex is illustrated in Fig.~\ref{fig:schematic} by comparison with the standard diffusion equation, which lacks particle entity.

The Doi-Peliti field theory and the response field field theory derived from Dean's equation can be mapped onto each other by means of a Cole-Hopf transformation of the fields \cite{lefevre_2007},
\begin{equation}
    \psi^\dagger \to e^{\tilde\rho}, \quad \psi \to \rho e^{-\tilde{\rho}}~.
\end{equation}
This equivalence implies that the two formalisms should be equally capable of capturing particle entity. The precise mechanisms by which each does so, however, turn out to be very different, as we will see in detail in \Srefs{dp_PI} and \ref{sec:dean_PI}.

\subsection{Example: the two-point density correlation function \seclabel{ex_calcl}}
To illustrate the similarities and differences between the two formalism introduced above, we now calculate the two-point correlation function of the particle number density for $n_0$ non-interacting diffusive particles in a flat potential, $\nabla V(\xvec) = 0$ and $\nabla U(\xvec) = 0$, all initialised at the same position $x_0$ and time $t_0$, first in the Doi-Peliti scheme and then using Dean's equation. While the result of this detailed calculation is somewhat trivial and can be derived by straightforward probabilistic arguments, its derivation elucidates certain formalism-specific cancellation mechanisms that will play an important role in the remainder of this work. The reader interested in the generic definition of particle entity but not in the details of the field theoretic approach can skip directly to \Sref{prob_PI}.

We first use the parameterisation of the field theories in $\kvec$ and $\omega$, which is very commonly used in field theories. 
In real-space and time, the two-point correlation function $C(\xvec_1,\xvec_2,t_1,t_2)$ in the Doi-Peliti framework is the observable \cite{tauber_2014,cardynotes}
\begin{align}\elabel{observable_naive_DP}
	C(\xvec_1,\xvec_2, t_1, t_2) &= \ave{
	\big(\psidagger(\xvec_2,t_2) \psi(\xvec_2,t_2)\big)
	\big(\psidagger(\xvec_1,t_1) \psi(\xvec_1,t_1)\big)
	\psi^{\dagger n_0}(\xvec_0,t_0)} \\
&=
\elabel{observable_naive_DP_simplified}
{n_0 \choose 1} 
\ave{\psi(\xvec_2,t_2)\psitilde(\xvec_1,t_1)}
\ave{\psi(\xvec_1,t_1)\psitilde(\xvec_0,t_0)}
\\ \nonumber &\quad +
{n_0 \choose 1} 
\ave{\psi(\xvec_1,t_1)\psitilde(\xvec_2,t_2)}
\ave{\psi(\xvec_2,t_2)\psitilde(\xvec_0,t_0)}
\\ &\quad +
2 {n_0 \choose 2}
\ave{\psi(\xvec_2,t_2)\psitilde(\xvec_0,t_0)}
\ave{\psi(\xvec_1,t_1)\psitilde(\xvec_0,t_0)}
\nonumber \\
& \hspace*{-2.8cm} \corresponds
{n_0 \choose 1} \tikz[baseline=-2.5pt]{
      \draw[Aactivity] (-1.0,0) node[above] {$\xvec_2,t_2$} -- (-0.2,0.0) node[above] {$ $};
      \draw[Aactivity] (0.2,0) node[above] {$ $} -- (1.0,0.0) node[above] {$\xvec_0,t_0$};
      \draw[thin] (0,0) circle (0.2) node[above=0.2] {$\xvec_1,t_1$};
      \draw[thin] (45:0.2) -- (45:-0.2);
      \draw[thin] (-45:0.2) -- (-45:-0.2);
      \draw[Aactivity] (90:0.2) -- (90:0.3);
    }
\quad+\quad
{n_0 \choose 1} \tikz[baseline=-2.5pt]{
      \draw[Aactivity] (-1.0,0) node[above] {$\xvec_1,t_1$} -- (-0.2,0.0) node[above] {$ $};
      \draw[Aactivity] (0.2,0) node[above] {$ $} -- (1.0,0.0) node[above] {$\xvec_0,t_0$};
      \draw[thin] (0,0) circle (0.2) node[above=0.2] {$\xvec_2,t_2$};
      \draw[thin] (45:0.2) -- (45:-0.2);
      \draw[thin] (-45:0.2) -- (-45:-0.2);
      \draw[Aactivity] (90:0.2) -- (90:0.3);
    }
\quad+\quad
2 {n_0 \choose 2}
    \tikz[baseline=-2.5pt]{
        \draw[Aactivity] (0,0.2) node[above] {$\xvec_1,t_1$} -- (1.4,0.2) node[above] {$\xvec_0,t_0$};
        \draw[Aactivity] (0,-0.2) node[below] {$\xvec_2,t_2$} -- (1.4,-0.2) node[below] {$\xvec_0,t_0$};
    }
\elabel{observable_naive_DP_diagrams}
\end{align}
where we assume $\xvec_1\ne\xvec_2$ to avoid the special case of non-commutation of the operators. The high number of terms in \Eref{observable_naive_DP} is due to the Doi-shift, which splits each daggered creator field in two terms, $\psidagger=1+\psitilde$. This turns the contribution of the initial particles into $\psi^{\dagger n_0}=\sum_k^{n_0} {n_0 \choose k} \psitilde^{n_0}$. 
The vertices made from a crossed circle in \Eref{observable_naive_DP_diagrams}
are meant to indicate an annihilation field at the indicated position and time with immediate re-creation. \Eref{observable_naive_DP} has the generic form of a two-point correlation function in the Doi-Peliti framework without interaction.

\Eref{observable_naive_DP_simplified} is still expressed in real space and direct time and needs to be Fourier-transformed to write it in the common $\kvec,\omega$ parameterisation. Each of the three terms in \Eref{observable_naive_DP_simplified} requires four integrals in $\kvec$ and four in $\omega$, for example
\begin{align}
&n_0 \tikz[baseline=-2.5pt]{
      \draw[Aactivity] (-1.0,0) node[above] {$\xvec_2,t_2$} -- (-0.2,0.0) node[above] {$ $};
      \draw[Aactivity] (0.2,0) node[above] {$ $} -- (1.0,0.0) node[above] {$\xvec_0,t_0$};
      \draw[thin] (0,0) circle (0.2) node[above=0.2] {$\xvec_1,t_1$};
      \draw[thin] (45:0.2) -- (45:-0.2);
      \draw[thin] (-45:0.2) -- (-45:-0.2);
      \draw[Aactivity] (90:0.2) -- (90:0.3);
    }
	\\ 
	&\corresponds
	n_0 \int 
	\ddintbar{k_2}\ddintbar{k'_1}\ddintbar{k_1}\ddintbar{k_0}
	\dintbar{\omega_2}\dintbar{\omega'_1}\dintbar{\omega_1}\dintbar{\omega_0}
	\frac{\deltabar(\kvec_2+\kvec'_1)\deltabar(\omega_2+\omega'_1)}{-\imag\omega_2+D\kvec_2^2+r}
	\frac{\deltabar(\kvec_1+\kvec_0) \deltabar(\omega_1+\omega_0)} {-\imag\omega_1+D\kvec_1^2+r}\\
	&\nonumber \quad\times\exp{\imag (\kvec_2\cdot\xvec_2 + \kvec'_1\cdot\xvec_1 + \kvec_1\cdot\xvec_1 + \kvec_0\cdot\xvec_0) }
	\exp{-\imag (\omega_2 t_2 + \omega'_1 t_1 + \omega_1 t_1 + \omega_0 t_0) }
\end{align}
drawing on the propagator introduced in \Eref{DP_propagator}. 
Using the $\delta$-functions, the integrals in each term are immediately reduced to only two, all differing solely in the arguments of the exponentials:
\begin{align}\elabel{observable_DP_done}
	&C(\xvec_1,\xvec_2, t_1, t_2) = 
\int 
	\ddintbar{k_2}\ddintbar{k_1}
	\dintbar{\omega_2}\dintbar{\omega_1}
	\frac{1}{-\imag\omega_2+D\kvec_2^2+r}
	\frac{1} {-\imag\omega_1+D\kvec_1^2+r}
	\\
	&\nonumber\quad\times\Big\{
n_0 
\exp{ \imag (\kvec_2\cdot(\xvec_2-\xvec_1) + \kvec_1\cdot(\xvec_1-\xvec_0) ) }	
\exp{-\imag (\omega_2 (t_2-t_1)        + \omega_1 (t_1-t_0) ) }	
\\
&\nonumber\!\!\qquad\quad+
n_0 
\exp{ \imag (\kvec_2\cdot(\xvec_2-\xvec_0) + \kvec_1\cdot(\xvec_1-\xvec_2) ) }	
\exp{-\imag (\omega_2 (t_2-t_0)        + \omega_1 (t_1-t_2) ) }	
\\
&\nonumber\!\!\qquad\quad+
n_0(n_0-1)
\exp{ \imag (\kvec_2\cdot(\xvec_2-\xvec_0) + \kvec_1\cdot(\xvec_1-\xvec_0) ) }	
\exp{-\imag (\omega_2 (t_2-t_0)        + \omega_1 (t_1-t_0) ) }	
	\Big\}
	\end{align}
with $r\to 0^+$ still to be taken. The first of the three terms in the integrand describes the propagation of any of $n_0$ particles from $\xvec_0$ at $t_0$ to $\xvec_1$ at $t_1$ and from there to $\xvec_2$ at $t_2$. This term will contribute only if $t_2\ge t_1\ge t_0$. The second term describes a similar process, from $\xvec_0$ at $t_0$ to $\xvec_2$ at $t_2$ and from there to $\xvec_1$ at $t_1$, contributing only if $t_1\ge t_2\ge t_0$. The last term describes the propagation of two independent particles from $\xvec_0$ at $t_0$ to $\xvec_1$ at $t_1$ and another one from $\xvec_0$ at $t_0$ to $\xvec_2$ at $t_2$. There are $n_0(n_0-1)$ such pairs. If $n_0\le1$, the last term vanishes, leaving only the first two terms, both of which vanish if $t_1=t_2$ and $\xvec_1\ne \xvec_2$ as we will show below, because a \emph{particle} cannot possibly be found at two different places simultaneously. 
\Eref{observable_DP_done} completes the derivation of the correlation function in the Doi-Peliti framework.

To derive the correlation function in Dean's framework, we use the action as stated in \Eref{dean_action} with both the interaction and the source treated perturbatively.
The role of the creator fields in the field theory of Dean's equation is very different from Doi-Peliti. In the Dean framework, the two-point correlation function is 
\begin{equation}\elabel{observable_naive_Dean}
	C(\xvec_1,\xvec_2, t_1, t_2) = \ave{\rho(\xvec_2,t_2) \rho(\xvec_1,t_1)} 
	\corresponds
    \tikz[baseline=-2.5pt]{
      \draw[Aactivity] (0,0.0) -- (0.85,0.0) node[above] {$\xvec_0,t_0$} -- ++(0,0) node[right] {$\ (n_0)$} +(2pt,0) circle (2pt)[fill];
      \draw[Aactivity] (220:0.85) node[below] {$\xvec_2,t_2$} -- (0,0.0);
      \draw[Aactivity] (140:0.85) node[above] {$\xvec_1,t_1$} -- (0,0.0);
      \draw[thin] ($(0.07,0.07)+(140:0.35)$) -- ($(-0.07,-0.07)+(140:0.35)$);
      \draw[thin] ($(-0.07,0.07)+(220:0.35)$) -- ($(0.07,-0.07)+(220:0.35)$);
      \draw[dotted,thin] (140:0.35) arc (140:220:0.35);
    }
    \quad+\quad
    \tikz[baseline=-2.5pt]{
        \draw[Aactivity] (0,0.2) node[above] {$\xvec_1,t_1$} -- (1.4,0.2) node[above] {$\xvec_0,t_0$} -- +(0,0) node[right] {$\ (n_0)$} +(2pt,0) circle (2pt)[fill];
        \draw[Aactivity] (0,-0.2) node[below] {$\xvec_2,t_2$} -- (1.4,-0.2) node[below] {$\xvec_0,t_0$} -- +(0,0) node[right] {$\ (n_0)$} +(2pt,0) circle (2pt)[fill];
    }
\end{equation}
as every field $\rho(\xvec,t)$ can be matched with a creator field from the perturbative part of the action, shown as a small filled circle at the right end of the incoming propagators.
Each such creator field appears with a \emph{coupling} $n_0$, which we have highlighted by writing it in brackets behind each source in the diagram.
While the second term in \Eref{observable_naive_Dean} is structurally identical to the last term in \Eref{observable_naive_DP_diagrams} and indeed captures the same process, the pre-factors of the two differ by $n_0$. The first two terms in \Eref{observable_naive_DP_diagrams} on the other hand seem to be absent from \Eref{observable_naive_Dean}. In turn, the first diagram of \Eref{observable_naive_Dean}, is solely due to the Dean-vertex \Eref{dean_vertex} and therefore absent in Doi-Peliti, \Eref{observable_naive_DP_diagrams}. Writing this term in $\kvec, \omega$ gives
\begin{align}
&
    \tikz[baseline=-2.5pt]{
      \draw[Aactivity] (0,0.0) -- (0.85,0.0) node[above] {$\xvec_0,t_0$} -- +(0,0) node[right] {$\ (n_0)$} +(2pt,0) circle (2pt)[fill];
      \draw[Aactivity] (220:0.85) node[below] {$\xvec_2,t_2$} -- (0,0.0);
      \draw[Aactivity] (140:0.85) node[above] {$\xvec_1,t_1$} -- (0,0.0);
      \draw[thin] ($(0.07,0.07)+(140:0.35)$) -- ($(-0.07,-0.07)+(140:0.35)$);
      \draw[thin] ($(-0.07,0.07)+(220:0.35)$) -- ($(0.07,-0.07)+(220:0.35)$);
      \draw[dotted,thin] (140:0.35) arc (140:220:0.35);
    }
\nonumber
 \\ & \nonumber 
 \corresponds
\int 
	\ddintbar{k_2}\ddintbar{k_1}
	\dintbar{\omega_2}\dintbar{\omega_1}
\exp{ \imag (\kvec_2\cdot (\xvec_2-\xvec_0) + \kvec_1\cdot (\xvec_1-\xvec_0) ) }	
\exp{-\imag (\omega_2 (t_2 - t_0) + \omega_1 (t_1 - t_0) ) }	
	\elabel{Dean_term_step1}
  \nonumber \\
	&\times 
	(-2 n_0 D \kvec_1\cdot\kvec_2)
	\frac{1}{-\imag\omega_2+D\kvec_2^2+r}
	\frac{1}{-\imag\omega_1+D\kvec_1^2+r}
	\frac{1}{-\imag(\omega_1+\omega_2)+D(\kvec_1+\kvec_2)^2+r}
	\end{align}
using \Eref{kw_prop} for the propagator and
where the factor $(-2 n_0 D \kvec_1\cdot\kvec_2)$ is due to the sign of the interaction term in the action \Eref{dean_action}, including a factor $2$ from symmetry.

The second term in \Eref{observable_naive_Dean} can be read off from \Erefs{observable_naive_DP_diagrams} and \eref{observable_DP_done}. Its pre-factor of $n_0^2$ has to be split into $n_0^2=n_0(n_0-1)+n_0$ to reveal the cancellation mechanism, 
\begin{align}
&
    \tikz[baseline=-2.5pt]{
      \draw[Aactivity] (0,0.0) -- (0.85,0.0) node[above] {$\xvec_0,t_0$} -- +(0,0) node[right] {$\ (n_0)$} +(2pt,0) circle (2pt)[fill];
      \draw[Aactivity] (220:0.85) node[below] {$\xvec_2,t_2$} -- (0,0.0);
      \draw[Aactivity] (140:0.85) node[above] {$\xvec_1,t_1$} -- (0,0.0);
      \draw[thin] ($(0.07,0.07)+(140:0.35)$) -- ($(-0.07,-0.07)+(140:0.35)$);
      \draw[thin] ($(-0.07,0.07)+(220:0.35)$) -- ($(0.07,-0.07)+(220:0.35)$);
      \draw[dotted,thin] (140:0.35) arc (140:220:0.35);
    }
\quad + \quad
    \tikz[baseline=-2.5pt]{
        \draw[Aactivity] (0,0.2) node[above] {$\xvec_1,t_1$} -- (1.4,0.2) node[above] {$\xvec_0,t_0$} -- +(0,0) node[right] {$\ (n_0)$} +(2pt,0) circle (2pt)[fill];
        \draw[Aactivity] (0,-0.2) node[below] {$\xvec_2,t_2$} -- (1.4,-0.2) node[below] {$\xvec_0,t_0$} -- +(0,0) node[right] {$\ (n_0)$} +(2pt,0) circle (2pt)[fill];
    }
\\
\nonumber&\corresponds
\int 
	\ddintbar{k_2}\ddintbar{k_1}
	\dintbar{\omega_2}\dintbar{\omega_1}
	\exp{ \imag (\kvec_2\cdot (\xvec_2-\xvec_0) + \kvec_1\cdot (\xvec_1-\xvec_0) ) }	
\exp{-\imag (\omega_2 (t_2 - t_0) + \omega_1 (t_1 - t_0) ) }	
\\
\elabel{Dean_cancellation}
& \times
	\frac{1}{-\imag\omega_2+D\kvec_2^2+r}
	\frac{1}{-\imag\omega_1+D\kvec_1^2+r}
	\left(
	\frac{-2 n_0 D \kvec_1\cdot\kvec_2}{-\imag(\omega_1+\omega_2)+D(\kvec_1+\kvec_2)^2+r}
	+n_0 + n_0(n_0-1)
	\right)
\\
&\nonumber = 
\int 
	\ddintbar{k_2}\ddintbar{k_1}
	\dintbar{\omega_2}\dintbar{\omega_1}
	\exp{ \imag (\kvec_2\cdot (\xvec_2-\xvec_0) + \kvec_1\cdot (\xvec_1-\xvec_0) ) }	
\exp{-\imag (\omega_2 (t_2 - t_0) + \omega_1 (t_1 - t_0) ) }\\
\elabel{Dean_cancellation_step_naive1}
&\qquad \times \Big\{
n_0
\frac{1}{-\imag(\omega_1+\omega_2)+D(\kvec_1+\kvec_2)^2+r}\\
&\qquad\qquad\nonumber\times 
\left(
	\frac{1}{-\imag\omega_2+D\kvec_2^2+r}
+
	\frac{1}{-\imag\omega_1+D\kvec_1^2+r}
-	
	\frac{r }{(-\imag\omega_1+D\kvec_1^2+r)(-\imag\omega_2+D\kvec_2^2+r)}
	\right)
\\
&\qquad\qquad\nonumber
+ n_0(n_0-1)
	\frac{1}{-\imag\omega_2+D\kvec_2^2+r}
	\frac{1}{-\imag\omega_1+D\kvec_1^2+r}
\Big\}
\ .
\end{align}
The term proportional to $r$ in the numerator eventually vanishes when $r\to0$. To see now that \Eref{Dean_cancellation_step_naive1} is in fact identical to the first two terms in \Erefs{observable_naive_DP_diagrams} and \eref{observable_DP_done} requires a simple substitution of the dummy variables, for example $\kvec_1+\kvec_2$ becoming $\kvec_1$,
\begin{align}
\nonumber
&\int 
	\ddintbar{k_2}\ddintbar{k_1}
	\dintbar{\omega_2}\dintbar{\omega_1}
	\exp{ \imag (\kvec_2\cdot (\xvec_2-\xvec_0) + \kvec_1\cdot (\xvec_1-\xvec_0) ) }	
\exp{-\imag (\omega_2 (t_2 - t_0) + \omega_1 (t_1 - t_0) ) }\\
&\qquad\qquad\times
\frac{1}{-\imag(\omega_1+\omega_2)+D(\kvec_1+\kvec_2)^2+r}
\frac{1}{-\imag\omega_2+D\kvec_2^2+r}\\
&\nonumber =
\int 
	\ddintbar{k_2}\ddintbar{k_1}
	\dintbar{\omega_2}\dintbar{\omega_1}
	\exp{ \imag (\kvec_2\cdot (\xvec_2-\xvec_1) + \kvec_1\cdot (\xvec_1-\xvec_0) ) }	
	\exp{-\imag (\omega_2 (t_2 - t_1) + \omega_1 (t_1 - t_0) ) }\\
&\qquad\qquad\times
\frac{1}{-\imag\omega_1+D\kvec_1^2+r}
\frac{1}{-\imag\omega_2+D\kvec_2^2+r}
\end{align}
In summary, after pairing in \Eref{Dean_cancellation_step_naive1} the interaction term of Dean's equation with the two independent propagators, the field theory of Dean's equation reproduces the two-point correlation function as the Doi-Peliti framework, \Eref{observable_DP_done}, except for a term proportional to $r$ which vanishes in the limit of $r\to0$:
\begin{align}\elabel{corr_Dean_final_diagrams}
&	C(\xvec_1,\xvec_2, t_1, t_2) = \ave{\rho(\xvec_2,t_2) \rho(\xvec_1,t_1)} 
	\corresponds
    \tikz[baseline=-2.5pt]{
      \draw[Aactivity] (0,0.0) -- (0.85,0.0) node[above] {$\xvec_0,t_0$} -- +(0,0) node[right] {$\ (n_0)$}  +(2pt,0) circle (2pt)[fill];
      \draw[Aactivity] (220:0.85) node[below] {$\xvec_2,t_2$} -- (0,0.0);
      \draw[Aactivity] (140:0.85) node[above] {$\xvec_1,t_1$} -- (0,0.0);
      \draw[thin] ($(0.07,0.07)+(140:0.35)$) -- ($(-0.07,-0.07)+(140:0.35)$);
      \draw[thin] ($(-0.07,0.07)+(220:0.35)$) -- ($(0.07,-0.07)+(220:0.35)$);
      \draw[dotted,thin] (140:0.35) arc (140:220:0.35);
    }
\quad + \quad
    \tikz[baseline=-2.5pt]{
        \draw[Aactivity] (0,0.2) node[above] {$\xvec_1,t_1$} -- (1.4,0.2) node[above] {$\xvec_0,t_0$} -- +(0,0) node[right] {$\ (n_0)$}  +(2pt,0) circle (2pt)[fill];
        \draw[Aactivity] (0,-0.2) node[below] {$\xvec_2,t_2$} -- (1.4,-0.2) node[below] {$\xvec_0,t_0$} -- +(0,0) node[right] {$\ (n_0)$}  +(2pt,0) circle (2pt)[fill];
    }
\\
&\elabel{observable_Dean_done} \corresponds
\int 
	\ddintbar{k_2}\ddintbar{k_1}
	\dintbar{\omega_2}\dintbar{\omega_1}
	\frac{1}{-\imag\omega_2+D\kvec_2^2+r}
	\frac{1} {-\imag\omega_1+D\kvec_1^2+r}
\\&\nonumber\qquad\times
	\Big\{
n_0 
\exp{ \imag (\kvec_2\cdot(\xvec_2-\xvec_1) + \kvec_1\cdot(\xvec_1-\xvec_0) ) }	
\exp{-\imag (\omega_2 (t_2-t_1)        + \omega_1 (t_1-t_0) ) }	
\\&\nonumber\qquad
+
n_0 
\exp{ \imag (\kvec_2\cdot(\xvec_2-\xvec_0) + \kvec_1\cdot(\xvec_1-\xvec_2) ) }	
\exp{-\imag (\omega_2 (t_2-t_0)        + \omega_1 (t_1-t_2) ) }	
\\&\nonumber\qquad
-
n_0 \frac{r}{-\imag(\omega_1+\omega_2)+D(\kvec_1+\kvec_2)^2+r}
\\&\nonumber\qquad\qquad\times
\exp{ \imag (\kvec_2\cdot (\xvec_2-\xvec_0) + \kvec_1\cdot (\xvec_1-\xvec_0) ) }	\exp{-\imag (\omega_2 (t_2 - t_0) + \omega_2 (t_1 - t_0) ) }
\\&\nonumber\qquad
+
n_0(n_0-1)
\exp{ \imag (\kvec_2\cdot(\xvec_2-\xvec_0) + \kvec_1\cdot(\xvec_1-\xvec_0) ) }	
\exp{-\imag (\omega_2 (t_2-t_0)        + \omega_1 (t_1-t_0) ) }	
\Big\} \ .
\end{align}
This concludes the demonstration that the Doi-Peliti framework and Dean's equation produce identical results for the two-point correlation function. \Eref{Dean_cancellation_step_naive1} illustrates the central cancellation mechanism, which we  generalise to the relevant observables below, in particular  Appendix~\ref{a:induction_conn}. 
As \Eref{dean_vertex} is a perturbative term, the resulting branching diagrams in \Eref{corr_Dean_final_diagrams} \emph{discount} contributions due to independent particle movement, shown as two parallel propagators in \Eref{corr_Dean_final_diagrams}, of which there are $n_0^2$ rather than $n_0(n_0-1)$.

Performing the calculation above immediately in direct time and real space is most easily done assuming a particular time ordering, say $t_2 > t_1 > t_0$. In that case, Doi-Peliti produces
\begin{align}\elabel{corr_direct_xt}
    C(\xvec_1,\xvec_2,t_1,t_2) = 
     n_0&\frac{e^{-\frac{(\xvec_2-\xvec_1)^2}{4D(t_2-t_1)}}}{(4\pi D(t_2-t_1))^{d/2}}\frac{e^{-\frac{(\xvec_1-\xvec_0)^2}{4D(t_1-t_0)}}}{(4\pi D(t_1-t_0))^{d/2}}\\
     \nonumber 
    + 
        n_0(n_0-1)&\frac{e^{-\frac{(\xvec_1-\xvec_0)^2}{4D(t_1-t_0)}}}{(4\pi D(t_1-t_0))^{d/2}}\frac{e^{-\frac{(\xvec_2-\xvec_0)^2}{4D(t_2-t_0)}}}{(4\pi D(t_2-t_0))^{d/2}}
\end{align}
directly from \Eref{observable_naive_DP_diagrams} 
using the propagator \Eref{dp_prop}. As $t_2>t_1$, only the first and the last diagrams of \Eref{observable_naive_DP_diagrams} contribute, the first due to a particle travelling from $\xvec_0$ to $\xvec_1$ and then to $\xvec_2$ and the last due to two particles travelling independently. 
In the limit of $t_2\downarrow t_1$ the first term, proportional to $n_0$, becomes $n_0 \delta(\xvec_2-\xvec_1) (4\pi D(t_1-t_0))^{-d/2}\exp{-(\xvec_1-\xvec_0)^2/(4\pi D (t_1-t_0)}$, vanishing if $\xvec_1\ne\xvec_2$ as the same particle cannot be at two different places simultaneously. Fig.~\ref{fig:schematic} provides a visual illustration of this property.

Although this approach no longer requires regularisation by a mass $r$, it is somewhat more demanding to perform the calculation of the correlation function within Dean's equation in direct time and real space using \Eref{dean_prop}, because the Dean-vertex requires a convolution over the time and the position where the virtual branching takes place,
\begin{align}\elabel{Dean_vertex_direct}
&        \tikz[baseline=-2.5pt]{
      \draw[Aactivity] (0,0.0) -- (0.85,0.0) node[above] {$\xvec_0,t_0$} -- +(0,0) node[right] {$\ (n_0)$} +(2pt,0) circle (2pt)[fill];
      \draw[Aactivity] (220:0.85) node[below] {$\xvec_2,t_2$} -- (0,0.0);
      \draw[Aactivity] (140:0.85) node[above] {$\xvec_1,t_1$} -- (0,0.0);
      \draw[thin] ($(0.07,0.07)+(140:0.35)$) -- ($(-0.07,-0.07)+(140:0.35)$);
      \draw[thin] ($(-0.07,0.07)+(220:0.35)$) -- ($(0.07,-0.07)+(220:0.35)$);
      \draw[dotted,thin] (140:0.35) arc (140:220:0.35);
    }
\\ \nonumber &    \corresponds
2 n_0 \int \ddint{x'}\dint{t'}
\bigg( \nabla_{\xvec'} G(\xvec_2-\xvec', t_2-t') \bigg)
\cdot
\bigg( \nabla_{\xvec'} G(\xvec_1-\xvec', t_1-t') \bigg) 
G(\xvec'-\xvec_0, t'-t_0)
\end{align}
After some algebra, Dean's equation produces of course the same correlation function \Eref{corr_direct_xt} as Doi-Peliti. 

In explicit calculations below, notably Appendix~\ref{a:induction_conn}, we will make use of a mixed momentum-time, $k,t$, parameterisation, for which we briefly outline the cancellation mechanism in the following. In Doi-Peliti, the diagrams \Eref{observable_naive_DP_diagrams} can immediately be written as
\begin{align}\elabel{observable_DP_mixed} 
	&C(\xvec_1,\xvec_2, t_1, t_2) = 
\int 
	\ddintbar{k_2}\ddintbar{k_1}
	\\
	&\nonumber\quad\times\Big\{
n_0 
\exp{ \imag (\kvec_2\cdot(\xvec_2-\xvec_1) + \kvec_1\cdot(\xvec_1-\xvec_0) ) }	
\theta(t_2-t_1)\theta(t_1-t_0) 
\exp{-D\kvec_2^2 (t_2-t_1)        -D\kvec_1^2 (t_1-t_0) ) }	
\\
&\nonumber\!\!\quad\quad+
n_0 
\exp{ \imag (\kvec_2\cdot(\xvec_2-\xvec_0) + \kvec_1\cdot(\xvec_1-\xvec_2) ) }	
\theta(t_1-t_2)\theta(t_2-t_0)
\exp{-D\kvec_1^2(t_1-t_2)        -D\kvec_2^2 (t_2-t_0) ) }	
\\
&\nonumber\!\!\!\!+
n_0(n_0-1)
\exp{ \imag (\kvec_2\cdot(\xvec_2-\xvec_0) + \kvec_1\cdot(\xvec_1-\xvec_0) ) }	
\theta(t_2-t_0)\theta(t_1-t_0)
\exp{-D\kvec_2^2(t_2-t_0)        -D\kvec_1^2 (t_1-t_0) ) }	
	\Big\}
	\end{align}
by replacing each of the bare propagators of \Eref{observable_naive_DP_simplified} by \Eref{DP_mixed_propagator} and making use of the $\delta$-functions on the momenta, or by direct interpretation of the diagrams. 

Dean's equation, \Eref{observable_naive_Dean}, on the other hand, produces 
\begin{align}\elabel{Dean_mixed_step1}
    &\nonumber C(\xvec_1,\xvec_2, t_1, t_2) =
    \int \ddintbar{k_2}\ddintbar{k_1}\ddintbar{k_0}
    \exp{\imag\kvec_2\cdot\xvec_2}
    \exp{\imag\kvec_1\cdot\xvec_1}
    \exp{\imag\kvec_0\cdot\xvec_0}\deltabar(\kvec_2+\kvec_1+\kvec_0)
     \\
    &\nonumber\times\Big\{
    (-2n_0D\kvec_1\cdot\kvec_2)
    \int_{-\infty}^\infty\dint{t'}
    \theta(t_2-t') \exp{-D\kvec_2^2(t_2-t')}
    \theta(t_1-t') \exp{-D\kvec_1^2(t_1-t')}
    \theta(t'-t_0) \exp{-D\kvec_0^2(t'-t_0)}
    \Big\}\\
    &+
    \int \ddintbar{k_2}\ddintbar{k_1}
    \exp{\imag\kvec_2\cdot(\xvec_2-\xvec_0)}
    \exp{\imag\kvec_1\cdot(\xvec_1-\xvec_0)}
    \\
    &\nonumber\qquad\qquad\times
    \Big\{
    n_0^2
    \theta(t_2-t_0) \exp{-D\kvec_2^2(t_2-t_0)}
    \theta(t_1-t_0) \exp{-D\kvec_1^2(t_1-t_0)}
    \Big\}\ ,
\end{align}
with the convolution over $t'$, the time of the virtual branching in the first diagram. While the lower limit of this integral is fixed to $t_0$ by $\theta(t'-t_0)$, the upper limit is $t_{\text{min}}=\min{(t_1,t_2)}$ via the product of two Heaviside $\theta$-functions. Its two possible values generate two terms as in \Eref{observable_naive_DP_diagrams}, conditioned by $\theta$-functions. Using the $\delta$-function in the first line of \Eref{Dean_mixed_step1} to eliminate the integral over $\kvec_0$, the $n_0$ branching terms each produce
\begin{multline}\elabel{Dean_cancellation_kt}
    \int_{t_0}^{t_{\text{min}}} 
    \dint{t'}
    \exp{-D\kvec_2^2(t_2-t')}
    \exp{-D\kvec_1^2(t_1-t')}
    \exp{-D(\kvec_1+\kvec_2)^2(t'-t_0)}
    \\=
    \exp{-D\kvec_2^2(t_2-t_0)}
    \exp{-D\kvec_1^2(t_1-t_0)}
    \left(1-\exp{-2D\kvec_1\cdot\kvec_2(t_{\text{min}}-t_0)}\right)
    \frac{1}
    {2D\kvec_1\cdot\kvec_2} \ .
\end{multline}
The $1$-term in the bracket is independent of $t_{\text{min}}$ and cancels
with $n_0$ of the $n_0^2$ disconnected terms. The remaining terms can be simplified using for example
\begin{equation}
    \exp{-D\kvec_1^2(t_1-t_0)} \exp{-2D\kvec_1\cdot\kvec_2(t_1-t_0)}
    =
    \exp{-D(\kvec_1+\kvec_2)^2(t_1-t_0)} \exp{D\kvec_2^2(t_1-t_0)}
\end{equation}
in the case of $t_{\text{min}}=t_1$ and, after a shift in $\kvec_i$, such as $\kvec_1+\kvec_2\to\kvec_1$ in the example above, reproduce the result from Doi-Peliti, \Eref{observable_DP_mixed}. This concludes the illustration.

To summarise this section, the correlation function of the particle position of $n_0$ non-interacting particles is not a single term, as it needs to capture multiple scenarios of particles moving, while keeping track of the particle nature of the constituent degrees of freedom. Both frameworks result in the same expressions, such \Erefs{observable_DP_done}, \eref{observable_Dean_done}, \eref{corr_direct_xt} and \eref{observable_DP_mixed}. A cancellation mechanism such as \Eref{Dean_cancellation} in the $\kvec,\omega$ parameterisation and the convolution in \Eref{Dean_cancellation_kt} for $\kvec,t$, connects Doi-Peliti and Dean, revealing  that the perturbative, virtual branching in Dean's framework is in fact a sum of sequential propagation of a single particle and independent propagation of two distinct ones.

The calculation in this preliminary section suggests that the interaction vertex \Eref{dean_vertex} in Dean's formalism contains the same information as the commutation relation of the Doi-Peliti ladder operators. The importance of this observation will become evident in \Srefs{dp_PI} and \ref{sec:dean_PI}, where we analyse the particle nature in greater detail.

\section{Probing for particle entity \seclabel{prob_PI}}
Within the Dean framework $\rho(\xvec,t)$ denotes the instantaneous particle number density in state $\xvec$ at time $t$. We define  particle entity as a property of the evolution equation for $\rho(\xvec,t)$ whereby this time-dependent random variable can be written as a finite sum of ``single particle densities'' with integer coefficients. In the case of a discrete phase space, the single-particle density for a particle in state $\bar{\xvec}$ is the Kronecker-delta with unit prefactor, $\delta_{\xvec,\bar{\xvec}}$. For continuous degrees of freedom, the single-particle density for a particle in state $\bar{\xvec}$ is the Dirac-delta distribution normalised to unity, $\delta(\xvec-\bar{\xvec})$. Correspondingly,
\begin{equation}
    \rho(\xvec,t) = 
    \begin{cases}
    \sum_{i} n_i(t) \delta_{\xvec,\bar{\xvec}_i}, & \text{for discrete states}  \\
    \sum_{i} n_i(t) \delta(\xvec-\bar{\xvec}_i), & \text{for continuous states} 
  \end{cases}
  \elabel{eq:discreteDensity}
\end{equation}
where $n_i(t) \in \mathbb{N}$. It follows from this requirement that the integral of the particle number density $\rho(\xvec,t)$ over any (sub-)volume $\Omega$ of the space is an integer-valued random variable,
\begin{equation} \elabel{eq:integer_integral}
    \forall \Omega \subset \mathbb{R}^d: \SumInt_\Omega \ddint{x} \rho(\xvec,t) \in \mathbb{N}~.
\end{equation}

For discrete states, \Eref{eq:integer_integral} also implies \Eref{eq:discreteDensity}, i.e. there can be no densities satisfying \Eref{eq:integer_integral} that are not a sum of Kronecker deltas with integer coefficient. We leave the proof that this equivalence also holds in the continuum for future work. Such a proof will surely draw on the arbitraryness of $\Omega$, which can be used to include or exclude from the integral in \Eref{eq:PI_signature} any part of $\rho(\xvec,t)$.
In the case of stochastic dynamics, it is convenient to re-express the condition \Eref{eq:integer_integral} in terms of an expectation value as
\begin{equation}
\elabel{eq:PI_signature}
    \left\langle \rm{exp}\left( 2\pi \imag \int_\Omega \ddint{x}  \rho(\xvec,t) \right) \right\rangle = 1 ~,
\end{equation}
which needs to be satisfied for any volume $\Omega$ and all times $t$. \Eref{eq:PI_signature} will play the role of a signature of particle entity in the following. 

Obviously, \Eref{eq:integer_integral} implies \Eref{eq:PI_signature}. Yet, expressing the particle entity condition as an expectation might appear less stringent than demanding it at the level of individual trajectories. 
However, rewriting \Eref{eq:PI_signature} as $\langle \cos(2\pi \int \ddint{x} \rho(\xvec, t))\rangle = 1$ on the basis of $\rho(\xvec,t)$ being real, shows that the integral must be integer valued almost surely, because $\Rset \ni \cos(x) \le1$ for $x\in\Rset$.

In order to ease the calculation of the left-hand side of \Eref{eq:PI_signature} for a particular field theory of interest, we can expand the complex exponential as a Taylor series and invoke linearity of the expectation to obtain the particle entity signature,
\begin{equation}
\elabel{eq:PI_mom_gen}
 \sum_{n=0}^\infty \frac{(2\pi \imag)^n}{n!} \
\tikz[baseline=-2.5pt]{
\begin{scope}[rotate=-30]
  \draw[Aactivity] (-130:0.5) -- (-130:1.3);
  \path [polarity]
  (-152:1.2cm) arc (-152:-130:1.2cm);
  \draw[Aactivity] (-160:0.5) -- (-160:1.3);
  \draw[Aactivity] (-170:0.5) -- (-170:1.3);
\end{scope}
\fill[pattern=north west lines,opacity=.6,draw] 
  (0,0) circle [radius=0.5cm];
\node [yshift=-0.5pt,xshift=0pt,fill=white] {$n$};
}   
 \corresponds
 \sum_{n=0}^\infty \frac{(2\pi \imag)^n}{n!} \left\langle \left( \int_\Omega \ddint{x}  \rho(\xvec,t) \right)^n \right\rangle 
=
\left\langle \rm{exp}\left( 2\pi \imag \int_\Omega \ddint{x}  \rho(\xvec,t) \right) \right\rangle
 = 1 ~,
\end{equation}
where the left-hand side is now a function of the $n$th full moment of the integrated particle number density, 
$\left\langle \left( \int_\Omega \ddint{x}  \rho(\xvec,t) \right)^n \right\rangle$. 
In \Eref{eq:PI_mom_gen} we have also introduced the diagrammatic notation for the $n$th full moment of the \emph{integrated} particle number density. The hatched vertex henceforth indicates generally a sum of possibly disconnected diagram involving an arbitrary amount of sources. More specifically, here it is the $n$-fold spatial integral of a sum of products of connected diagrams. An example for such a sum is \Eref{observable_naive_Dean}. 
These moments are perfectly suited for being calculated in both Doi-Peliti and response field field theories, as done in the following. 

An alternative form of our particle entity signature can be obtained by recognising that the left hand side of \Eref{eq:PI_signature} is also the moment-generating function $\langle \exp{z X}\rangle$ of the random variable $X = \int_\Omega \dint{\xvec}  \rho(\xvec,t)$ evaluated at $z = 2 \pi \imag$. It is a well-known result from the field-theoretic literature on equilibrium critical phenomena \cite{lebellac1991,binney1992} that the generating function of the full moments $\langle X^n \rangle$ can be expressed as the exponential of the generating function of the so-called connected moments, denoted $\langle X^n \rangle_c$. Outside the field-theoretic literature, connected moments are usually referred to as cumulants \cite{van1992stochastic}. While one would normally expect source fields $j(\xvec,t)$ to be introduced corresponding to a conjugate variable $z$ at each point in space and time, in the present context a single variable suffices, as in \Eref{eq:PI_mom_gen} every field $\rho(\xvec,t)$ is integrated over the same volume $\Omega$. Diagrammatically,
\begin{multline}
\elabel{eq:gen_fn_rel}
    \sum_{n=0}^\infty \frac{z^n}{n!} 
    \tikz[baseline=-2.5pt]{
\begin{scope}[rotate=-30]
  \draw[Aactivity] (-130:0.5) -- (-130:1.3);
  \path [polarity]
  (-152:1.2cm) arc (-152:-130:1.2cm);
  \draw[Aactivity] (-160:0.5) -- (-160:1.3);
  \draw[Aactivity] (-170:0.5) -- (-170:1.3);
\end{scope}
\fill[pattern=north west lines,opacity=.6,draw] 
  (0,0) circle [radius=0.5cm];
\node [yshift=-0.5pt,xshift=0pt,fill=white] {$n$};
}  
\corresponds
\sum_{n=0}^\infty \frac{z^n}{n!} \ave{\left( 
\int_\Omega \ddint{x} \rho(\xvec,t)\right)^n}=
\ave{\Exp{z\int_\Omega\ddint{x}\rho(\xvec,t)}}
\\
=
\Exp{
\sum_{n=1}^\infty \frac{z^n}{n!} \left\langle\left( 
\int_\Omega \ddint{x} \rho(\xvec,t)\right)^n\right\rangle_c
}
=
 \Exp{ \sum_{n=1}^\infty \frac{z^n}{n!} 
    \tikz[baseline=-2.5pt]{
\begin{scope}[rotate=-30]
  \draw[Aactivity] (-130:0.5) -- (-130:1.3);
  \path [polarity]
  (-152:1.2cm) arc (-152:-130:1.2cm);
  \draw[Aactivity] (-160:0.5) -- (-160:1.3);
  \draw[Aactivity] (-170:0.5) -- (-170:1.3);
\end{scope}
\draw[thick] (0,0) circle (0.5cm);
\draw[Aactivity] (0:0.5) -- (0:0.8) +(2pt,0) circle (2pt)[fill];
\node [yshift=-0.5pt,xshift=0pt] {$n$};
}  
    } ~,
\end{multline}
where 
we have introduced the notation for the $n$th {\it connected} moment of the integrated particle number density, shown as a circular vertex on the right, which differs from that of the corresponding full moment also by the presence of an explicit, single, ingoing propagator emerging from a single source, shown as a filled circle, the only possible form of a connected diagram contributing to  moments of the density.
In equilibrium statistical mechanics, this relationship provides the connection between the partition function and the Helmholtz free energy \cite{lebellac1991,binney1992}. 
Generating functions of observables such as $n$-point correlation functions of $\rho(\xvec,t)$ can be reduced to those of connected diagrams as long as the observables can be written as (functional) derivatives of an exponential and provided that each resulting diagram can be written as a product of connected diagrams. Under these conditions, \Eref{eq:gen_fn_rel} does all the right accounting. 

Evaluating \Eref{eq:gen_fn_rel} at $z=2\pi\imag$, according to \Eref{eq:PI_mom_gen}
one can write
\begin{equation}\elabel{eq:gen_fn_rel_criterion}
    \rm{exp}\left( \sum_{n=1}^\infty \frac{(2\pi \imag)^n}{n!} \left\langle \left( \int_\Omega \ddint{x}  \rho(\xvec,t) \right)^n \right\rangle_c \right) = 1 \ ,
\end{equation}
or, equivalently,
\begin{equation}
\elabel{eq:PI_connected}
\sum_{n=1}^\infty \frac{(2\pi \imag)^n}{n!} \ 
\tikz[baseline=-2.5pt]{
\begin{scope}[rotate=-30]
  \draw[Aactivity] (-130:0.5) -- (-130:1.3);
  \path [polarity]
  (-152:1.2cm) arc (-152:-130:1.2cm);
  \draw[Aactivity] (-160:0.5) -- (-160:1.3);
  \draw[Aactivity] (-170:0.5) -- (-170:1.3);
\end{scope}
\draw[thick] (0,0) circle (0.5cm);
\draw[Aactivity] (0:0.5) -- (0:0.8) +(2pt,0) circle (2pt)[fill];
\node [yshift=-0.5pt,xshift=0pt] {$n$};
}  
\corresponds
    \sum_{n=1}^\infty \frac{(2\pi \imag)^n}{n!} \left\langle \left( \int_\Omega \ddint{x}  \rho(\xvec,t) \right)^n \right\rangle_c  
   = 2\pi \imag \ell
\end{equation}
for some integer $\ell \in \mathbb{Z}$, on the basis of the connected moments of the particle number, to be compared to the particle signature on the basis of the full moments, \Eref{eq:PI_mom_gen}.

\section{Particle entity in Doi-Peliti \seclabel{dp_PI}}
Doi-Peliti field theories are designed to respect particle entity and they do indeed do so on a rather fundamental level. To probe for particle entity, we want to use \Eref{eq:PI_signature} with $\rho(\xvec,t)$ replaced by an object suitable for a Doi-Peliti field theory. In such a field theory, the instantaneous particle number at any position $\xvec$ is probed by the number operator $\hat{n}(\xvec) = a^\dagger(\xvec)a(\xvec)$. The expected particle number at position $\xvec$ and time $t$
is therefore
\begin{equation}
    \langle n(\xvec,t) \rangle = \bra{\abyss} a^\dagger(\xvec)a(\xvec) \ket{\Psi (t)} \ ,
\end{equation}
using a continuum version of the notation introduced in \Sref{setup_dp_dean}, in particular \Eref{eq:def_DP_state}.While this expectation might be any non-negative real, the instantaneous $a^\dagger(\xvec)a(\xvec)$ is an integer. As already discussed in Section \Sref{prob_PI}, \Eref{eq:PI_signature}, we therefore expect 
\begin{equation}\elabel{def_OC}
    \bra{\abyss} \Exp{2\pi\imag  a^\dagger(\xvec)a(\xvec)} \ket{\Psi (t)} = 1
\end{equation}
to hold for every $\xvec$, as $\exp{2\pi\imag n}=1$ for any $n\in\Zset$. If this holds for every point $\xvec$, it also holds for every patch $\Omega$, since
\begin{equation}
    \bra{\abyss} \Exp{2\pi\imag  \sum_{\xvec \in \Omega} a^\dagger(\xvec)a(\xvec)} \ket{\Psi (t)} =  \bra{\abyss} \prod_{\xvec \in \Omega}  \Exp{2\pi\imag  a^\dagger(\xvec)a(\xvec)} \ket{\Psi (t)} ~,
    \elabel{eq:PI_doi_peliti}
\end{equation}
where we have used that operators at different $\xvec$ commute.
In the continuum, one might argue that the particle number at $\xvec$ can only ever be $0$ or $1$, possibly leading to some simplifications, but on the lattice occupation is not bound to be sparse in this sense.

To show that \Eref{def_OC} is indeed satisfied in general Doi-Peliti field theories, we follow the standard procedure, outlined in \Eref{eq:dp_path_in}, to express the operator $\Exp{2\pi\imag  a^\dagger(\xvec)a(\xvec)}$ in terms of scalar fields $\psi(\xvec,t)$ and $\psi^\dagger(\xvec,t)$. The simple mapping of operator to field applies as soon as the operators are normal ordered,
\begin{align}
    \Exp{z  a^\dagger(\xvec)a(\xvec)} &= \sum_{n=0}^\infty \frac{1}{n!} z^n  \left(a^\dagger(\xvec)a(\xvec)\right)^n\\
    &=\sum_{n=0}^\infty \frac{1}{n!} z^n \sum_{k=0}^n \stirling{n}{k} \left( a^{\dagger}(\xvec)\right)^k a(\xvec)^k
\end{align}
where we have replaced $2\pi\imag$ by $z$ to improve readability and
used \cite{pausch2019topics} to normal order $\left(a^\dagger(\xvec)a(\xvec)\right)^n$. In terms of fields, the observable \Eref{eq:PI_doi_peliti} is thus
\begin{align}
   \OC&= \bra{\abyss} \Exp{z \sum_{x\in \Omega} \left(a^{\dagger}(\xvec)\right)^k a(\xvec)^k} \ket{\Psi(\xvec,t)}\\
   &=\bra{\abyss} \prod_{x\in \Omega} \sum_{n=0}^{\infty} \frac{1}{n!} z^n \sum_{k=0}^{n} \stirling{n}{k} \left(a^{\dagger}(\xvec)\right)^k a(\xvec)^k \ket{\Psi(\xvec,t)} \\
   \elabel{OC_DP_pre_final}
   &=\ave{\prod_{x\in \Omega} \sum_{n=0}^{\infty} \frac{1}{n!} z^n \sum_{k=0}^{n} \stirling{n}{k} \psi^k(\xvec,t)}\\
   \elabel{OC_DP_final}
   &=\ave{\Exp{\sum_{x\in \Omega}\psi(\xvec,t) (e^z-1)}}
\end{align}
as $\bra{\abyss} (a^{\dagger}(\xvec))^k = \bra{\abyss}$ \cite{gunnarnotes}. 
From \Eref{OC_DP_pre_final} to \eref{OC_DP_final}, we draw on the the mixed bivariate generating function for the Stirling numbers of the second kind \cite{comtet2012},
\begin{equation}\elabel{Stirling_id_DP}
    \sum_{n=0}^\infty \sum_{k=0}^n \stirling{n}{k} \frac{1}{n!} z^n y^k = \Exp{y(\exp{z}-1)} \ .
\end{equation}
For $z=2\pi\imag$, and any integer multiple thereof, \Eref{OC_DP_final} indeed produces $\OC=1$ as required by \Eref{eq:PI_signature}. 
Because this calculation never draws on any particular action, but rather on the fundamentals of normal ordering, Doi-Peliti field theories respect particle entity universally in the presence of any interactions and potentials.

\section{Particle entity in response field theories: Dean's equation \seclabel{dean_PI}}
Demonstrating that the response field theory derived from Dean's equation possesses particle entity turns out to be a much more challenging task, which requires us to compute explicitly the connected moments of the integrated particle number density to arbitrary order. This calculation draws on the specific action 
\Eref{dean_action} and \eref{Dean_action_simple_komega} as
we perform it here for the case of non-interacting diffusive particles without external potential.
This is done most conveniently by first computing the connected moments of the  density in the mixed momentum-time representation, where Dean's action reads 
\begin{align}
    A[\rho,\tilde{\rho}] 
    &= \int \ddintbar{k} \dint{t} \tilde{\rho}(\kvec,t) (\partial_t + D \kvec^2) \rho(-\kvec,t) 
    - \tilde{\rho}(\kvec,t) \sum_i n_i e^{\imag \kvec \cdot \xvec_i} \delta(t-t_i) \nonumber \\
    &\quad + \int \ddintbar{k} \ddintbar{k'} \dint{t} D (\kvec \cdot \kvec') \tilde{\rho}(\kvec,t) \tilde{\rho}(\kvec',t) \rho(-(\kvec+\kvec'),t).
    \elabel{Dean_action_mixed}
\end{align}
In this parametrisation, we find (Appendix~\ref{a:induction_conn}, \Eref{resum_rule_final})
\begin{align}\elabel{conjecture_Connected_k_t}
&\langle \rho(\kvec_1,t)...\rho(\kvec_n,t)
\rangle_c
= n_0 \theta(t-t_0) 
\knullfix{\kvec_1+...\kvec_n}
\sum_{m=1}^{n} (-1)^{m-1} (m-1)! \\
&\qquad
\times
\sum_{\{\Pset{P}_1,\dots,\Pset{P}_m\}\in \partitionX{\{1,\dots,n\}}{m}} 
\exp{-T(t-t_0) \sum_{i=1}^m \setvecX{\Pset{P}_i}^2}
\nonumber
\,.
\elabel{conjecture_Connected_k_t}
\end{align}
with $\partitionX{\{1,\dots,n\}}{m}$ the set of all partitions of the
set $\{1,\dots,n\}$ into $m$ non-empty, distinct subsets $\Pset{P}_i$ with
$i=1,...,m$, \ie $\union_{i=1}^m \Pset{P}_i=\{1,2,\dots,n\}$ and $\Pset{P}_i \cap \Pset{P}_j = \emptyset$ for $i\ne j$.
The sum thus runs over all possible partitions of \{1,2,\dots,n\}
into $m$ non-empty sets. There is one partition for $m=n$ and $n$ for
$m=1$.
The vector featuring in the right-most exponential of \Eref{conjecture_Connected_k_t} 
\begin{equation}\elabel{def_setvec}
\setvecX{\Pset{P}_i} = \sum_{p\in\Pset{P}_i} \kvec_p
\end{equation}
is the total momentum given by the
indices in the subset $\Pset{P}_i$, \ie it is the total momentum of
the subset $\Pset{P}_i$, and by linearity, $\setvec{A}+\setvec{B}=\setvecX{\Pset{A}\union\Pset{B}}$.
For example, one partition into two of
$\{1,2,3,4\}$ is  
$\{\Pset{P}_1=\{1,2,4\},\Pset{P}_2=\{3\}\}$, 
which is one of $7$ elements of $\partitionX{\{1,2,3,4\}}{2}$.
In this example, the momenta of the subsets are
$\setvecX{\Pset{P}_1} = \kvec_1+\kvec_2+\kvec_4$
and
$\setvecX{\Pset{P}_2} = \kvec_3$. OCT 2021 
Diagrammatically, the right-hand side of \Eref{conjecture_Connected_k_t}
is obtained by summing over all connected, topologically distinct
diagrams with a single incoming propagator and $n$ outgoing propagators labelled by the external momenta $\kvec_i$ ($i=1,...,n$), where we need to account for all non-equivalent permutations of the latter. 

The connected moments of the integrated particle number density in a patch $\Omega$ are then obtained by Fourier back-transforming \Eref{conjecture_Connected_k_t} into position-time representation and integrating over the probing locations $\xvec_i \in \Omega$,
\begin{align}
&\left\langle \int_\Omega \ddint{x}_1...\ \ddint{x}_n \rho(\xvec_1,t)...\rho(\xvec_n,t)
\right\rangle_c \\
=& \int_\Omega \prod_{i=1}^n \ddint{x_i} \int \prod_{j=0}^n \ddintbar{k_j} \Exp{-\imag \sum_{\ell=1}^n \kvec_\ell\cdot\xvec_\ell }\langle \rho(\kvec_1,t)...\rho(\kvec_n,t)
\rangle_c \elabel{ref_for_app} \\
\elabel{corr_int_stp}
=&\int_\Omega \prod_{i=1}^n \ddint{x_i} \int \prod_{j=1}^n \ddintbar{k_j}
\Exp{-\imag \sum_{\ell=1}^n \kvec_\ell\cdot(\xvec_\ell-\xvec_0) }
\\
\nonumber
&\qquad\times
n_0 \theta(t-t_0) \sum_{m=1}^{n} (-1)^{m-1} (m-1)!
\sum_{\{\Pset{P}_1,\dots,\Pset{P}_m\}\in \partitionX{\{1,\dots,n\}}{m}} 
\exp{-D(t-t_0) \sum_{p=1}^m \setvecX{\Pset{P}_p}^2}~.
\end{align}

The integrals in \Eref{corr_int_stp} can be carried out partition by partition, by taking the integration inside the summation over $m=1,\dots,n$ and the partitions $\{\Pset{P}_1,\dots\}\in \partitionX{\{1,\dots,n\}}{m}$. As $\xvec_i$ and $\kvec_j$ are both dummy variables, we may think of subset $\Pset{P}_p$ containing indices $1,\dots,a$ with $a=\cardinalityX{\Pset{P}_p}$, so that the integrals to be carried out for each $p=1,\dots,m$ are
\begin{equation}
J_p=
    \int_\Omega \prod_{i=1}^a \ddint{x_i} \int \prod_{j=1}^a \ddintbar{k_j}
    \Exp{-\imag \sum_{\ell=1}^a \kvec_\ell\cdot(\xvec_\ell-\xvec_0) }
\exp{-D(t-t_0) \setvecX{\Pset{P}_p}^2}~.
\end{equation}
In this indexing we have $\setvecX{\Pset{P}_p}=\kvec_1+\dots+\kvec_a$ which simplifies to $\tilde{\kvec}_1$ after suitable shifting of the origin in the integration over $\kvec_1=\tilde{\kvec}_1-(\kvec_2+\dots+\kvec_a)$, so that
\begin{subequations}
\begin{align}
J_p&=
    \int_\Omega \prod_{i=1}^a \ddint{x_i} \int 
    \ddintbar{\tilde{k}_1}
    \exp{-\imag \tilde{\kvec}_1\cdot(\xvec_1-\xvec_0) }
    \exp{-D(t-t_0) \tilde{\kvec}_1^2}\\
    \nonumber&\qquad\times
\int    \prod_{j=2}^a \ddintbar{k_j}
    \Exp{-\imag \sum_{\ell=2}^a \kvec_\ell\cdot(\xvec_\ell-\xvec_1) }\\
    \elabel{J_p_using_G_Fourier}
    &=
    \int_\Omega \prod_{i=1}^a \ddint{x_i}
    G(\xvec_1 - \xvec_0,t-t_0)
    \delta(\xvec_2-\xvec_1) ... \delta(\xvec_a-\xvec_1)\\
    \elabel{J_p_final}
    &=
    I_\Omega(t-t_0)
\end{align}
\end{subequations}
where we have used \Erefs{dean_prop} and \eref{Dean_mixed_propagator} in \Eref{J_p_using_G_Fourier} and introduced 
\begin{equation}\elabel{def_I_Omega}
    I_\Omega(t-t_0)=\int_{\Omega} \ddint{x} G(\xvec - \xvec_0,t-t_0) ~,
\end{equation}
in \Eref{J_p_final}, which is the probability to find a particle at time $t$ within the volume $\Omega$ that had at time $t_0$ been placed at $\xvec_0$. We may drop the time-dependence of $I_\Omega$ where that improves readability.

As $J_p$ is independent of the specific partition,
the sum over 
partitions $\{\Pset{P}_1,\dots,\Pset{P}_m\}\in\partitionX{\{1,\dots,n\}}{m}$
in \Eref{corr_int_stp} amounts to multiplying a product of $m$  such integrals by the number of partitions, given by the Stirling numbers of the second kind. Overall,
\begin{equation}
\elabel{n_point_connected}
    \tikz[baseline=-2.5pt]{
\begin{scope}[rotate=-30]
  \draw[Aactivity] (-130:0.5) -- (-130:1.3);
  \path [polarity]
  (-152:1.2cm) arc (-152:-130:1.2cm);
  \draw[Aactivity] (-160:0.5) -- (-160:1.3);
  \draw[Aactivity] (-170:0.5) -- (-170:1.3);
\end{scope}
\draw[thick] (0,0) circle (0.5cm);
\draw[Aactivity] (0:0.5) -- (0:0.8) +(2pt,0) circle (2pt)[fill];
\node [yshift=-0.5pt,xshift=0pt] {$n$};
}
\corresponds
\left\langle \left( \int_\Omega \ddint{x}  \rho(\xvec,t)  \right)^n \right\rangle_{c} 
= - n_0 
\theta(t-t_0)
\sum_{m=1}^n \big(-I_\Omega\big)^m  \stirling{n}{m} (m-1)!\ ,
\end{equation}
which provides us with the information we need to probe the theory for particle entity. It is a key result of the present work. Its derivation is generalised to distinct initial positions in Appendix~\ref{a:mult_strt_pt}.

Using the particle entity signature based on connected diagrams, \Eref{eq:PI_connected}, confronts us with some undesirable hurdles due to convergence.
We therefore turn our attention to the full moments, which can be constructed from the connected moments via \Eref{eq:gen_fn_rel}, so that for $t>t_0$,
\begin{align}
\tikz[baseline=-2.5pt]{
\begin{scope}[rotate=-30]
  \draw[Aactivity] (-130:0.5) -- (-130:1.3);
  \path [polarity]
  (-152:1.2cm) arc (-152:-130:1.2cm);
  \draw[Aactivity] (-160:0.5) -- (-160:1.3);
  \draw[Aactivity] (-170:0.5) -- (-170:1.3);
\end{scope}
\fill[pattern=north west lines,opacity=.6,draw] 
  (0,0) circle [radius=0.5cm];
\node [yshift=-0.5pt,xshift=0pt,fill=white] {$n$};
} 
&\corresponds
\ave{\left( 
\int_\Omega \ddint{x} \rho(\xvec,t)\right)^n}\\
&= \left. \frac{d^n}{dz^n} \right|_{z=0} \Exp{ \sum_{m=1}^\infty \frac{z^m}{m!} 
    \left\langle\left( 
    \int_\Omega \ddint{x} \rho(\xvec,t)\right)^m\right\rangle_c
    } \elabel{con_der_1}\\
&= \left. \frac{d^n}{dz^n} \right|_{z=0} \Exp{\sum_{m=1}^\infty \frac{z^m}{m!} (-n_0)  \sum_{\ell=1}^{m} (-I_\Omega)^\ell  \stirling{m}{\ell} (\ell-1)!
    } \elabel{con_der_2} \\
\elabel{con_der_2b}
&= \left. \frac{d^n}{dz^n} \right|_{z=0} \Exp{-n_0 
\sum_{\ell=1}^{\infty} (-I_\Omega)^\ell (\ell-1)! \sum_{m=\ell}^\infty \frac{z^m}{m!}  \stirling{m}{\ell}} \, ,
\end{align}
where we have changed the order of summation in the exponential to arrive at the final equality. This step deserves further scrutiny below. 
Using in \Eref{con_der_2b} the generating function of the Stirling numbers \cite{comtet2012}  in the form
\begin{equation}
\elabel{stirl_gen_f}
    \sum_{m=\ell}^\infty \frac{z^m}{m!} 
    \stirling{m}{\ell} 
    = \frac{(e^z - 1)^\ell}{\ell!}~,
\end{equation}
as used previously in the Doi-Peliti field theory, \Eref{Stirling_id_DP},
leads to 
\begin{align}\elabel{con_der_2c}
\tikz[baseline=-2.5pt]{
\begin{scope}[rotate=-30]
  \draw[Aactivity] (-130:0.5) -- (-130:1.3);
  \path [polarity]
  (-152:1.2cm) arc (-152:-130:1.2cm);
  \draw[Aactivity] (-160:0.5) -- (-160:1.3);
  \draw[Aactivity] (-170:0.5) -- (-170:1.3);
\end{scope}
\fill[pattern=north west lines,opacity=.6,draw] 
  (0,0) circle [radius=0.5cm];
\node [yshift=-0.5pt,xshift=0pt,fill=white] {$n$};
} 
&\corresponds\left. \frac{d^n}{dz^n} \right|_{z=0} \Exp{-n_0 \sum_{\ell=1}^{\infty}(-I_\Omega )^\ell  \frac{(e^z-1)^\ell}{\ell}} \, .
\end{align}
We briefly return to the change of the order of summartion from \Eref{con_der_2} to \eref{con_der_2b}. Top justify thius, we require \emph{absolute} convergence
\begin{equation}
    \sum_{\ell=1}^\infty \sum_{m=\ell}^\infty \frac{|z|^m}{m!} I_\Omega^\ell \stirling{m}{\ell} (\ell-1)!<\infty
\end{equation}
for $I_\Omega\in[0,1]$ and $z$ within a finite vicinity around the origin, given the repeated differentiation in \Eref{con_der_2c}.
As \Eref{stirl_gen_f} holds for all $z\in\Cset$, we require $\sum_{\ell=1}^\infty I_\Omega^\ell (\Exp{|z|}-1)^\ell/\ell<\infty$ and thus $|z|<\ln(2)$ by the ratio test. 

Rewriting the exponent in \Eref{con_der_2c} for any $|z|<\ln(2)$ as a logarithm,
\begin{align}
\sum_{l=1}^\infty \frac{(-x)^l}{l}  = - \log(1+x) \, ,
\end{align}
we have
\begin{align}\elabel{con_der_3_top}
\tikz[baseline=-2.5pt]{
\begin{scope}[rotate=-30]
  \draw[Aactivity] (-130:0.5) -- (-130:1.3);
  \path [polarity]
  (-152:1.2cm) arc (-152:-130:1.2cm);
  \draw[Aactivity] (-160:0.5) -- (-160:1.3);
  \draw[Aactivity] (-170:0.5) -- (-170:1.3);
\end{scope}
\fill[pattern=north west lines,opacity=.6,draw] 
  (0,0) circle [radius=0.5cm];
\node [yshift=-0.5pt,xshift=0pt,fill=white] {$n$};
} 
&\corresponds\left. \frac{d^n}{dz^n} \right|_{z=0} \left( 1+  (e^z-1) I_\Omega \right)^{n_0} \\
& = \left. \frac{d^n}{dz^n} \right|_{z=0} \sum_{k=0}^{n_0} \binom{n_0}{k}  (e^z-1)^k I_\Omega^k \\
&=  \sum_{k=0}^{n_0} \binom{n_0}{k} I_\Omega^k \left. \frac{d^n}{dz^n} \right|_{z=0} \sum_{j = 0}^k \binom{k}{j} (-1)^j e^{z(k-j)}\\
&= \sum_{k=0}^{n_0} \binom{n_0}{k} I_\Omega^k \sum_{j = 0}^k \binom{k}{j} (-1)^j(k-j)^n \, ,
\elabel{con_der_3}
\end{align}
and using the definition of the Stirling numbers of the second kind \cite{comtet2012}
\begin{equation}\elabel{bin2stir}
     \sum_{g=0}^k {k \choose g} (-1)^g (k-g)^n = k! \stirling{n}{k} \, ,
\end{equation}
we finally arrive at
\begin{align}
\tikz[baseline=-2.5pt]{
\begin{scope}[rotate=-30]
  \draw[Aactivity] (-130:0.5) -- (-130:1.3);
  \path [polarity]
  (-152:1.2cm) arc (-152:-130:1.2cm);
  \draw[Aactivity] (-160:0.5) -- (-160:1.3);
  \draw[Aactivity] (-170:0.5) -- (-170:1.3);
\end{scope}
\fill[pattern=north west lines,opacity=.6,draw] 
  (0,0) circle [radius=0.5cm];
\node [yshift=-0.5pt,xshift=0pt,fill=white] {$n$};
} 
&\corresponds \sum_{k=0}^{n_0} \binom{n_0}{k} I_\Omega^k k! \stirling{n}{k}  \,.
\elabel{con_der_4}
\end{align}
This is the central result of the present section.

The $n$th full moment, in the form of the right hand side \Eref{con_der_4}, has an instructive physical interpretation drawing on $n_0$ being integer. To see this, we write the $n$th moment of the particle number as 
\begin{equation}\elabel{moment_from_rho}
\ave{\left( 
\int_\Omega \ddint{x} \rho(\xvec,t)\right)^n}
=
\int_\Omega \ddint{x'_1}\ddint{x'_2}\ldots\ddint{x'_n}
\ave{
\rho(\xvec'_1,t) \rho(\xvec'_2,t)\ldots \rho(\xvec'_n,t)
}
\end{equation}
with each density $\rho(\xvec'_i,t)$ considered a random variable as a function of the positions $x_j(t)$ of $n_0$ particles indexed by $j=1,2,\ldots,n_0$
\begin{equation}\elabel{rho_is_a_sum_of_deltas}
\rho(\xvec'_i,t) = \sum_{j=1}^{n_0} \delta(\xvec_i'-\xvec_j(t)) \ ,
\end{equation}
generating $n_0^n$ terms of products of $\delta$-functions in \Eref{moment_from_rho}.
To calculate the right-hand side of \Eref{moment_from_rho} on the basis of \Eref{rho_is_a_sum_of_deltas} using \Eref{def_I_Omega} in the form 
\begin{equation}
    \int_\Omega \ddint{x'} \ave{\delta(\xvec'-\xvec_j(t)} = I_\Omega(t-t_0)
\end{equation}

requires careful bookkeeping of how often each particle coorrdinate $\xvec_j(t)$ is repeated. For example
\begin{subequations}
\elabel{expectation_I_power}
\begin{align}
\elabel{expectation_two_indep}
\int_\Omega \ddint{x'_1}\ddint{x'_2} \ave{
\delta(\xvec'_1-\xvec_1(t))\delta(\xvec'_2-\xvec_2(t))
}
&= I_\Omega^2\\
\elabel{expectation_one_indep}
\int_\Omega \ddint{x'_1}\ddint{x'_2} \ave{
\delta(\xvec'_1-\xvec_1(t))\delta(\xvec'_2-\xvec_1(t))
}
&= I_\Omega
\end{align}
\end{subequations}
as particle coordinates are independent in \Eref{expectation_two_indep}, but not in \Eref{expectation_one_indep}, where in fact 
$\delta(\xvec'_1-\xvec_1(t))\delta(\xvec'_2-\xvec_1(t))=
\delta(\xvec'_1-\xvec_1(t))\delta(\xvec'_2-\xvec'_1)$.
To calculate the right-hand side of \Eref{moment_from_rho} thus is a matter of allowing for $k=1,2,\ldots,n$ distinct particle coordinates $x_j(t)$ from each of the $n$ sums \Eref{rho_is_a_sum_of_deltas}. There are 
$n_0(n_0-1)\cdot\ldots\cdot(n_0-k+1)={n_0\choose k}k!$ 
such choices. As each of the $\rho(\xvec_i',t)$ is a function of a different dummy variable, they have to be distributed among the $k$ distinct particle coordinates. There are 
$\stirling{n}{k}$ choices for that. The integration produces $I_\Omega^k$ depending on the number $k$ of distinct particle coordinates following the reasoning for \Eref{expectation_I_power}. 
In summary, we arrive at
\begin{equation}\elabel{con_der_4_by_hand}
    \ave{\left( 
\int_\Omega \ddint{x} \rho(\xvec,t)\right)^n}
=
\sum_{k=0}^n 
{n_0\choose k} I_\Omega^k k! \stirling{n}{k} 
\ ,
\end{equation}
including $k=0$ in the summation to cover the special case of $n=0$. \Eref{con_der_4_by_hand}
is subtly different from \Eref{con_der_4}, as the upper limit of the sum in  \Eref{con_der_4} is $n_0$, while it is $n$ in \Eref{con_der_4_by_hand}.
However, $\stirling{n}{k}$ in \Eref{con_der_4} vanishes if $k$ exceeds $n$ and ${n_0\choose k}$ in \Eref{con_der_4} vanishes if \emph{integer} $k$ exceeds \emph{integer} $n_0$, \ie in both sums the upper limit may be replaced by $\min{(n,n_0)}$, provided only $n_0$ is integer. We conclude that the full moment \Eref{con_der_4} has a sensible interpretation in terms of particle numbers only if $n_0$ is integer.

A second form of \Eref{con_der_4} can be derived,
which offers a more immediate physical interpretation. To this end, we draw on the identity 
\begin{equation}\elabel{binomial_rewrite}
    \sum_{k=0}^{n_0} {n_0 \choose k} (1-x)^{n_0-k} x^k k^n
    =
    \sum_{\ell=0}^{n_0} {n_0\choose \ell} x^\ell \ell! \stirling{n}{\ell}
\end{equation}
which can be obtained by resolving the binomial $(1-x)^{n_0-k}$ on the left into a sum, collecting terms of order $x^{\ell}$ and then using ${n_0 \choose k}{n_0-k \choose \ell-k}={n_0 \choose \ell}{\ell \choose k}$ to arrive at an expression that simplifies to the final sum by means of \Eref{bin2stir}.

Using \Eref{binomial_rewrite} in \Eref{con_der_4} gives 
\begin{equation}\elabel{eq:phys_pict}
    \ave{\left( 
\int_\Omega \ddint{x} \rho(\xvec,t)\right)^n}
=
\sum_{k=0}^{n_0}
{n_0 \choose k} 
\bigg(1-I_\Omega(t-t_0)\bigg)^{n_0-k}
\bigg(I_\Omega(t-t_0) \bigg)^k
k^n \ ,
\end{equation}
which has a rather simple physical interpretation: of the (integer) $n_0$ particles, $k$ can be found in the volume $\Omega$, each independently with probability $I_\Omega$, so that the probability of such a configuration is that of a repeated Bernoulli trial, ${n_0 \choose k} 
(1-I_\Omega)^{n_0-k}
I_\Omega^k$. As $k$ particles are in the relevant volume, the contribution to the $n$th particle number moment is $k^n$.

In the form \Eref{eq:phys_pict}, the particle number moments can be used in our particle entity signature \Eref{eq:PI_mom_gen},
\begin{align}
\elabel{particle_entity_full_moments_first}
    \sum_{n=0}^\infty \frac{(2\pi\imag)^n}{n!} 
    \tikz[baseline=-2.5pt]{
\begin{scope}[rotate=-30]
  \draw[Aactivity] (-130:0.5) -- (-130:1.3);
  \path [polarity]
  (-152:1.2cm) arc (-152:-130:1.2cm);
  \draw[Aactivity] (-160:0.5) -- (-160:1.3);
  \draw[Aactivity] (-170:0.5) -- (-170:1.3);
\end{scope}
\fill[pattern=north west lines,opacity=.6,draw] 
  (0,0) circle [radius=0.5cm];
\node [yshift=-0.5pt,xshift=0pt] {$n$};
}  
    &= \sum_{n=0}^\infty \frac{(2\pi\imag)^n}{n!}  \sum^{n_0}_{k=0}\binom{n_0}{k }(1-I_\Omega)^{n_0-k}
I_\Omega^k k^n  \\
    \elabel{particle_entity_full_moments}
    &=  \sum_{k=0}^{n_0} \binom{n_0}{k } (1-I_\Omega)^{n_0-k}
I_\Omega^k \left( \sum_{n=0}^\infty \frac{(2\pi\imag)^n}{n!} k^n \right)  = 1~,
\end{align}
where we have used $e^{2\pi\imag k} = 1$ for $k \in \mathbb{Z}$ in the last bracket and the normalisation of a binomial distribution. To change the order of summation going from \Eref{particle_entity_full_moments_first} to \eref{particle_entity_full_moments} we draw on the absolute convergence of the last line \Eref{particle_entity_full_moments}. This concludes the proof of particle entity on the basis of \Eref{eq:PI_mom_gen} in the Dean formalism.

We have seen how the falling factorials $n_0(n_0-1)\cdot\ldots\cdot(n_0-k+1)={n_0\choose k}k!$
emerge in Dean's theory by computing the connected moments, \Eref{con_der_4} and also \Eref{con_der_4_by_hand}, explicitly. This structure is another signature of particle entity, in the sense that it highlights the special role played by the integer nature of the initial particle number $n_0$. Unsurprisingly, it similarly emerges in Doi-Peliti. In particular, the full moments of the particle number at position $\xvec$ are given by
\begin{equation} \elabel{eq:full_m_dp}
    \tikz[baseline=-2.5pt]{
\begin{scope}[rotate=-30]
  \draw[Aactivity] (-130:0.5) -- (-130:1.3);
  \path [polarity]
  (-152:1.2cm) arc (-152:-130:1.2cm);
  \draw[Aactivity] (-160:0.5) -- (-160:1.3);
  \draw[Aactivity] (-170:0.5) -- (-170:1.3);
\end{scope}
\fill[pattern=north west lines,opacity=.6,draw] 
  (0,0) circle [radius=0.5cm];
\node [yshift=-0.5pt,xshift=0pt] {$n$};
}  \corresponds \bra{\abyss}  \big(a^\dagger(\xvec) a(\xvec)\big)^n
\exp{\hat{A}(t-t_0)}
\big(a^\dagger(\xvec_0)\big)^{n_0} 
\ket{0}
~,
\end{equation}
where $\hat{A}$ denotes the time evolution operator of \Eref{eq:shroed_im_t}.
We then use 
\begin{equation}
   \big(a^\dagger(\xvec_0)\big)^{n_0} = \big(\tilde{a}(\xvec_0) + 1\big)^{n_0} = \sum_{k=0}^{n_0} {n_0 \choose k} \big(\tilde{a}(\xvec_0)\big)^k
\end{equation}
together with \cite{garcia2018field}
\begin{equation}
    \bra{\abyss} (a^\dagger(\xvec) a(\xvec))^n = \sum_{\ell=0}^n \stirling{n}{\ell} \bra{\abyss} \big(a(\xvec) \big)^\ell
\end{equation}
to rewrite the right-hand side of \Eref{eq:full_m_dp} as
\begin{align}
    \bra{\abyss} (a^\dagger(\xvec) a(\xvec))^n
    \exp{\hat{A}(t-t_0)}
    (a^\dagger(\xvec_0))^{n_0} \ket{0} &= \sum_{\ell=0}^n \stirling{n}{\ell} \sum_{k=0}^{n_0} {n_0 \choose k} \bra{\abyss} \big(a(\xvec)\big)^\ell
    \exp{\hat{A}(t-t_0)}
    \big(\tilde{a}(\xvec_0)\big)^k \ket{0} \\
    &= \sum_{\ell=0}^{n_0} \stirling{n}{\ell} {n_0 \choose \ell} \ell! \bra{\abyss} a(\xvec) 
    \exp{\hat{A}(t-t_0)} 
    \tilde{a}(\xvec_0) \ket{0}^\ell 
\end{align}
where we have used $\bra{\abyss} (a(\xvec))^\ell \ExpNB(\hat{A}(t-t_0))(\tilde{a}(\xvec_0))^k \ket{0} = \delta_{k\ell} \ell!  \bra{\abyss} a(\xvec) \ExpNB(\hat{A}(t-t_0)) \tilde{a}(\xvec_0) \ket{0}^\ell$ in the absence of interactions. The final expression corresponds to the one we found via the Dean route, 
\Eref{con_der_4},
with $I_\Omega(t-t_0)$ replaced by the Doi-Peliti propagator $\bra{\abyss} a(\xvec) \ExpNB\hat{A}(t-t_0)) \tilde{a}(\xvec_0) \ket{0}$.

\section{Conclusion \seclabel{conclusion}}
To the best of our knowledge, this paper presents the first formalisation and systematic study of the concept of
particle entity in the context of statistical field theory. Focusing on two well-known field theoretic formalisms applied to the study of stochastic processes, namely the Doi-Peliti \cite{tauber_2014} and the Martin-Siggia-Rose-Janssen-De Dominicis \cite{MartinSiggiaRose:1973,Janssen:1976,DeDominicis:1976} response field theories, we have demonstrated that particle entity is enforced in a formalism-specific way. In Doi-Peliti field theories, particle entity is built into its foundation, namely in the commutation relation of the ladder operators, \Eref{commutation_rel}. In the response field field theory derived from Dean's equation, particle entity is a perturbative feature that relies on the precise form of the interaction vertex, \Eref{dean_vertex}. This "Dean vertex" originates from the It{\^o}-multiplicative noise term in the original Langevin equation \Eref{eq:deans}. It compensates for some overcounting that occurs in the bilinear part of the field theory \Eref{bilin_Dean}, a mechanism that was already identified in an earlier work on the statistics of the non-interacting Brownian gas \cite{Velenich_2008}. As a result, one is faced with more complicated branching diagrams in the response field formalism equipped with particle entity via Dean's equation compared to the Doi-Peliti formalism, \cf \Erefs{observable_naive_DP} and \eref{observable_naive_Dean} or \Erefs{corr_direct_xt} and \eref{Dean_vertex_direct}. 

To test for particle entity, we introduced the condition \Eref{eq:PI_signature}, that we rewrote in terms of particle number moments, \Eref{eq:PI_mom_gen}, and, on the basis of the identity \Eref{eq:gen_fn_rel}, in terms of connected moments, \Eref{eq:PI_connected}.

In \Sref{dp_PI}, we were able to show in a few lines that particle entity according to \Eref{eq:PI_mom_gen} generally holds in Doi-Peliti field theories, \Eref{OC_DP_final}. This finding is independent of the specifics of the action.
To demonstrate particle entity for non-interacting, diffusive field theories on the basis of Dean's equation, we used in \Sref{dean_PI} our key result 
on the connected particle number moments, \Eref{n_point_connected}, 
before constructing the main result \Eref{particle_entity_full_moments} on the basis of the full moments,
with some of the more cumbersome calculations relegated to Appendix~\ref{a:induction_conn}.

It is interesting to speculate whether our derivation simplifies further by exploiting the well-known identity \cite{lebellac1991} relating the Legendre transform of the generating function of the connected moments and the effective action, which only depends on the one-particle irreducible (1PI) diagrams. Since 1PIs represent a relatively small subset of all connected diagrams, a particle entity signature of this type might be more easily applicable to theories involving pair interactions, which are beyond the scope of the present analysis.

\section*{Acknowledgements}
GP would like to thank Rohit Jain for interesting discussions and sparking his interest in Dean's equation.
LC acknowledges support from the Francis Crick Institute, which receives its core funding from Cancer Research UK (FC001317), the UK Medical Research Council (FC001317), and the Wellcome Trust (FC001317).

\bibliography{PE_refs}

\begin{thebibliography}{29}%
\makeatletter
\providecommand \@ifxundefined [1]{%
 \@ifx{#1\undefined}
}%
\providecommand \@ifnum [1]{%
 \ifnum #1\expandafter \@firstoftwo
 \else \expandafter \@secondoftwo
 \fi
}%
\providecommand \@ifx [1]{%
 \ifx #1\expandafter \@firstoftwo
 \else \expandafter \@secondoftwo
 \fi
}%
\providecommand \natexlab [1]{#1}%
\providecommand \enquote  [1]{``#1''}%
\providecommand \bibnamefont  [1]{#1}%
\providecommand \bibfnamefont [1]{#1}%
\providecommand \citenamefont [1]{#1}%
\providecommand \href@noop [0]{\@secondoftwo}%
\providecommand \href [0]{\begingroup \@sanitize@url \@href}%
\providecommand \@href[1]{\@@startlink{#1}\@@href}%
\providecommand \@@href[1]{\endgroup#1\@@endlink}%
\providecommand \@sanitize@url [0]{\catcode `\\12\catcode `\$12\catcode
  `\&12\catcode `\#12\catcode `\^12\catcode `\_12\catcode `\%12\relax}%
\providecommand \@@startlink[1]{}%
\providecommand \@@endlink[0]{}%
\providecommand \url  [0]{\begingroup\@sanitize@url \@url }%
\providecommand \@url [1]{\endgroup\@href {#1}{\urlprefix }}%
\providecommand \urlprefix  [0]{URL }%
\providecommand \Eprint [0]{\href }%
\providecommand \doibase [0]{https://doi.org/}%
\providecommand \selectlanguage [0]{\@gobble}%
\providecommand \bibinfo  [0]{\@secondoftwo}%
\providecommand \bibfield  [0]{\@secondoftwo}%
\providecommand \translation [1]{[#1]}%
\providecommand \BibitemOpen [0]{}%
\providecommand \bibitemStop [0]{}%
\providecommand \bibitemNoStop [0]{.\EOS\space}%
\providecommand \EOS [0]{\spacefactor3000\relax}%
\providecommand \BibitemShut  [1]{\csname bibitem#1\endcsname}%
\let\auto@bib@innerbib\@empty
\bibitem [{\citenamefont {Nardini}\ \emph {et~al.}(2017)\citenamefont
  {Nardini}, \citenamefont {Fodor}, \citenamefont {Tjhung}, \citenamefont {van
  Wijland}, \citenamefont {Tailleur},\ and\ \citenamefont
  {Cates}}]{nardini_entropy_2017}%
  \BibitemOpen
  \bibfield  {author} {\bibinfo {author} {\bibfnamefont {C.}~\bibnamefont
  {Nardini}}, \bibinfo {author} {\bibfnamefont {{\'E}.}~\bibnamefont {Fodor}},
  \bibinfo {author} {\bibfnamefont {E.}~\bibnamefont {Tjhung}}, \bibinfo
  {author} {\bibfnamefont {F.}~\bibnamefont {van Wijland}}, \bibinfo {author}
  {\bibfnamefont {J.}~\bibnamefont {Tailleur}},\ and\ \bibinfo {author}
  {\bibfnamefont {M.~E.}\ \bibnamefont {Cates}},\ }\bibfield  {title} {\bibinfo
  {title} {Entropy {Production} in {Field} {Theories} without {Time}-{Reversal}
  {Symmetry}: {Quantifying} the {Non}-{Equilibrium} {Character} of {Active}
  {Matter}},\ }\href {https://doi.org/10.1103/PhysRevX.7.021007} {\bibfield
  {journal} {\bibinfo  {journal} {Physical Review X}\ }\textbf {\bibinfo
  {volume} {7}},\ \bibinfo {pages} {021007} (\bibinfo {year}
  {2017})}\BibitemShut {NoStop}%
\bibitem [{\citenamefont {Cocconi}\ \emph {et~al.}(2020)\citenamefont
  {Cocconi}, \citenamefont {Garcia-Millan}, \citenamefont {Zhen}, \citenamefont
  {Buturca},\ and\ \citenamefont {Pruessner}}]{cocconi_en}%
  \BibitemOpen
  \bibfield  {author} {\bibinfo {author} {\bibfnamefont {L.}~\bibnamefont
  {Cocconi}}, \bibinfo {author} {\bibfnamefont {R.}~\bibnamefont
  {Garcia-Millan}}, \bibinfo {author} {\bibfnamefont {Z.}~\bibnamefont {Zhen}},
  \bibinfo {author} {\bibfnamefont {B.}~\bibnamefont {Buturca}},\ and\ \bibinfo
  {author} {\bibfnamefont {G.}~\bibnamefont {Pruessner}},\ }\bibfield  {title}
  {\bibinfo {title} {Entropy production in exactly solvable systems},\
  }\bibfield  {journal} {\bibinfo  {journal} {Entropy}\ }\textbf {\bibinfo
  {volume} {22}},\ \href {https://doi.org/10.3390/e22111252}
  {10.3390/e22111252} (\bibinfo {year} {2020})\BibitemShut {NoStop}%
\bibitem [{\citenamefont {Garcia-Millan}\ and\ \citenamefont
  {Pruessner}(2021)}]{garcia2020rnt}%
  \BibitemOpen
  \bibfield  {author} {\bibinfo {author} {\bibfnamefont {R.}~\bibnamefont
  {Garcia-Millan}}\ and\ \bibinfo {author} {\bibfnamefont {G.}~\bibnamefont
  {Pruessner}},\ }\bibfield  {title} {\bibinfo {title} {Run-and-tumble motion
  in a harmonic potential: field theory and entropy production},\ }\href
  {https://doi.org/10.1088/1742-5468/ac014d} {\bibfield  {journal} {\bibinfo
  {journal} {Journal of Statistical Mechanics: Theory and Experiment}\ }\textbf
  {\bibinfo {volume} {2021}},\ \bibinfo {pages} {063203} (\bibinfo {year}
  {2021})}\BibitemShut {NoStop}%
\bibitem [{\citenamefont {Busiello}\ \emph {et~al.}(2019)\citenamefont
  {Busiello}, \citenamefont {Hidalgo},\ and\ \citenamefont
  {Maritan}}]{Busiello_2019}%
  \BibitemOpen
  \bibfield  {author} {\bibinfo {author} {\bibfnamefont {D.~M.}\ \bibnamefont
  {Busiello}}, \bibinfo {author} {\bibfnamefont {J.}~\bibnamefont {Hidalgo}},\
  and\ \bibinfo {author} {\bibfnamefont {A.}~\bibnamefont {Maritan}},\
  }\bibfield  {title} {\bibinfo {title} {Entropy production for coarse-grained
  dynamics},\ }\href {https://doi.org/10.1088/1367-2630/ab29c0} {\bibfield
  {journal} {\bibinfo  {journal} {New Journal of Physics}\ }\textbf {\bibinfo
  {volume} {21}},\ \bibinfo {pages} {073004} (\bibinfo {year}
  {2019})}\BibitemShut {NoStop}%
\bibitem [{\citenamefont {Fodor}\ \emph {et~al.}(2021)\citenamefont {Fodor},
  \citenamefont {Jack},\ and\ \citenamefont {Cates}}]{fodor2021}%
  \BibitemOpen
  \bibfield  {author} {\bibinfo {author} {\bibfnamefont {{\'E}.}~\bibnamefont
  {Fodor}}, \bibinfo {author} {\bibfnamefont {R.~L.}\ \bibnamefont {Jack}},\
  and\ \bibinfo {author} {\bibfnamefont {M.~E.}\ \bibnamefont {Cates}},\
  }\bibfield  {title} {\bibinfo {title} {Irreversibility and biased ensembles
  in active matter: Insights from stochastic thermodynamics},\ }\href@noop {}
  {\bibfield  {journal} {\bibinfo  {journal} {arXiv preprint arXiv:2104.06634}\
  } (\bibinfo {year} {2021})}\BibitemShut {NoStop}%
\bibitem [{\citenamefont {Gompper}\ \emph {et~al.}(2020)\citenamefont
  {Gompper}, \citenamefont {Winkler}, \citenamefont {Speck}, \citenamefont
  {Solon}, \citenamefont {Nardini}, \citenamefont {Peruani}, \citenamefont
  {L{\"o}wen}, \citenamefont {Golestanian}, \citenamefont {Kaupp},
  \citenamefont {Alvarez} \emph {et~al.}}]{gompper20202020}%
  \BibitemOpen
  \bibfield  {author} {\bibinfo {author} {\bibfnamefont {G.}~\bibnamefont
  {Gompper}}, \bibinfo {author} {\bibfnamefont {R.~G.}\ \bibnamefont
  {Winkler}}, \bibinfo {author} {\bibfnamefont {T.}~\bibnamefont {Speck}},
  \bibinfo {author} {\bibfnamefont {A.}~\bibnamefont {Solon}}, \bibinfo
  {author} {\bibfnamefont {C.}~\bibnamefont {Nardini}}, \bibinfo {author}
  {\bibfnamefont {F.}~\bibnamefont {Peruani}}, \bibinfo {author} {\bibfnamefont
  {H.}~\bibnamefont {L{\"o}wen}}, \bibinfo {author} {\bibfnamefont
  {R.}~\bibnamefont {Golestanian}}, \bibinfo {author} {\bibfnamefont {U.~B.}\
  \bibnamefont {Kaupp}}, \bibinfo {author} {\bibfnamefont {L.}~\bibnamefont
  {Alvarez}}, \emph {et~al.},\ }\bibfield  {title} {\bibinfo {title} {The 2020
  motile active matter roadmap},\ }\href@noop {} {\bibfield  {journal}
  {\bibinfo  {journal} {Journal of Physics: Condensed Matter}\ }\textbf
  {\bibinfo {volume} {32}},\ \bibinfo {pages} {193001} (\bibinfo {year}
  {2020})}\BibitemShut {NoStop}%
\bibitem [{\citenamefont {Soto}\ and\ \citenamefont
  {Golestanian}(2015)}]{soto2015self}%
  \BibitemOpen
  \bibfield  {author} {\bibinfo {author} {\bibfnamefont {R.}~\bibnamefont
  {Soto}}\ and\ \bibinfo {author} {\bibfnamefont {R.}~\bibnamefont
  {Golestanian}},\ }\bibfield  {title} {\bibinfo {title} {Self-assembly of
  active colloidal molecules with dynamic function},\ }\href@noop {} {\bibfield
   {journal} {\bibinfo  {journal} {Physical Review E}\ }\textbf {\bibinfo
  {volume} {91}},\ \bibinfo {pages} {052304} (\bibinfo {year}
  {2015})}\BibitemShut {NoStop}%
\bibitem [{\citenamefont {Slowman}\ \emph {et~al.}(2016)\citenamefont
  {Slowman}, \citenamefont {Evans},\ and\ \citenamefont
  {Blythe}}]{slowman2016jamming}%
  \BibitemOpen
  \bibfield  {author} {\bibinfo {author} {\bibfnamefont {A.}~\bibnamefont
  {Slowman}}, \bibinfo {author} {\bibfnamefont {M.}~\bibnamefont {Evans}},\
  and\ \bibinfo {author} {\bibfnamefont {R.}~\bibnamefont {Blythe}},\
  }\bibfield  {title} {\bibinfo {title} {Jamming and attraction of interacting
  run-and-tumble random walkers},\ }\href@noop {} {\bibfield  {journal}
  {\bibinfo  {journal} {Physical review letters}\ }\textbf {\bibinfo {volume}
  {116}},\ \bibinfo {pages} {218101} (\bibinfo {year} {2016})}\BibitemShut
  {NoStop}%
\bibitem [{\citenamefont {{Le Bellac}}(1991)}]{lebellac1991}%
  \BibitemOpen
  \bibfield  {author} {\bibinfo {author} {\bibfnamefont {M.}~\bibnamefont {{Le
  Bellac}}},\ }\href@noop {} {\emph {\bibinfo {title} {Quantum and Statistical
  Field Theory [Phenomenes critiques aux champs de jauge, English]}}}\
  (\bibinfo  {publisher} {Oxford University Press},\ \bibinfo {address} {New
  York, NY, USA},\ \bibinfo {year} {1991})\ \bibinfo {note} {translated by G.
  Barton}\BibitemShut {NoStop}%
\bibitem [{\citenamefont {Hohenberg}\ and\ \citenamefont
  {Halperin}(1977)}]{Hohenberg_1977}%
  \BibitemOpen
  \bibfield  {author} {\bibinfo {author} {\bibfnamefont {P.~C.}\ \bibnamefont
  {Hohenberg}}\ and\ \bibinfo {author} {\bibfnamefont {B.~I.}\ \bibnamefont
  {Halperin}},\ }\bibfield  {title} {\bibinfo {title} {Theory of dynamic
  critical phenomena},\ }\href {https://doi.org/10.1103/RevModPhys.49.435}
  {\bibfield  {journal} {\bibinfo  {journal} {Rev. Mod. Phys.}\ }\textbf
  {\bibinfo {volume} {49}},\ \bibinfo {pages} {435} (\bibinfo {year}
  {1977})}\BibitemShut {NoStop}%
\bibitem [{\citenamefont {Täuber}(2014)}]{tauber_2014}%
  \BibitemOpen
  \bibfield  {author} {\bibinfo {author} {\bibfnamefont {U.~C.}\ \bibnamefont
  {Täuber}},\ }\href {https://doi.org/10.1017/CBO9781139046213} {\emph
  {\bibinfo {title} {Critical Dynamics: A Field Theory Approach to Equilibrium
  and Non-Equilibrium Scaling Behavior}}}\ (\bibinfo  {publisher} {Cambridge
  University Press},\ \bibinfo {year} {2014})\BibitemShut {NoStop}%
\bibitem [{\citenamefont {Cardy}(2008)}]{cardynotes}%
  \BibitemOpen
  \bibfield  {author} {\bibinfo {author} {\bibfnamefont {J.}~\bibnamefont
  {Cardy}},\ }\bibfield  {title} {\bibinfo {title} {Reaction-diffusion
  processes},\ }in\ \href@noop {} {\emph {\bibinfo {booktitle} {Non-equilibrium
  Statistical Mechanics and Turbulence}}},\ \bibinfo {editor} {edited by\
  \bibinfo {editor} {\bibfnamefont {S.}~\bibnamefont {Nazarenko}}\ and\
  \bibinfo {editor} {\bibfnamefont {O.~V.}\ \bibnamefont {Zaboronski}}}\
  (\bibinfo  {publisher} {Cambridge University Press},\ \bibinfo {address}
  {Cambridge, UK},\ \bibinfo {year} {2008})\ pp.\ \bibinfo {pages}
  {108--161}\BibitemShut {NoStop}%
\bibitem [{\citenamefont {Pruessner}(2011)}]{gunnarnotes}%
  \BibitemOpen
  \bibfield  {author} {\bibinfo {author} {\bibfnamefont {G.}~\bibnamefont
  {Pruessner}},\ }\href@noop {} {\bibinfo {title} {Lecture notes on
  non-equilibrium statistical mechanics}} (\bibinfo {year} {2011})\BibitemShut
  {NoStop}%
\bibitem [{\citenamefont {Martin}\ \emph {et~al.}(1973)\citenamefont {Martin},
  \citenamefont {Siggia},\ and\ \citenamefont {Rose}}]{MartinSiggiaRose:1973}%
  \BibitemOpen
  \bibfield  {author} {\bibinfo {author} {\bibfnamefont {P.~C.}\ \bibnamefont
  {Martin}}, \bibinfo {author} {\bibfnamefont {E.~D.}\ \bibnamefont {Siggia}},\
  and\ \bibinfo {author} {\bibfnamefont {H.~A.}\ \bibnamefont {Rose}},\
  }\bibfield  {title} {\bibinfo {title} {Statistical dynamics of classical
  systems},\ }\href {https://doi.org/10.1103/PhysRevA.8.423} {\bibfield
  {journal} {\bibinfo  {journal} {Phys. Rev. A}\ }\textbf {\bibinfo {volume}
  {8}},\ \bibinfo {pages} {423} (\bibinfo {year} {1973})}\BibitemShut {NoStop}%
\bibitem [{\citenamefont {Janssen}(1976)}]{Janssen:1976}%
  \BibitemOpen
  \bibfield  {author} {\bibinfo {author} {\bibfnamefont {H.-K.}\ \bibnamefont
  {Janssen}},\ }\bibfield  {title} {\bibinfo {title} {On a lagrangean for
  classical field dynamics and renormalization group calculations of dynamical
  critical properties},\ }\href@noop {} {\bibfield  {journal} {\bibinfo
  {journal} {Zeitschrift f{\"u}r Physik B Condensed Matter}\ }\textbf {\bibinfo
  {volume} {23}},\ \bibinfo {pages} {377} (\bibinfo {year} {1976})}\BibitemShut
  {NoStop}%
\bibitem [{\citenamefont {Dominicis}(1976)}]{DeDominicis:1976}%
  \BibitemOpen
  \bibfield  {author} {\bibinfo {author} {\bibfnamefont {C.~d.}\ \bibnamefont
  {Dominicis}},\ }\bibfield  {title} {\bibinfo {title} {Technics of field
  renormalization and dynamics of critical phenomena},\ }in\ \href@noop {}
  {\emph {\bibinfo {booktitle} {J. Phys.(Paris), Colloq}}}\ (\bibinfo {year}
  {1976})\ pp.\ \bibinfo {pages} {C1--247}\BibitemShut {NoStop}%
\bibitem [{\citenamefont {Hertz}\ \emph {et~al.}(2016)\citenamefont {Hertz},
  \citenamefont {Roudi},\ and\ \citenamefont {Sollich}}]{hertz_path_2016}%
  \BibitemOpen
  \bibfield  {author} {\bibinfo {author} {\bibfnamefont {J.~A.}\ \bibnamefont
  {Hertz}}, \bibinfo {author} {\bibfnamefont {Y.}~\bibnamefont {Roudi}},\ and\
  \bibinfo {author} {\bibfnamefont {P.}~\bibnamefont {Sollich}},\ }\bibfield
  {title} {\bibinfo {title} {Path integral methods for the dynamics of
  stochastic and disordered systems},\ }\href
  {https://doi.org/10.1088/1751-8121/50/3/033001} {\bibfield  {journal}
  {\bibinfo  {journal} {Journal of Physics A: Mathematical and Theoretical}\
  }\textbf {\bibinfo {volume} {50}},\ \bibinfo {pages} {033001} (\bibinfo
  {year} {2016})},\ \bibinfo {note} {publisher: IOP Publishing}\BibitemShut
  {NoStop}%
\bibitem [{\citenamefont {Dean}(1996)}]{Dean_1996}%
  \BibitemOpen
  \bibfield  {author} {\bibinfo {author} {\bibfnamefont {D.~S.}\ \bibnamefont
  {Dean}},\ }\bibfield  {title} {\bibinfo {title} {Langevin equation for the
  density of a system of interacting langevin processes},\ }\href
  {https://doi.org/10.1088/0305-4470/29/24/001} {\bibfield  {journal} {\bibinfo
   {journal} {Journal of Physics A: Mathematical and General}\ }\textbf
  {\bibinfo {volume} {29}},\ \bibinfo {pages} {L613} (\bibinfo {year}
  {1996})}\BibitemShut {NoStop}%
\bibitem [{\citenamefont {Gelimson}\ and\ \citenamefont
  {Golestanian}(2015)}]{gemlinson_2015}%
  \BibitemOpen
  \bibfield  {author} {\bibinfo {author} {\bibfnamefont {A.}~\bibnamefont
  {Gelimson}}\ and\ \bibinfo {author} {\bibfnamefont {R.}~\bibnamefont
  {Golestanian}},\ }\bibfield  {title} {\bibinfo {title} {Collective dynamics
  of dividing chemotactic cells},\ }\href
  {https://doi.org/10.1103/PhysRevLett.114.028101} {\bibfield  {journal}
  {\bibinfo  {journal} {Phys. Rev. Lett.}\ }\textbf {\bibinfo {volume} {114}},\
  \bibinfo {pages} {028101} (\bibinfo {year} {2015})}\BibitemShut {NoStop}%
\bibitem [{\citenamefont {Velenich}\ \emph {et~al.}(2008)\citenamefont
  {Velenich}, \citenamefont {Chamon}, \citenamefont {Cugliandolo},\ and\
  \citenamefont {Kreimer}}]{Velenich_2008}%
  \BibitemOpen
  \bibfield  {author} {\bibinfo {author} {\bibfnamefont {A.}~\bibnamefont
  {Velenich}}, \bibinfo {author} {\bibfnamefont {C.}~\bibnamefont {Chamon}},
  \bibinfo {author} {\bibfnamefont {L.~F.}\ \bibnamefont {Cugliandolo}},\ and\
  \bibinfo {author} {\bibfnamefont {D.}~\bibnamefont {Kreimer}},\ }\bibfield
  {title} {\bibinfo {title} {On the brownian gas: a field theory with a
  poissonian ground state},\ }\href
  {https://doi.org/10.1088/1751-8113/41/23/235002} {\bibfield  {journal}
  {\bibinfo  {journal} {Journal of Physics A: Mathematical and Theoretical}\
  }\textbf {\bibinfo {volume} {41}},\ \bibinfo {pages} {235002} (\bibinfo
  {year} {2008})}\BibitemShut {NoStop}%
\bibitem [{\citenamefont {Lef{\`{e}}vre}\ and\ \citenamefont
  {Biroli}(2007)}]{lefevre_2007}%
  \BibitemOpen
  \bibfield  {author} {\bibinfo {author} {\bibfnamefont {A.}~\bibnamefont
  {Lef{\`{e}}vre}}\ and\ \bibinfo {author} {\bibfnamefont {G.}~\bibnamefont
  {Biroli}},\ }\bibfield  {title} {\bibinfo {title} {Dynamics of interacting
  particle systems: stochastic process and field theory},\ }\href
  {https://doi.org/10.1088/1742-5468/2007/07/p07024} {\bibfield  {journal}
  {\bibinfo  {journal} {Journal of Statistical Mechanics: Theory and
  Experiment}\ }\textbf {\bibinfo {volume} {2007}},\ \bibinfo {pages} {P07024}
  (\bibinfo {year} {2007})}\BibitemShut {NoStop}%
\bibitem [{\citenamefont {T{\"a}uber}\ \emph {et~al.}(2005)\citenamefont
  {T{\"a}uber}, \citenamefont {Howard},\ and\ \citenamefont
  {Vollmayr-Lee}}]{tauber2005applications}%
  \BibitemOpen
  \bibfield  {author} {\bibinfo {author} {\bibfnamefont {U.~C.}\ \bibnamefont
  {T{\"a}uber}}, \bibinfo {author} {\bibfnamefont {M.}~\bibnamefont {Howard}},\
  and\ \bibinfo {author} {\bibfnamefont {B.~P.}\ \bibnamefont {Vollmayr-Lee}},\
  }\bibfield  {title} {\bibinfo {title} {Applications of field-theoretic
  renormalization group methods to reaction--diffusion problems},\ }\href@noop
  {} {\bibfield  {journal} {\bibinfo  {journal} {Journal of Physics A:
  Mathematical and General}\ }\textbf {\bibinfo {volume} {38}},\ \bibinfo
  {pages} {R79} (\bibinfo {year} {2005})}\BibitemShut {NoStop}%
\bibitem [{\citenamefont {Pausch}(2019)}]{pausch2019topics}%
  \BibitemOpen
  \bibfield  {author} {\bibinfo {author} {\bibfnamefont {J.}~\bibnamefont
  {Pausch}},\ }\emph {\bibinfo {title} {Topics in statistical mechanics}},\
  \href@noop {} {Ph.D. thesis},\ \bibinfo  {school} {Imperial College}
  (\bibinfo {year} {2019})\BibitemShut {NoStop}%
\bibitem [{\citenamefont {Honkonen}(2011)}]{honkonen}%
  \BibitemOpen
  \bibfield  {author} {\bibinfo {author} {\bibfnamefont {J.}~\bibnamefont
  {Honkonen}},\ }\bibfield  {title} {\bibinfo {title} {{Ito} and {Stratonovich}
  calculuses in stochastic field theory},\ }\href@noop {} {\bibfield  {journal}
  {\bibinfo  {journal} {Invited talk presented at The 12th Small Triangle
  Meeting on Theoretical Physics}\ } (\bibinfo {year} {2011})}\BibitemShut
  {NoStop}%
\bibitem [{\citenamefont {Binney}\ \emph {et~al.}(1998)\citenamefont {Binney},
  \citenamefont {Dowrick}, \citenamefont {Fisher},\ and\ \citenamefont
  {Newman}}]{binney1992}%
  \BibitemOpen
  \bibfield  {author} {\bibinfo {author} {\bibfnamefont {J.~J.}\ \bibnamefont
  {Binney}}, \bibinfo {author} {\bibfnamefont {N.~J.}\ \bibnamefont {Dowrick}},
  \bibinfo {author} {\bibfnamefont {A.~J.}\ \bibnamefont {Fisher}},\ and\
  \bibinfo {author} {\bibfnamefont {M.~E.~J.}\ \bibnamefont {Newman}},\
  }\href@noop {} {\emph {\bibinfo {title} {The Theory of Critical Phenomena}}}\
  (\bibinfo  {publisher} {Oxford University Press},\ \bibinfo {address}
  {Oxford, UK},\ \bibinfo {year} {1998})\BibitemShut {NoStop}%
\bibitem [{\citenamefont {{van Kampen}}(1992)}]{van1992stochastic}%
  \BibitemOpen
  \bibfield  {author} {\bibinfo {author} {\bibfnamefont {N.~G.}\ \bibnamefont
  {{van Kampen}}},\ }\href@noop {} {\emph {\bibinfo {title} {Stochastic
  Processes in Physics and Chemistry}}}\ (\bibinfo  {publisher} {Elsevier
  Science B. V.},\ \bibinfo {address} {Amsterdam, The Netherlands},\ \bibinfo
  {year} {1992})\ \bibinfo {note} {third impression 2001, enlarged and
  revised}\BibitemShut {NoStop}%
\bibitem [{\citenamefont {Comtet}(2012)}]{comtet2012}%
  \BibitemOpen
  \bibfield  {author} {\bibinfo {author} {\bibfnamefont {L.}~\bibnamefont
  {Comtet}},\ }\href@noop {} {\emph {\bibinfo {title} {Advanced Combinatorics:
  The art of finite and infinite expansions}}}\ (\bibinfo  {publisher}
  {Springer Science \& Business Media},\ \bibinfo {year} {2012})\BibitemShut
  {NoStop}%
\bibitem [{\citenamefont {Garcia-Millan}\ \emph {et~al.}(2018)\citenamefont
  {Garcia-Millan}, \citenamefont {Pausch}, \citenamefont {Walter},\ and\
  \citenamefont {Pruessner}}]{garcia2018field}%
  \BibitemOpen
  \bibfield  {author} {\bibinfo {author} {\bibfnamefont {R.}~\bibnamefont
  {Garcia-Millan}}, \bibinfo {author} {\bibfnamefont {J.}~\bibnamefont
  {Pausch}}, \bibinfo {author} {\bibfnamefont {B.}~\bibnamefont {Walter}},\
  and\ \bibinfo {author} {\bibfnamefont {G.}~\bibnamefont {Pruessner}},\
  }\bibfield  {title} {\bibinfo {title} {Field-theoretic approach to the
  universality of branching processes},\ }\href@noop {} {\bibfield  {journal}
  {\bibinfo  {journal} {Phys. Rev. E}\ }\textbf {\bibinfo {volume} {98}},\
  \bibinfo {pages} {062107} (\bibinfo {year} {2018})}\BibitemShut {NoStop}%
\bibitem [{\citenamefont {Gradshteyn}\ and\ \citenamefont
  {Ryzhik}(2007)}]{GradshteynRyzhik:2007}%
  \BibitemOpen
  \bibfield  {author} {\bibinfo {author} {\bibfnamefont {I.~S.}\ \bibnamefont
  {Gradshteyn}}\ and\ \bibinfo {author} {\bibfnamefont {I.~M.}\ \bibnamefont
  {Ryzhik}},\ }\href@noop {} {\emph {\bibinfo {title} {Table of integrals,
  series and products}}},\ \bibinfo {edition} {7th}\ ed.\ (\bibinfo
  {publisher} {Academic Press},\ \bibinfo {address} {San Diego, CA, USA},\
  \bibinfo {year} {2007})\BibitemShut {NoStop}%
\end{thebibliography}%
\newpage

\appendix

\section{Induction over connected diagrams \label{a:induction_conn}}
We want to prove that the connected moments of the particle number density in the response field field theory derived from Dean's equation obey \Eref{conjecture_Connected_k_t} to all orders of $n$, restated here for convenience:
\begin{align}
&\langle \rho(\kvec_1,t)...\rho(\kvec_n,t)
\rangle_c
= n_0 \theta(t-t_0) 
\knullfix{\kvec_1+...\kvec_n}
\sum_{m=1}^{n} (-1)^{m-1} (m-1)! \\
&\qquad
\times
\sum_{\{\Pset{P}_1,\dots,\Pset{P}_m\}\in \partitionX{\{1,\dots,n\}}{m}} 
\exp{-D(t-t_0) \sum_{i=1}^m \setvecX{\Pset{P}_i}^2}
\nonumber
\,.
\end{align}
For this we have to consider all diagrams with a single incoming leg and an arbitrary number $n$ of outgoing legs, the first four orders of which are depicted in Eqs.~\eref{connected_diagramms_1}, \eref{connected_diagramms_2}, \eref{connected_diagramms_3} and \eref{connected_diagramms_4}. 
\begingroup
\addtolength{\jot}{0.2cm}
\begin{align}
    \ave{\rho(\kvec_1,t)}_c \corresponds
    \tikz[baseline=-2.5pt]{
    \begin{scope}[rotate=-30]
    \draw[Aactivity] (-150:0.5) -- (-150:1.);
    \end{scope}
    \draw[thick,pattern=dots] 
    (0,0) circle (0.5cm);
    \draw[Aactivity] (0:0.5) -- (0:0.8) +(2pt,0) circle (2pt)[fill];
    \node [yshift=-0.5pt,xshift=0pt,fill=white] {$1$};
}    &= 
\tikz[baseline=-2.5pt]{
    \begin{scope}
      \draw[Aactivity] (0,0.0) -- (1.2,0.0) +(2pt,0) circle (2pt)[fill];
     \end{scope} \elabel{connected_diagramms_1}
     \node [yshift=0cm,xshift=-0.19cm] {$\kvec_1$};
    }\\[0.1cm]
\ave{\rho(\kvec_1,t)\rho(\kvec_2,t)}_c \corresponds \tikz[baseline=-2.5pt]{
    \begin{scope}[rotate=-30]
    \draw[Aactivity] (-140:0.5) -- (-140:1.);
     \draw[Aactivity] (-160:0.5) -- (-160:1.);
    \end{scope}
    \draw[thick,pattern=dots] (0,0) circle (0.5cm);
    \draw[Aactivity] (0:0.5) -- (0:0.8) +(2pt,0) circle (2pt)[fill];
    \node [yshift=-0.5pt,xshift=0pt,fill=white] {$2$};
}    &= 
\tikz[baseline=-2.5pt]{
    \begin{scope}
      \draw[Aactivity] (0,0.0) -- (0.7,0.0) +(2pt,0) circle (2pt)[fill];
      \draw[Aactivity] (220:0.7) -- (0,0.0);
      \draw[Aactivity] (140:0.7) -- (0,0.0);
      \draw[Aactivity] ($(0.07,0.07)+(140:0.2)$) -- ($(-0.07,-0.07)+(140:0.2)$);
      \draw[Aactivity] ($(-0.07,0.07)+(220:0.2)$) -- ($(0.07,-0.07)+(220:0.2)$);
     \end{scope}
    \node [yshift=0.45cm,xshift=-0.75cm] {$\kvec_1$};
    \node [yshift=-0.4cm,xshift=-0.75cm] {$\kvec_2$};
    \draw[dotted,thin] (140:0.25) arc (140:220:0.25);
    } \elabel{connected_diagramms_2}\\[0.1cm]
\ave{\rho(\kvec_1,t)\rho(\kvec_2,t) \rho(\kvec_3,t)}_c \corresponds  \tikz[baseline=-2.5pt]{
    \begin{scope}[rotate=-30]
    \draw[Aactivity] (-130:0.5) -- (-130:1.);
     \draw[Aactivity] (-150:0.5) -- (-150:1.);
     \draw[Aactivity] (-170:0.5) -- (-170:1.);
    \end{scope}
    \draw[thick,pattern=dots] (0,0) circle (0.5cm);
    \draw[Aactivity] (0:0.5) -- (0:0.8) +(2pt,0) circle (2pt)[fill];
    \node [yshift=-0.5pt,xshift=0pt,fill=white] {$3$};
}    &=
\tikz[baseline=-2.5pt]{
      \draw[Aactivity] (0,0.0) -- (0.7,0.0) +(2pt,0) circle (2pt)[fill];
      \begin{scope}[rotate=0]
      \draw[Aactivity] (220:1) -- (0,0.0);
      \draw[Aactivity] ($(-0.07,0.07)+(220:0.2)$) -- ($(0.07,-0.07)+(220:0.2)$);
      \draw[dotted,thin] (140:0.25) arc (140:220:0.25);
      \end{scope}
      \draw[Aactivity] (140:0.5) -- (0,0.0);
      \draw[Aactivity] ($(0.07,0.07)+(140:0.2)$) -- ($(-0.07,-0.07)+(140:0.2)$);
      \begin{scope}[xshift=-0.35cm,yshift=0.3cm]
      \draw[Aactivity] (220:0.6) -- (0,0.0);
      \draw[Aactivity] (140:0.6) -- (0,0.0);
      \draw[Aactivity] ($(0.07,0.07)+(140:0.2)$) -- ($(-0.07,-0.07)+(140:0.2)$);
      \draw[Aactivity] ($(-0.07,0.07)+(220:0.2)$) -- ($(0.07,-0.07)+(220:0.2)$);
      \draw[dotted,thin] (140:0.25) arc (140:220:0.25);
     \end{scope}
     \node [yshift=0.8cm,xshift=-1cm] {$\kvec_1$};
     \node [yshift=0.cm,xshift=-1cm] {$\kvec_2$};
     \node [yshift=-0.6cm,xshift=-1cm] {$\kvec_3$};
    } +
    \tikz[baseline=-2.5pt]{
      \draw[Aactivity] (0,0.0) -- (0.7,0.0) +(2pt,0) circle (2pt)[fill];
      \begin{scope}[rotate=0]
      \draw[Aactivity] (220:1) -- (0,0.0);
      \draw[Aactivity] ($(-0.07,0.07)+(220:0.2)$) -- ($(0.07,-0.07)+(220:0.2)$);
      \draw[dotted,thin] (140:0.25) arc (140:220:0.25);
      \end{scope}
      \draw[Aactivity] (140:0.5) -- (0,0.0);
      \draw[Aactivity] ($(0.07,0.07)+(140:0.2)$) -- ($(-0.07,-0.07)+(140:0.2)$);
      \begin{scope}[xshift=-0.35cm,yshift=0.3cm]
      \draw[Aactivity] (220:0.6) -- (0,0.0);
      \draw[Aactivity] (140:0.6) -- (0,0.0);
      \draw[Aactivity] ($(0.07,0.07)+(140:0.2)$) -- ($(-0.07,-0.07)+(140:0.2)$);
      \draw[Aactivity] ($(-0.07,0.07)+(220:0.2)$) -- ($(0.07,-0.07)+(220:0.2)$);
      \draw[dotted,thin] (140:0.25) arc (140:220:0.25);
     \end{scope}
     \node [yshift=0.8cm,xshift=-1cm] {$\kvec_1$};
     \node [yshift=0.cm,xshift=-1cm] {$\kvec_3$};
     \node [yshift=-0.6cm,xshift=-1cm] {$\kvec_2$};
    } + 
    \tikz[baseline=-2.5pt]{
      \draw[Aactivity] (0,0.0) -- (0.7,0.0) +(2pt,0) circle (2pt)[fill];
      \begin{scope}[rotate=0]
      \draw[Aactivity] (220:1) -- (0,0.0);
      \draw[Aactivity] ($(-0.07,0.07)+(220:0.2)$) -- ($(0.07,-0.07)+(220:0.2)$);
      \draw[dotted,thin] (140:0.25) arc (140:220:0.25);
      \end{scope}
      \draw[Aactivity] (140:0.5) -- (0,0.0);
      \draw[Aactivity] ($(0.07,0.07)+(140:0.2)$) -- ($(-0.07,-0.07)+(140:0.2)$);
      \draw[dotted,thin] (140:0.25) arc (140:220:0.25);
      \begin{scope}[xshift=-0.35cm,yshift=0.3cm]
      \draw[Aactivity] (220:0.6) -- (0,0.0);
      \draw[Aactivity] (140:0.6) -- (0,0.0);
      \draw[Aactivity] ($(0.07,0.07)+(140:0.2)$) -- ($(-0.07,-0.07)+(140:0.2)$);
      \draw[Aactivity] ($(-0.07,0.07)+(220:0.2)$) -- ($(0.07,-0.07)+(220:0.2)$);
      \draw[dotted,thin] (140:0.25) arc (140:220:0.25);
     \end{scope}
     \node [yshift=0.8cm,xshift=-1cm] {$\kvec_3$};
     \node [yshift=0.cm,xshift=-1cm] {$\kvec_2$};
     \node [yshift=-0.6cm,xshift=-1cm] {$\kvec_1$};
    }
    \elabel{connected_diagramms_3} 
\end{align}
\begin{align}
    &\ave{\rho(\kvec_1,t)\rho(\kvec_2,t) \rho(\kvec_3,t) \rho(\kvec_4,t)}_c \corresponds \tikz[baseline=-2.5pt]{
    \begin{scope}[rotate=-30]
    \draw[Aactivity] (-128:0.5) -- (-128:1.);
    \draw[Aactivity] (-142:0.5) -- (-142:1.);
     \draw[Aactivity] (-158:0.5) -- (-158:1.);
     \draw[Aactivity] (-172:0.5) -- (-172:1.);
    \end{scope}
    \draw[thick,pattern=dots] (0,0) circle (0.5cm);
    \draw[Aactivity] (0:0.5) -- (0:0.8) +(2pt,0) circle (2pt)[fill];
    \node [yshift=-0.5pt,xshift=0pt,fill=white] {$4$};
}    \nonumber 
    \fourPointConnectedA{\kvec_1}{\kvec_2}{\kvec_3}{\kvec_4}
     +\fourPointConnectedA{\kvec_1}{\kvec_4}{\kvec_3}{\kvec_2} +\fourPointConnectedA{\kvec_1}{\kvec_3}{\kvec_2}{\kvec_4}  
     \nonumber\\
    &\qquad +
    \fourPointConnectedB{\kvec_1}{\kvec_2}{\kvec_3}{\kvec_4}  
    + \fourPointConnectedB{\kvec_1}{\kvec_3}{\kvec_2}{\kvec_4}  
    + \fourPointConnectedB{\kvec_3}{\kvec_2}{\kvec_1}{\kvec_4}  
    + \fourPointConnectedB{\kvec_1}{\kvec_2}{\kvec_4}{\kvec_3} 
    \nonumber \\
    &\qquad + 
    \fourPointConnectedB{\kvec_1}{\kvec_4}{\kvec_2}{\kvec_3}  
    + \fourPointConnectedB{\kvec_4}{\kvec_2}{\kvec_1}{\kvec_3} 
    + \fourPointConnectedB{\kvec_1}{\kvec_4}{\kvec_3}{\kvec_2}  
    + \fourPointConnectedB{\kvec_1}{\kvec_3}{\kvec_4}{\kvec_2}  
    \nonumber \\
    &\qquad + 
    \fourPointConnectedB{\kvec_3}{\kvec_4}{\kvec_1}{\kvec_2}   
    + \fourPointConnectedB{\kvec_4}{\kvec_2}{\kvec_3}{\kvec_1}  
    + \fourPointConnectedB{\kvec_4}{\kvec_3}{\kvec_2}{\kvec_1}  
    + \fourPointConnectedB{\kvec_3}{\kvec_2}{\kvec_4}{\kvec_1}  
    \elabel{connected_diagramms_4}
\end{align}
\endgroup

The shading of the circular vertices above is meant as a visual reminder that we are now dealing with connected moments of the local number {\it density}, which depend on an unordered set of $n$ external momenta $\kvec_1,...,\kvec_n$, as opposed to connected moments of the {\it integrated} number density in a patch $\Omega$ (cf.\ \Eref{conjecture_Connected_k_t} and \Eref{corr_int_stp} ). The shaded diagrams are by construction invariant under permutations of the momenta $\kvec_1,...,\kvec_n$.

We proceed by determining some of the shaded diagrams. The trivial case, $n=1$ shown in \Eref{connected_diagramms_1}, is given by the propagator \Eref{Dean_mixed_propagator} and the perturbative contribution from the source \Eref{Dean_action_simple_komega} with coupling $n_0$. 
Through direct computation in the mixed momentum-time representation, see \Eref{Dean_action_mixed}, we find for the $n=2$ case \Eref{connected_diagramms_2}:
\begin{align} \elabel{2pt_kt_dean}
\tikz[baseline=-2.5pt]{
    \begin{scope}[rotate=-30]
    \draw[Aactivity] (-140:0.5) -- (-140:1.);
     \draw[Aactivity] (-160:0.5) -- (-160:1.);
    \end{scope}
    \draw[thick,pattern=dots] (0,0) circle (0.5cm) ;
    \draw[Aactivity] (0:0.5) -- (0:0.8) +(2pt,0) circle (2pt)[fill];
    \node [yshift=-0.5pt,xshift=0pt,fill=white] {$2$};
} 
&\corresponds n_0 \Theta(t-t_0)
\knullfix{\kvec_1+\kvec_2}
\int_{t_0}^t \dint{t'} (-2D \kvec_1\cdot \kvec_2) \exp{-D (t-t')\kvec_1^2} \exp{-D (t-t')\kvec_2^2} \exp{-D (t'-t_0)\kvec_0^2} \\
&= n_0 
\knullfix{\kvec_1+\kvec_2}
\left[ \exp{-D(t-t_0)(\kvec_1+\kvec_2)^2}-\exp{-D(t-t_0)(\kvec_1^2+\kvec_2^2)} \right] ~,
\elabel{2pt_kt_dean_done}
\end{align}
where the integral over $t$ arises from the representing the Dean vertex \Eref{dean_vertex} in k-t-space. Each Dean vertex comes with a symmetry factor of 2. A factor of $(-2\kvec_1 \cdot \kvec_2)^{-1}$ that arises in the $t$ integration precisely cancels with the vertex prefactor $-2D \kvec_1 \cdot \kvec_2$, which simplifies the result \Eref{2pt_kt_dean_done} considerably.
A similar calculation for the $n=3$ case \Eref{connected_diagramms_3} yields
\begin{align} \elabel{3pt_kt_dean}
\tikz[baseline=-2.5pt]{
    \begin{scope}[rotate=-30]
    \draw[Aactivity] (-130:0.5) -- (-130:1.);
     \draw[Aactivity] (-150:0.5) -- (-150:1.);
     \draw[Aactivity] (-170:0.5) -- (-170:1.);
    \end{scope}
    \draw[thick,pattern=dots] (0,0) circle (0.5cm);
    \draw[Aactivity] (0:0.5) -- (0:0.8)  +(2pt,0) circle (2pt)[fill];
    \node [yshift=-0.5pt,xshift=0pt,fill=white] {$3$};
}    \corresponds& 
n_0 \Theta(t-t_0)
\knullfix{\kvec_3+\kvec_2+\kvec_1}
\left[ \exp{-D(t-t_0)(\kvec_1+\kvec_2+\kvec_3)^2} -\exp{-D(t-t_0)((\kvec_1+\kvec_2)^2+\kvec_3^2)} \right. \nonumber \\ 
& \left. -\exp{-D(t-t_0)((\kvec_1+\kvec_3)^2+\kvec_2^2)}-\exp{-D(t-t_0)((\kvec_3+\kvec_2)^2+\kvec_1^2)}+2 \exp{-D(t-t_0)(\kvec_1^2+\kvec_2^2+\kvec_3^2)} \right] \,.
\end{align}
The left-hand side of \Eref{3pt_kt_dean} contains the sum over all distinct ways to assign the external momenta to the outgoing legs, as shown in \eref{connected_diagramms_3}. The whole sum of the terms is necessary for the coefficients of the Dean vertices to cancel with the $k$-dependent factor coming down from the $t$-integrations. We will see that this mechanism is instrumental in performing the induction later on.

Based on \Eref{2pt_kt_dean_done} and \eref{3pt_kt_dean} we conjecture
and indeed show below that a general connected moment has the form
\Eref{conjecture_Connected_k_t},
\begin{align} \elabel{a_induc_hyp}
\tikz[baseline=-2.5pt]{
    \begin{scope}[rotate=-30]
    \draw[Aactivity] (-130:0.5) -- (-130:1.3);
    \path [polarity={decorate,decoration={raise=0ex,text along path,
     text={|\large|...}}}]
     (-152:1.2cm) arc (-152:-130:1.2cm);
     \draw[Aactivity] (-160:0.5) -- (-160:1.3);
     \draw[Aactivity] (-170:0.5) -- (-170:1.3);
    \end{scope}
    \draw[thick,pattern=dots] (0,0) circle (0.5cm);
    \draw[Aactivity] (0:0.5) -- (0:0.8) +(2pt,0) circle (2pt)[fill];
    \node [yshift=-0.5pt,xshift=0pt,fill=white] {$n$};
} 
&\corresponds 
\ave{\rho(\kvec_1,t)\rho(\kvec_1,t)\ldots\rho(\kvec_n,t)}
=
n_0 \theta(t-t_0) 
\knullfix{\kvec_1+\ldots}
\sum_{m=1}^{n} (-1)^{m-1} (m-1)! \\
\nonumber
&\qquad
\times
\sum_{\{\Pset{P}_1,\dots,\Pset{P}_m\}\in \partitionX{\{1,\dots,n\}}{m}} 
\exp{-D(t-t_0) \sum_{i=1}^m \setvecX{\Pset{P}_i}^2}\,,
\end{align}
with $\partitionX{\{1,\dots,n\}}{m}$ the set of all partitions of the
set $\{1,\dots,n\}$ into $m$ non-empty, distinct subsets $\Pset{P}_i$ with $i=1,2,\dots,m$ so that
$\union_{i=1}^m \Pset{P}_i=\{1,2,\dots,n\}$, as introduced after
\Eref{conjecture_Connected_k_t}. The second sum
$\sum_{\{\Pset{P}_1,\dots,\Pset{P}_m\}\in\partitionX{\{1,\dots,n\}}{m}}$ in \Eref{a_induc_hyp}
thus runs over all distinct partitions of $\{1,2,\dots,n\}$ into $m$ subsets
$\Pset{P}_1,\dots,\Pset{P}_m$. There is no order to the subsets,
so that the partition $\{\{1\},\{2\}\}$ of $\{1,2\}$ is identical to
$\{\{2\},\{1\}\}$ and thus considered the same in $\partitionX{\{1,2\}}{2}$.
We use $\setvec{P}$ to denote sums over momenta given by the indices in set $\Pset{P}$, \Eref{def_setvec}, 
$\setvec{P} = \sum_{p\in\Pset{P}} \kvec_p$.
\Eref{a_induc_hyp} is a function of the \emph{set} of momenta $\{\kvec_1,\dots,\kvec_n\}$, or simply the indices $\{1,\dots,n\}$ alone, and invariant under their permutation.

\Eref{a_induc_hyp} will be our induction hypothesis,
with the induction to be taken in $n$, the number of outgoing legs. The base cases $n=1$, $n=2$ and $n=3$ are immediately verified, 
as
$\partitionX{\{1\}}{1}=\{\,\{\{1\}\}\,\}$ reduces \Eref{a_induc_hyp} trivially to \Eref{Dean_mixed_propagator},
$\partitionX{\{1,2\}}{1}=\{\,\{\{1,2\}\}\,\}$ and $\partitionX{\{1,2\}}{2}=\{\,\{\{1\},\{2\}\}\,\}$ to \Eref{2pt_kt_dean_done}
and
$\partitionX{\{1,2,3\}}{1}=\{\,\{\{1,2,3\}\}\,\}$, $\partitionX{\{1,2,3\}}{2}=\{\,\{\{1\},\{2,3\}\},\{\{2\},\{3,1\}\},\{\{3\},\{1,2\}\}\,\}$ and $\partitionX{\{1,2,3\}}{3}=\{\,\{\{\{1\},\{2\},\{3\}\}\}\,\}$ to \Eref{3pt_kt_dean}.

We want to show that if \Eref{a_induc_hyp} holds for all strictly positive $n\le m-1$ then it also holds for $n=m$. 
To this end we consider two distinct subsets of indices $\Pset{A}$ and $\Pset{B}$ with cardinality $\cardinality{A}>0$ and $\cardinality{B}>0$
respectively, so that $\Pset{A}\intersect\Pset{B}=\emptyset$, $\Pset{A}\union\Pset{B}=\{1,\dots,n\}$ and thus $\cardinality{A}+\cardinality{B}=n$. 
Each of these sets enters as the argument of diagrams 
\[
\tikz[baseline=-2.5pt]{
    \begin{scope}[rotate=-30]
    \draw[Aactivity] (-130:0.5) -- (-130:1.3);
    \path [polarity={decorate,decoration={raise=0ex,text along path,
     text={|\large|...}}}]
     (-152:1.2cm) arc (-152:-130:1.2cm);
     \draw[Aactivity] (-160:0.5) -- (-160:1.3);
     \draw[Aactivity] (-170:0.5) -- (-170:1.3);
    \end{scope}
    \draw[thick,pattern=dots] (0,0) circle (0.5cm);
    \draw[Aactivity] (0:0.5) -- (0:0.8);
    \node [yshift=-0.5pt,xshift=0pt,fill=white] {$\cardinality{A}$};
}
\quad
\text{ and }
\quad
\tikz[baseline=-2.5pt]{
    \begin{scope}[rotate=-30]
    \draw[Aactivity] (-130:0.5) -- (-130:1.3);
    \path [polarity={decorate,decoration={raise=0ex,text along path,
     text={|\large|...}}}]
     (-152:1.2cm) arc (-152:-130:1.2cm);
     \draw[Aactivity] (-160:0.5) -- (-160:1.3);
     \draw[Aactivity] (-170:0.5) -- (-170:1.3);
    \end{scope}
    \draw[thick,pattern=dots] (0,0) circle (0.5cm);
    \draw[Aactivity] (0:0.5) -- (0:0.8);
    \node [yshift=-0.5pt,xshift=0pt,fill=white] {$\cardinality{B}$};
}
\]
that have $\cardinality{A}$ and $\cardinality{B}$ external legs respectively, each parameterised by
the momenta given by the subsets, $\{\kvec_q | q\in \Pset{A}\}$ and $\{\kvec_q | q\in \Pset{B}\}$ respectively.
These diagrams can be ``stitched together'' via the Dean vertex, \Eref{dean_vertex}, so that
\begin{align}
\tikz[baseline=-2.5pt]{
    \begin{scope}
      \draw[Aactivity] (0,0.0) -- (0.5,0.0) +(2pt,0) circle (2pt)[fill];
      \draw[Aactivity] (220:0.7) -- (0,0.0);
      \draw[Aactivity] (140:0.7) -- (0,0.0);
      \draw[dotted,thin] (140:0.3) arc (140:220:0.3);
      \draw[Aactivity] ($(0.07,0.07)+(140:0.3)$) -- ($(-0.07,-0.07)+(140:0.3)$);
      \draw[Aactivity] ($(-0.07,0.07)+(220:0.3)$) -- ($(0.07,-0.07)+(220:0.3)$);
     \end{scope}
    \begin{scope}[xshift=-0.99cm,yshift=0.5cm]
    \begin{scope}[rotate=-30]
      \draw[Aactivity] (-130:0.45) -- (-130:1);
      \path [polarity={decorate,decoration={raise=0ex,text along path,
      text={|\large|...}}}]
      (-152:0.8cm) arc (-152:-130:0.8cm);
      \draw[Aactivity] (-160:0.45) -- (-160:1);
      \draw[Aactivity] (-170:0.45) -- (-170:1);
    \end{scope}
    \draw[thick,pattern=dots] (0,0) circle (0.45cm);
    \node [yshift=-0.5pt,xshift=0pt,fill=white] {$\cardinality{A}$};
    \end{scope}
    \begin{scope}[xshift=-0.99cm,yshift=-0.5cm]
    \begin{scope}[rotate=-30]
      \draw[Aactivity] (-130:0.45) -- (-130:1);
      \path [polarity={decorate,decoration={raise=0ex,text along path,
      text={|\large|...}}}]
      (-152:0.8cm) arc (-152:-130:0.8cm);
      \draw[Aactivity] (-160:0.45) -- (-160:1);
      \draw[Aactivity] (-170:0.45) -- (-170:1);
    \end{scope}
    \draw[thick,pattern=dots] (0,0) circle (0.45cm);
    \node [yshift=-0.5pt,xshift=0pt,fill=white] {$\cardinality{B}$};
    \end{scope}
}   
\corresponds& n_0\theta(t-t_0) 
\knullfix{\kvec_1+\ldots+\kvec_n}
\int_{t_0}^{t} \dint{t'} \, (-2 D \setvec{A}  \cdot \setvec{B} )  e^{-D(t'-t_0)(\kvec_1+...+\kvec_n)^2}\nonumber\\
& \, \times \left[ \sum_{a=1}^{\cardinality{A}} (-1)^{a-1} (a-1)! \sum_{\{\Pset{P}_1,\ldots,\Pset{P}_a\}\in \partition{A}{a}} \exp{-D(t-t') \sum_{i=1}^a \setvecX{\Pset{P}_i}^2}\right]\elabel{stitching}\\
& \, \times \left[ \sum_{b=1}^{\cardinality{B}} (-1)^{b-1} (b-1)! \sum_{\{\Pset{Q}_1,\ldots,\Pset{Q}_b\}\in \partition{B}{b}} \exp{-D(t-t') \sum_{j=1}^b \setvecX{\Pset{Q}_j}^2}\right] \nonumber\\
=& n_0\theta(t-t_0) 
\knullfix{\kvec_1+\ldots+\kvec_n}
\sum_{a=1}^{\cardinality{A}} \sum_{b=1}^{\cardinality{B}} (-1)^{a+b} (a-1)! (b-1)! (-2 D \setvec{A}  \cdot \setvec{B} ) \nonumber\\
& \, \times \!\!\sum_{\{\Pset{P}_1,\ldots\}\in \partition{A}{a}} \!\! \sum_{\{\Pset{Q}_1,\ldots\}\in \partition{B}{b}} \!\! \exp{D(\kvec_1+...+\kvec_n)^2 t_0} 
   \exp{-Dt \sum_{i=1}^a \setvecX{\Pset{P}_i}^2 } \exp{-Dt \sum_{j=1}^b \setvecX{\Pset{Q}_j}^2 } \nonumber \\
& \, \times \int_{t_0}^t \dint{t'} \, \operatorname{exp}\left[ Dt'\left( \sum_{i=1}^a \setvecX{\Pset{P}_i}^2  + \sum_{j=1}^b \setvecX{\Pset{Q}_j}^2  - (\kvec_1 + \dots + \kvec_n)^2 \right)\right]  \nonumber ~.
\end{align}
On the left is a diagram with $\cardinality{A}+\cardinality{B}$ legs and on the right we use \Eref{a_induc_hyp} for diagrams with fewer legs, because neither $\Pset{A}$ nor $\Pset{B}$ can be empty. If we can show that the sum of all diagrams on the left obeys \Eref{a_induc_hyp}, then the induction step is completed.

Since $\union_{i=1}^a\Pset{P}_i=\Pset{A}$, we have from \Eref{def_setvec} that
$\sum_{i=1}^a \setvecX{\Pset{P}_i} = \setvec{A}$ and similarly 
$\sum_{j=1}^b \setvecX{\Pset{Q}_j} = \setvec{B}$ and further 
$\setvec{A}+\setvec{B}=\setvecX{\{1,\dots,n\}}=\kvec_1+\dots\kvec_n$, so that 
the exponent in square brackets appearing within the $t'$ integral in the last line of \Eref{stitching}
can be rearranged as follows:
\begin{multline}\elabel{vector_cross-terms}
   \sum_{i=1}^a \setvecX{\Pset{P}_i}^2  + \sum_{j=1}^b \setvecX{\Pset{Q}_j}^2  - (\kvec_1 + \dots + \kvec_n)^2
   \\
   =
   -2 \left(  \setvec{A} \cdot \setvec{B} 
            + \sum_{i=1}^a\sum_{e=i+1}^a \setvecX{\Pset{P}_i}\cdot\setvecX{\Pset{P}_e}
            + \sum_{j=1}^b\sum_{f=j+1}^b \setvecX{\Pset{Q}_j}\cdot\setvecX{\Pset{Q}_f}
	    \right) \ ,
\end{multline}
with the nested double summations generating all cross-terms once. In fact, the bracket on the right hand side of \Eref{vector_cross-terms} is the sum of the vector products of all $ab+a(a-1)/2+b(b-1)/2=(a+b)(a+b-1)/2$ distinct pairs of vectors generated with \Eref{def_setvec} from the $a+b$ sets $\{\Pset{P}_1,\dots,\Pset{P}_a,\Pset{Q}_1,\dots,\Pset{Q}_b\}$.
With \Eref{vector_cross-terms} we arrive at
\begin{align}
&\tikz[baseline=-2.5pt]{
    \begin{scope}
      \draw[Aactivity] (0,0.0) -- (0.7,0.0) +(2pt,0) circle (2pt)[fill];
      \draw[Aactivity] (220:0.7) -- (0,0.0);
      \draw[Aactivity] (140:0.7) -- (0,0.0);
      \draw[dotted,thin] (140:0.3) arc (140:220:0.3);
      \draw[Aactivity] ($(0.07,0.07)+(140:0.3)$) -- ($(-0.07,-0.07)+(140:0.3)$);
      \draw[Aactivity] ($(-0.07,0.07)+(220:0.3)$) -- ($(0.07,-0.07)+(220:0.3)$);
     \end{scope}
    \begin{scope}[xshift=-0.99cm,yshift=0.5cm]
    \begin{scope}[rotate=-30]
      \draw[Aactivity] (-130:0.45) -- (-130:1);
      \path [polarity={decorate,decoration={raise=0ex,text along path,
      text={|\large|...}}}]
      (-152:0.8cm) arc (-152:-130:0.8cm);
      \draw[Aactivity] (-160:0.45) -- (-160:1);
      \draw[Aactivity] (-170:0.45) -- (-170:1);
    \end{scope}
    \draw[thick,pattern=dots] (0,0) circle (0.45cm);
    \node [yshift=-0.5pt,xshift=0pt,fill=white] {$\cardinality{A}$};
    \end{scope}
    \begin{scope}[xshift=-0.99cm,yshift=-0.5cm]
    \begin{scope}[rotate=-30]
      \draw[Aactivity] (-130:0.45) -- (-130:1);
      \path [polarity={decorate,decoration={raise=0ex,text along path,
      text={|\large|...}}}]
      (-152:0.8cm) arc (-152:-130:0.8cm);
      \draw[Aactivity] (-160:0.45) -- (-160:1);
      \draw[Aactivity] (-170:0.45) -- (-170:1);
    \end{scope}
    \draw[thick,pattern=dots] (0,0) circle (0.45cm);
    \node [yshift=-0.5pt,xshift=0pt,fill=white] {$\cardinality{B}$};
    \end{scope}
} 
\corresponds  n_0 \theta(t-t_0) 
\knullfix{\kvec_1+\ldots+\kvec_n}
\sum_{a=1}^{\cardinality{A}} \sum_{b=1}^{\cardinality{B}} (-1)^{a+b} (a-1)! (b-1)! 
\nonumber \\
& \times  
\!\!\!\!\!\!\!\!\!\sum_{\{\Pset{P}_1,\ldots\}\in \partition{A}{a}} \sum_{\{\Pset{Q}_1,\ldots\}\in \partition{B}{b}}
\frac
{\setvec{A} \cdot \setvec{B}}
{\setvec{A} \cdot \setvec{B}
            + \sum\limits_{i=1}^a\sum\limits_{e=i+1}^a \setvecX{\Pset{P}_i}\cdot\setvecX{\Pset{P}_e}
            + \sum\limits_{j=1}^b\sum\limits_{f=j+1}^b \setvecX{\Pset{Q}_j}\cdot\setvecX{\Pset{Q}_f}}
\elabel{proof4} \\
& \qquad\qquad \times  
\left[ e^{-D (t-t_0) (\kvec_1+...+\kvec_n)^2} - e^{-D (t-t_0) \left(   \sum_{i=1}^a \setvecX{\Pset{P}_i}^2  + \sum_{j=1}^b \setvecX{\Pset{Q}_j}^2 \right) } \right]  \nonumber \,.
\end{align}
The main obstacle to simplify \Eref{proof4} further at this point is the fraction of scalar products of $\setvec{\cdot}$'s. Similar to the 3-point case, \Eref{3pt_kt_dean},  this will simplify only once we consider the sum over all diagrams with non-equivalent permutations of the external momenta. Since the expressions inserted for sub-diagrams already take care of permutations within each subdiagram we need to consider only different partitionings of the indices $1,...,n$ into subsets $\Pset{A}$ and $\Pset{B}$. Diagrammatically, for $n\ge2$,
\begin{align}
\tikz[baseline=-2.5pt]{
\begin{scope}[rotate=-30]
  \draw[Aactivity] (-130:0.5) -- (-130:1.3);
  \path [polarity]
  (-152:1.2cm) arc (-152:-130:1.2cm);
  \draw[Aactivity] (-160:0.5) -- (-160:1.3);
  \draw[Aactivity] (-170:0.5) -- (-170:1.3);
\end{scope}
\draw[thick,pattern=dots] (0,0) circle (0.5cm);
\draw[Aactivity] (0:0.5) -- (0:1.2) +(2pt,0) circle (2pt)[fill];
\node [yshift=-0.5pt,xshift=0pt,fill=white] {$n$};
}
= 
\sum_{\{\Pset{A},\Pset{B}\} \in \partitionX{\{1,\dots,n\}}{2}}
\tikz[baseline=-2.5pt]{
    \begin{scope}
      \draw[Aactivity] (0,0.0) -- (0.7,0.0) +(2pt,0) circle (2pt)[fill];
      \draw[Aactivity] (220:0.7) -- (0,0.0);
      \draw[Aactivity] (140:0.7) -- (0,0.0);
      \draw[dotted,thin] (140:0.3) arc (140:220:0.3);
      \draw[Aactivity] ($(0.07,0.07)+(140:0.3)$) -- ($(-0.07,-0.07)+(140:0.3)$);
      \draw[Aactivity] ($(-0.07,0.07)+(220:0.3)$) -- ($(0.07,-0.07)+(220:0.3)$);
     \end{scope}
    \begin{scope}[xshift=-0.99cm,yshift=0.5cm]
    \begin{scope}[rotate=-30]
      \draw[Aactivity] (-130:0.45) -- (-130:1);
      \path [polarity={decorate,decoration={raise=0ex,text along path,
      text={|\large|...}}}]
      (-152:0.8cm) arc (-152:-130:0.8cm);
      \draw[Aactivity] (-160:0.45) -- (-160:1);
      \draw[Aactivity] (-170:0.45) -- (-170:1);
    \end{scope}
    \draw[thick,pattern=dots] (0,0) circle (0.45cm);
    \node [yshift=-0.5pt,xshift=0pt,fill=white] {$\cardinality{A}$};
    \end{scope}
    \begin{scope}[xshift=-0.99cm,yshift=-0.5cm]
    \begin{scope}[rotate=-30]
      \draw[Aactivity] (-130:0.45) -- (-130:1);
      \path [polarity={decorate,decoration={raise=0ex,text along path,
      text={|\large|...}}}]
      (-152:0.8cm) arc (-152:-130:0.8cm);
      \draw[Aactivity] (-160:0.45) -- (-160:1);
      \draw[Aactivity] (-170:0.45) -- (-170:1);
    \end{scope}
    \draw[thick,pattern=dots] (0,0) circle (0.45cm);
    \node [yshift=-0.5pt,xshift=0pt,fill=white] {$\cardinality{B}$};
    \end{scope}
}   \elabel{proof3} \,,
\end{align}
where $\Pset{A}$ and $\Pset{B}$ are again the sets of indices of the momenta associated with each part of the partition.
The external fields of the sub-diagram labelled $\cardinality{A}$ have momenta $\kvec_a$ with $a\in\Pset{A}$ and correspondingly for the other sub-diagram. 
The cardinality $\cardinality{A}$ of the non-empty partition $\Pset{A}$, ranges from $1$ to $n-1$. The cardinality of the non-empty partition $\Pset{B}$ is then given by $\cardinality{B} = n - \cardinality{A}$.
Using \Eref{proof4} for the diagrams summed over in \Eref{proof3}, we re-organise the partitioning, as explained below, and rewrite it as 
\begin{subequations}
\elabel{resum_rule}
\begin{align}
&
\tikz[baseline=-2.5pt]{
\begin{scope}[scale=1.]
\begin{scope}[rotate=-30]
  \draw[Aactivity] (-130:0.5) -- (-130:1.3);
  \path [polarity]
  (-152:1.2cm) arc (-152:-130:1.2cm);
  \draw[Aactivity] (-160:0.5) -- (-160:1.3);
  \draw[Aactivity] (-170:0.5) -- (-170:1.3);
\end{scope}
\draw[thick,pattern=dots] (0,0) circle (0.5cm);
\draw[Aactivity] (0:0.5) -- (0:1.2) +(2pt,0) circle (2pt)[fill];
\node [yshift=-0.5pt,xshift=0pt,fill=white] {$n$};
\end{scope}
} \nonumber
\\
\elabel{resum_rule_line2}
 &\corresponds 
 \!\!\!\!\!\!
\sum_{\{\Pset{A},\Pset{B}\} \in \partitionX{\{1,\dots,n\}}{2}}
\sum_{a=1}^{\cardinality{A}} 
\sum_{\{\Pset{P}_1,\ldots\}\in \partition{A}{a}} 
\sum_{b=1}^{\cardinality{B}}
\sum_{\{\Pset{Q}_1,\ldots\}\in \partition{B}{b}}
 \!\!\!\!\!\!
\fguts(\{\Pset{P}_1,\dots,\Pset{P}_a\},\{\Pset{Q}_1,\dots,\Pset{Q}_b\})
\\
\elabel{resum_rule_line3}
&=  \sum_{m=2}^n 
\sum_{\{\Pset{W}_1,\dots,\Pset{W}_m\} \in \partitionX{\{1,\dots,n\}}{m}} 
\left[ \sum_{\{\Pset{T}_A,\Pset{T}_B\} \in \partitionX{\{1,\dots,m\}}{2}} 
\fguts\left( \bigcup\limits_{t\in\Pset{T}_A} \{\Pset{W}_t\},  \bigcup\limits_{t\in\Pset{T}_B} \{\Pset{W}_t\} \right)
\right] \,,
\end{align}
\end{subequations}
where $\fguts$ is given by
\begin{multline}\elabel{def_fguts}
\fguts(\{\Pset{P}_1,\dots,\Pset{P}_a\},\{\Pset{Q}_1,\dots,\Pset{Q}_b\}) 
\\
=
  \frac{ n_0 \theta(t-t_0) 
  \knullfix{\kvec_1+\ldots+\kvec_n}
  (-1)^{a+b}\,(a-1)!\,(b-1)!\,
\setvecX{\bigcup\limits_{i=1}^a\Pset{P}_i} \cdot \setvecX{\bigcup\limits_{i=1}^b\Pset{Q}_i}
}
{
\setvecX{\bigcup\limits_{i=1}^a\Pset{P}_i} \cdot \setvecX{\bigcup\limits_{i=1}^b\Pset{Q}_i}
+ \sum\limits_{i=1}^a\sum\limits_{e=i+1}^a \setvecX{\Pset{P}_i}\cdot\setvecX{\Pset{P}_e}
+ \sum\limits_{j=1}^b\sum\limits_{f=j+1}^b \setvecX{\Pset{Q}_j}\cdot\setvecX{\Pset{Q}_f}}\\
\times \left[ e^{-D (t-t_0) (\kvec_1+...+\kvec_n)^2} - e^{-D (t-t_0) \left(   \sum_{i=1}^a \setvecX{\Pset{P}_i}^2  + \sum_{j=1}^b \setvecX{\Pset{Q}_j}^2 \right) } \right]  \,.
\end{multline}
The parameters $a$ and $b$ are the cardinalities of the first and the second partition in the argument of $\fguts$. This function depends on two partitions of two sets, $\Pset{A}$ and $\Pset{B}$, and it is invariant under exchange of its two arguments, which are sets of sets. On the basis of the partitions and the globally known $\kvec_1,\dots,\kvec_n$, all the vectors on the right of \Eref{def_fguts} can be constructed, so that $\fguts$ is solely a function of the two partitions.

Both sides of \Eref{resum_rule} are performing the same summation, based on the five sums from \Erefs{proof4} and \eref{proof3}. Both summations generate all possible ways of partitioning the $n$ external legs into two or more subsets. In fact there is a one-to-one correspondence between every term in the two sums, as we will demonstrate below.
The first sum in \Eref{resum_rule_line2} considers all partitions $\partitionX{\{1,\dots,n\}}{2}$ of the full set of indices $\{1,\dots,n\}$ into two sets, $\Pset{A}$ and $\Pset{B}$. These indicate the momenta the two subdiagrams shown in \Erefs{proof4} and \eref{proof3} depend on. To calculate these two subdiagrams all partitions of $\Pset{A}$ and $\Pset{B}$ need to be summed over, which is done in the remaining four sums. The right-hand side \Eref{resum_rule_line3} of \Eref{resum_rule} performs the same summation, but first produces all partitions of all $\{1,\dots,n\}$ into $m=2,\dots,n$ non-empty subsets $\{\Pset{W}_1,\dots,\Pset{W}_m\}$. In the rightmost sum, these subsets are distributed among the upper and the lower subdiagram by perfoming a partition into two subsets $\Pset{T}_A$ and $\Pset{T}_B$ on the indexing $\{1,\dots,m\}$ of the subsets $\Pset{W}_t$. These selections of subsets enter into the function $\fguts$, 
with, for example, the upper diagram being parameterised by the collection of sets
\begin{equation}
    \bigcup_{t\in\Pset{T}_A} \{\Pset{W}_t\}
    =
    \big\{
    \Pset{W}_{t_1}, \Pset{W}_{t_2},\ldots 
    \big\}
    \ne
    \bigcup_{t\in\Pset{T}_A} \Pset{W}_t
    \quad\text{ for }\quad
    \Pset{T}_A=\{t_1,t_2,\ldots\} \ .
\end{equation}

Both summations of \Eref{resum_rule} generate all
partitions of the indices and their division into an upper and a lower subdiagram. Any term appearing on the left can be found on the right and vice versa. A set of parameters $\{\Pset{P}_1,\dots,\Pset{P}_a\}$ and $\{\Pset{Q}_1,\dots,\Pset{Q}_b\}$ on the left
is found on the right when $m=a+b$ and $\{\Pset{W}_1,\dots,\Pset{W}_m\}=\{\Pset{P}_1,\dots,\Pset{P}_a\} \union \{\Pset{Q}_1,\dots,\Pset{Q}_b\}$ are the same partition of $\{1,\dots,n\}$, with exactly one of the partitions $\Pset{T}_A,\Pset{T}_B$ of the elements of $\{\Pset{W}_1,\dots\}$ such that 
$\union_{t\in\Pset{T}_A} \{\Pset{W}_t\}=\{\Pset{P}_1,\dots,\Pset{P}_a\}$
and
$\union_{t\in\Pset{T}_B} \{\Pset{W}_t\}=\{\Pset{Q}_1,\dots,\Pset{Q}_b\}$
or equivalenty
$\union_{t\in\Pset{T}_A} \{\Pset{W}_t\}=\{\Pset{Q}_1,\dots,\Pset{Q}_b\}$
and
$\union_{t\in\Pset{T}_B} \{\Pset{W}_t\}=\{\Pset{P}_1,\dots,\Pset{P}_a\}$.
Similarly, the term generated by the partition $\union_{t\in\Pset{T}_A} \{\Pset{W}_t\}, \union_{t\in\Pset{T}_B} \{\Pset{W}_t\}$
on the right, can be identified on the left, as the one where 
$\Pset{A}=\union_{t\in\Pset{T}_A} \Pset{W}_t$ and $\Pset{B}=\union_{t\in\Pset{T}_B} \Pset{W}_t$
or vice versa, which are both sets, not partitions. They need to be partitioned subsequently, for example
$\Pset{A}$ into $\{\Pset{P}_1,\dots,\Pset{P}_a\}=\union_{t\in\Pset{T}_A}\Pset{W}_t$ and
$\Pset{B}$ into $\{\Pset{P}_1,\dots,\Pset{P}_b\}=\union_{t\in\Pset{T}_B}\Pset{W}_t$
or equally
$\Pset{A}$ into $\{\Pset{P}_1,\dots,\Pset{P}_a\}=\union_{t\in\Pset{T}_B}\Pset{W}_t$ and
$\Pset{B}$ into $\{\Pset{P}_1,\dots,\Pset{P}_b\}=\union_{t\in\Pset{T}_A}\Pset{W}_t$.

Writing 
$\Pset{T}_A=\{\alpha_1, \alpha_2, \dots, \alpha_a\}$ and $\Pset{T}_B=\{\beta_1, \beta_2, \dots, \beta_b\}$,
the parameterisation of the right-hand side of \Eref{resum_rule} allows us to express the denominator of $\fguts$, \Eref{def_fguts}, succinctly in terms of the new partition  $\{\Pset{W}_1,\dots,\Pset{W}_m\}$,
\begin{multline}\elabel{sum_of_crosses_simplified}
\setvecX{ \union_{t\in\Pset{T}_A} \{\Pset{W}_t\}}
\cdot
\setvecX{ \union_{t\in\Pset{T}_B} \{\Pset{W}_t\}}
+ 
\sum_{i=1}^a \sum_{e=i+1}^a \setvecX{\Pset{W}_{\alpha_i}}\cdot\setvecX{\Pset{W}_{\alpha_e}}
+
\sum_{j=1}^b \sum_{f=j+1}^b \setvecX{\Pset{W}_{\beta_j}} \cdot\setvecX{\Pset{W}_{\beta_f}}\\
=
\sum_{u=1}^m \sum_{v=u+1}^m \setvecX{\Pset{W}_u}\cdot\setvecX{\Pset{W}_v}
\end{multline}
as this is the sum of all cross-terms in the square of $\sum_{t=1}^m \setvecX{\Pset{W}_t}$, as
alluded to after \Eref{vector_cross-terms}. Further, the sum of the squares in the exponent of the right-most exponential in \Eref{def_fguts} can be written as
\begin{equation}\elabel{sum_of_squares_simplified}
\sum_{i=1}^a \setvecX{\Pset{W}_{\alpha_i}}^2
+
\sum_{j=1}^b \setvecX{\Pset{W}_{\beta_j}}^2
=
\sum_{u=1}^m \setvecX{\Pset{W}_u}^2
\end{equation}
as $\Pset{T}_A\union\Pset{T}_B=\{1,\dots,m\}$. Because the right-hand sides of \Erefs{sum_of_crosses_simplified} and \eref{sum_of_squares_simplified} are independent of the partitioning of $\{\Pset{W}_1,\dots,\Pset{W}_m\}$ via $\Pset{T}_A$ and $\Pset{T}_B$, they can be taken outside the sum over these partitions together with 
$(-1)^{\cardinalityX{\Pset{T}_A}+\cardinalityX{\Pset{T}_B}}=(-1)^m$,
\begin{align} \elabel{resum_rule_improved}
&\tikz[baseline=-2.5pt]{
\begin{scope}[scale=1.]
\begin{scope}[rotate=-30]
  \draw[Aactivity] (-130:0.5) -- (-130:1.3);
  \path [polarity]
  (-152:1.2cm) arc (-152:-130:1.2cm);
  \draw[Aactivity] (-160:0.5) -- (-160:1.3);
  \draw[Aactivity] (-170:0.5) -- (-170:1.3);
\end{scope}
\draw[thick,pattern=dots] (0,0) circle (0.5cm);
\draw[Aactivity] (0:0.5) -- (0:1.2) +(2pt,0) circle (2pt)[fill];
\node [yshift=-0.5pt,xshift=0pt,fill=white] {$n$};
\end{scope}
}
\corresponds  n_0 \theta(t-t_0)
\knullfix{\kvec_1+\ldots+\kvec_n}
\\
\nonumber
&\times \sum_{m=2}^n 
(-1)^m
\sum_{\{\Pset{W}_1,\dots,\Pset{W}_m\} \in \partitionX{\{1,\dots,n\}}{m}} 
\frac
{e^{-D (t-t_0) (\kvec_1+...+\kvec_n)^2} - e^{-D (t-t_0) \left(   \sum_{u=1}^m \setvecX{\Pset{W}_u}^2 \right) }}
{\sum_{u=1}^m \sum_{v=u+1}^m \setvecX{\Pset{W}_u}\cdot\setvecX{\Pset{W}_v}}
\\
\nonumber
&\times\left[ \sum_{\{\Pset{T}_A,\Pset{T}_B\} \in \partitionX{\{1,\dots,m\}}{2}} 
\!\!\!\!\!
\left(\cardinalityX{\Pset{T}_A}-1\right)!\,\left(\cardinalityX{\Pset{T}_B}-1\right)!\, 
\KBbar_A \cdot \KBbar_B
\right] \,
,
\end{align}
where we use the shorthands
\begin{subequations}
\begin{align}
\KBbar_A
&= \setvecX{ \bigcup\limits_{t\in\Pset{T}_A} \{\Pset{W}_t\}} = \sum_{t\in\Pset{T}_A} \setvecX{\Pset{W}_t} \\ 
\KBbar_B
&= \setvecX{ \bigcup\limits_{t\in\Pset{T}_B} \{\Pset{W}_t\}} = \sum_{t\in\Pset{T}_B} \setvecX{\Pset{W}_t} 
\end{align}
\end{subequations}

Next we want to simplify the sum in square brackets in \Eref{resum_rule_improved}. 
Since it is over all ways to partition $\{1,\dots,m\}$ into the two distinct sets that define $\KBbar_A$ and $\KBbar_B$,
we know that the sum over the products $\KBbar_A\cdots\KBbar_B$ 
will involve every cross-term $\setvecX{\Pset{W}_u}\cdot\setvecX{\Pset{W}_v}$ with $u\ne v$ at least once and by symmetry equally often.
The sum will therefore cancel with the denominator up to a pre-factor. In order to determine it,
we pick a particular scalar product, $\setvecX{\Pset{W}_u}\cdot\setvecX{\Pset{W}_v}$ for some fixed $u\ne v$
and consider those terms in the sum that contain $\setvecX{\Pset{W}_u}\cdot\setvecX{\Pset{W}_v}$:
\begin{multline}
C_{u,v}=
\setvecX{\Pset{W}_u}\cdot\setvecX{\Pset{W}_v}\\
\times\sum_{\{\Pset{T}_A,\Pset{T}_B\} \in \partitionX{\{1,\dots,m\}}{2}}
\mathcal{I}\Big(
(u\in\Pset{T}_A \land v\in\Pset{T}_B) 
\lor 
(u\in\Pset{T}_B \land v\in\Pset{T}_A) 
\Big)
\,(\cardinalityX{\Pset{T}_A}-1)!\,(\cardinalityX{\Pset{T}_B}-1)! \ ,
\end{multline}
where $\mathcal{I}(\ldots)$ is an indicator function that is $1$ only if
indices $u$ and $v$ are in different subsets and $0$ otherwise,
so that the square bracket in \Eref{resum_rule_improved} becomes $\sum_{u=1}^m \sum_{v=u+1}^m C_{u,v}$.
We may therefore simply sum over all partitions 
where $u$ and $v$ 
indeed \emph{are} in different subsets, for example $u$ in $\Pset{T}_A$ and $v$ in $\Pset{T}_B$ ---
there is no need to separately consider the case $u\in\Pset{T}_B$ and $v\in\Pset{T}_A$, as the resulting partitions are identical.
The make-up of the subsets enters only in as far as their cardinalities are concerned, which feature in the factorial.
If $n_A=\cardinalityX{\Pset{T}_A}>0$ is the cardinality of $\Pset{T}_A$, that leaves $n_B=m-n_A>0$ elements for $\Pset{T}_B$.
With one ``seat'' in $\Pset{T}_A$ given to $u$, 
there are $n_A-1$ further elements to be chosen from the $m-2$ elements in $\{1,\dots,m\}\setminus\{u,v\}$,
\begin{align}
C_{u,v}
&=\setvecX{\Pset{W}_u}\cdot\setvecX{\Pset{W}_v}
\sum_{n_A=1}^{m-1} {m-2 \choose n_A-1} \, (n_A-1)!\, (m-n_A-1)! \nonumber\\
&=\setvecX{\Pset{W}_u}\cdot\setvecX{\Pset{W}_v} \sum_{n_A=1}^{m-1} (m-2)!\,
=(m-1)!\,\setvecX{\Pset{W}_u}\cdot\setvecX{\Pset{W}_v} \ ,
\end{align}
which means that the square bracket in \Eref{resum_rule_improved} cancels with the denominator in the preceding fraction up to a factor $(m-1)!$, which can be taken outside the second sum,
\begin{align} \elabel{resum_rule_improved2}
&\tikz[baseline=-2.5pt]{
\begin{scope}[scale=1.]
\begin{scope}[rotate=-30]
  \draw[Aactivity] (-130:0.5) -- (-130:1.3);
  \path [polarity]
  (-152:1.2cm) arc (-152:-130:1.2cm);
  \draw[Aactivity] (-160:0.5) -- (-160:1.3);
  \draw[Aactivity] (-170:0.5) -- (-170:1.3);
\end{scope}
\draw[thick,pattern=dots] (0,0) circle (0.5cm);
\draw[Aactivity] (0:0.5) -- (0:1.2) +(2pt,0) circle (2pt)[fill];
\node [yshift=-0.5pt,xshift=0pt,fill=white] {$n$};
\end{scope}
}
\corresponds  n_0 \theta(t-t_0) 
\knullfix{\kvec_1+\ldots+\kvec_n}
\\
\nonumber
&\times \sum_{m=2}^n 
(-1)^m\,(m-1)!\!\!\!\!
\sum_{\{\Pset{W}_1,\dots,\Pset{W}_m\} \in \partitionX{\{1,\dots,n\}}{m}} 
\left(
e^{-D (t-t_0) (\kvec_1+...+\kvec_n)^2} - e^{-D (t-t_0) \left(   \sum_{u=1}^m \setvecX{\Pset{W}_u}^2 \right) }
\right) \ .
\end{align}
The first exponential in the final bracket is independent of the partition, so that the sum degenerates into the count of the ways a set of $n$ elements can be partitioned into $m$ non-empty sets, given by the Stirling number of the second kind, $\stirling{n}{m}$. 
We find with the help of
\cite[9.745.1]{GradshteynRyzhik:2007}
\begin{equation}\elabel{sum_of_partitions_with_wild_summand}
\sum_{m=2}^n (-1)^m\,(m-1)!\!\!\!\! \sum_{\{\Pset{W}_1,\dots,\Pset{W}_m\} \in \partitionX{\{1,\dots,n\}}{m}} 1
=
\sum_{m=2}^n (-1)^m\,(m-1)!\, \stirling{n}{m} 
=
1 \ .
\end{equation}

With that in place, we rewrite \Eref{resum_rule_improved2} as
\begin{align} \elabel{resum_rule_improved3}
&\tikz[baseline=-2.5pt]{
\begin{scope}[scale=1.]
\begin{scope}[rotate=-30]
  \draw[Aactivity] (-130:0.5) -- (-130:1.3);
  \path [polarity]
  (-152:1.2cm) arc (-152:-130:1.2cm);
  \draw[Aactivity] (-160:0.5) -- (-160:1.3);
  \draw[Aactivity] (-170:0.5) -- (-170:1.3);
\end{scope}
\draw[thick,pattern=dots] (0,0) circle (0.5cm);
\draw[Aactivity] (0:0.5) -- (0:1.2) +(2pt,0) circle (2pt)[fill];
\node [yshift=-0.5pt,xshift=0pt,fill=white] {$n$};
\end{scope}
}
\corresponds n_0 \theta(t-t_0) 
\knullfix{\kvec_1+\ldots+\kvec_n}
\\
\nonumber
&\times \left\{
e^{-D (t-t_0) (\kvec_1+...+\kvec_n)^2}
+
\sum_{m=2}^n 
(-1)^{m-1}\,(m-1)!\!\!\!\!
\sum_{\{\Pset{W}_1,\dots,\Pset{W}_m\} \in \partitionX{\{1,\dots,n\}}{m}} 
e^{-D (t-t_0) \left(   \sum_{u=1}^m \setvecX{\Pset{W}_u}^2 \right) }
\right\} \ .
\end{align}
The exponential of $-D (t-t_0) (\kvec_1+...+\kvec_n)^2$ in the curly brackets is the summand of the subsequent sum running over $m\ge2$ evaluated for $m=1$, because the only partition of $\{1,\dots,n\}$ into $m=1$ subsets is 
$\Pset{W}_1=\{1,\dots,n\}$ which produces the vector $\setvecX{\Pset{W}_1}=\kvec_1+...+\kvec_n$. It follows that
\begin{align} \elabel{resum_rule_final}
&\tikz[baseline=-2.5pt]{
\begin{scope}[scale=1.]
\begin{scope}[rotate=-30]
  \draw[Aactivity] (-130:0.5) -- (-130:1.3);
  \path [polarity]
  (-152:1.2cm) arc (-152:-130:1.2cm);
  \draw[Aactivity] (-160:0.5) -- (-160:1.3);
  \draw[Aactivity] (-170:0.5) -- (-170:1.3);
\end{scope}
\draw[thick,pattern=dots] (0,0) circle (0.5cm);
\draw[Aactivity] (0:0.5) -- (0:1.2) +(2pt,0) circle (2pt)[fill];
\node [yshift=-0.5pt,xshift=0pt,fill=white] {$n$};
\end{scope}
}
\corresponds  n_0 \theta(t-t_0) 
\knullfix{\kvec_1+\ldots+\kvec_n}
\\
\nonumber
&\times 
\sum_{m=1}^n 
(-1)^{m-1}\,(m-1)!\!\!\!\!
\sum_{\{\Pset{W}_1,\dots,\Pset{W}_m\} \in \partitionX{\{1,\dots,n\}}{m}} 
\left(
e^{-D (t-t_0) \left(   \sum_{u=1}^m \setvecX{\Pset{W}_u}^2 \right) }
\right) \ ,
\end{align}
which is \Eref{a_induc_hyp}. We have thus demonstrated that if \Eref{a_induc_hyp} holds for all diagrams with fewer than $n\ge2$ legs, as they enter into \Eref{stitching}, then \Eref{a_induc_hyp} also holds for the diagrams with $n$ legs. This concludes the induction step and together with the base case $n=1$ proves \Erefs{a_induc_hyp} and \eref{conjecture_Connected_k_t} for all $n\ge1$.

\section{Multiple starting points \label{a:mult_strt_pt}}
We generalise our calculation of the connected and full moments of the integrated particle number density in Dean's formalism, \Sref{dean_PI}, to the case where a total of $n_0$ particles are initialised at $H \leq n_0$ distinct sites. We upgrade the previous derivation by replacing in the action \Erefs{Dean_action_simple_komega} and \eref{Dean_action_mixed} and correspondingly in \Erefs{def_knullfix} and \eref{conjecture_Connected_k_t}
\begin{align}
n_0 \exp{i\kvec_0 \xvec_0} \qquad \text{by} \qquad \sum_{h=1}^{H} n_{0,h} \exp{i \kvec_0 \xvec_{0,h}}\elabel{replacing}
\end{align}
where $x_{0,h}$ for $h=1,...,H$ denote the $n_{0,h}\in \mathbb{N}$ particles' initial positions such that 
\begin{equation}
    \sum_{h=1}^H n_{0,h} = n_0~.
\end{equation}

In \Sref{dean_PI} and Appendix~\ref{a:induction_conn} we were entirely concerned with connected diagrams, where $n_0\exp{\imag \xvec_0\cdot\kvec_0}$ only ever enters linearly. Replacing it according to \Eref{replacing} renders each such diagram a sum over the $H$ distinct locations, each such sum still to be considered a single \emph{connected} diagram. 
This equally applies to the central result \Eref{n_point_connected}, which now reads
\begin{align}
\sum_{h=1}^{H}
\tikz[baseline=-2.5pt]{
\begin{scope}[rotate=-30]
  \draw[Aactivity] (-130:0.5) -- (-130:1.3);
  \path [polarity]
  (-152:1.2cm) arc (-152:-130:1.2cm);
  \draw[Aactivity] (-160:0.5) -- (-160:1.3);
  \draw[Aactivity] (-170:0.5) -- (-170:1.3);
\end{scope}
\draw[thick] (0,0) circle (0.5cm);
\draw[Aactivity] (0:0.5) -- (0:0.8)+(2pt,0) circle (2pt)[fill];
\node [yshift=-0.5pt,xshift=0pt] {$n$};
\node [yshift=5pt,xshift=1cm] {$n_{0,h}$};
}
=-\sum^H_{h=1}n_{0,h}\theta(t-t_0)\sum^n_{m=1}\left(I_{\Omega,h}(t-t_0)\right)^m(m-1)!\stirling{n}{m}
\elabel{eq:conn_mom_multisource}
\end{align}
where $I_{\Omega,h}(t-t_0)$ denotes the transition probability from the starting point $x_{0,h}$ into the set $\Omega$. 

The full moments of the integrated particle number density for distinct starting points can also be derived straighforwardly following the calculation in  \Erefs{con_der_1}--\eref{con_der_3}. Using \Eref{eq:conn_mom_multisource} for the associated connected moments, we arrive at
\begin{align}
    \tikz[baseline=-2.5pt]{
\begin{scope}[rotate=-30]
  \draw[Aactivity] (-130:0.5) -- (-130:1.3);
  \path [polarity]
  (-152:1.2cm) arc (-152:-130:1.2cm);
  \draw[Aactivity] (-160:0.5) -- (-160:1.3);
  \draw[Aactivity] (-170:0.5) -- (-170:1.3);
\end{scope}
\fill[pattern=north west lines,opacity=.6,draw] 
  (0,0) circle [radius=0.5cm];
\node [yshift=-0.5pt,xshift=0pt,fill=white] {$n$};
}
\corresponds& \left. \frac{d^n}{dz^n} \right|_{z=0} \prod_{h=1}^H \left[1 + (e^z-1)I_{\Omega,h})\right]^{n_{0,h}} \elabel{con_der_3a}\\
    =& \sum_{j_1+...+j_H=n} \binom{n}{j_1,...,j_H} \prod_{h=1}^H \left. \frac{d^{j_h}}{dz^{j_h}} \right|_{z=0} \left[1 + (e^z-1)I_{\Omega,h}\right]^{n_{0,h}} \elabel{con_der_4a}\\
    =& \sum_{j_1+...+j_H=n} \binom{n}{j_1,...,j_H} \prod_{h=1}^H \sum_{k_h=0}^{n_{0,h}}
    {n_{0,h} \choose k_h} \left(I_{\Omega,h}\right)^{k_h} k_h! \stirling{j_h}{k_h}
    \ ,\elabel{con_der_5a}
\end{align}
where we have used the generalised product rule to go from \Eref{con_der_3a} to \Eref{con_der_4a} by swapping the differential operator into the product, as well as the intermediate result \Erefs{con_der_3_top}--\eref{con_der_4} for the last step. One can easily verify that \Eref{con_der_5a} reduces to \Eref{con_der_4} when $H=1$, \ie when all particles are initialised at the same point.

To show particle entity we use a similar procedure as in the single source case from \Eref{con_der_4} to \eref{particle_entity_full_moments},
\begin{align}
    \sum_{n=0}^\infty \frac{(2\pi\imag)^n}{n!} 
    \tikz[baseline=-2.5pt]{
\begin{scope}[rotate=-30]
  \draw[Aactivity] (-130:0.5) -- (-130:1.3);
  \path [polarity]
  (-152:1.2cm) arc (-152:-130:1.2cm);
  \draw[Aactivity] (-160:0.5) -- (-160:1.3);
  \draw[Aactivity] (-170:0.5) -- (-170:1.3);
\end{scope}
\fill[pattern=north west lines,opacity=.6,draw] 
  (0,0) circle [radius=0.5cm];
\node [yshift=-0.5pt,xshift=0pt,fill=white] {$n$};
}  \corresponds \sum_{n=0}^\infty \sum_{j_1+...+j_H=n} \frac{(2 \pi \imag)^n}{j_1!...j_H!} \prod_{h=1}^H \sum_{k_h=0}^{n_{0,h}}
{n_{0,h} \choose k_h} \left(I_{\Omega,h}\right)^{k_h} k_h! \stirling{j_h}{k_h} \ ,
\end{align}
where the second sum on the right runs over all non-negative integers $j_1,j_2,\ldots,j_H$ which sum to $n$.
We now use that $\sum_{n=0}^\infty \sum_{j_1+...+j_H=n} = \sum_{j_1=0}^\infty ... \sum_{j_H=0}^\infty$ as the sums converge individually and absolutely, and \Eref{binomial_rewrite} to obtain
\begin{align}
    \sum_{n=0}^\infty \frac{(2\pi\imag)^n}{n!} 
    \tikz[baseline=-2.5pt]{
\begin{scope}[rotate=-30]
  \draw[Aactivity] (-130:0.5) -- (-130:1.3);
  \path [polarity]
  (-152:1.2cm) arc (-152:-130:1.2cm);
  \draw[Aactivity] (-160:0.5) -- (-160:1.3);
  \draw[Aactivity] (-170:0.5) -- (-170:1.3);
\end{scope}
\fill[pattern=north west lines,opacity=.6,draw] 
  (0,0) circle [radius=0.5cm];
\node [yshift=-0.5pt,xshift=0pt,fill=white] {$n$};
}  &\corresponds  \sum_{j_1=0}^\infty ... \sum_{j_H=0}^\infty \frac{(2 \pi \imag)^{j_1+...+j_H}}{j_1!...j_H!} \prod_{h=1}^H \sum_{k_h=0}^{n_{0,h}} \binom{n_{0,h}}{k_h} 
(1-I_{\Omega,h} )^{n_{0,h}-k_h} (I_{\Omega,h} )^{k_h} k_h^{j_h} \\
& = \prod_{h=1}^H \left( \sum_{j_h = 0}^\infty \sum_{k_h=0}^{n_{0,h}} \binom{n_{0,h}}{k_h} 
(1-I_{\Omega,h})^{n_{0,h}-k_h} (I_{\Omega,h} )^{k_h} \frac{(2 \pi \imag k_h)^{j_h}}{j_h !} \right)\\
& = \prod_{h=1}^H \left(  \sum_{k_h=0}^{n_{0,h}} \binom{n_{0,h}}{k_h} 
(1-I_{\Omega,h})^{n_{0,h}-k_h} (I_{\Omega,h} )^{k_h} e^{2 \pi \imag k_h} \right)\\
& = \prod_{h=1}^H \left( 1\right)\\
& = 1 \, .
\end{align}
This completes the derivation of particle entity according to the criterion \Erefs{eq:PI_mom_gen} for particles initialised as multiple origins, \cf \Eref{particle_entity_full_moments}.
\end{document}